\documentclass[fleqn,usenatbib]{mnras}

\usepackage{newtxtext,newtxmath}

\usepackage[T1]{fontenc}

\usepackage{graphicx}	
\usepackage{amsmath}	

\usepackage{CJK}
\usepackage{color}
\usepackage[svgnames]{xcolor}
\usepackage{color}
\usepackage{soul}
\usepackage{amsmath} 
\usepackage{bm} 

\usepackage{pdflscape} 
\usepackage{mathrsfs}
\usepackage{wasysym}
\usepackage{amsmath}
\usepackage{threeparttable}
\usepackage{multirow} 
\usepackage{array}
\usepackage{lineno}

\DeclareRobustCommand{\VAN}[3]{#2}
\let\VANthebibliography\thebibliography
\def\thebibliography{\DeclareRobustCommand{\VAN}[3]{##3}\VANthebibliography}

\usepackage{graphicx}	
\usepackage{amsmath}	

\title[Compare methods in constraining $H_0$]{Comparative Analysis of EMCEE, Gaussian Process, and Masked Autoregressive Flow in Constraining the Hubble Constant Using Cosmic Chronometers Dataset}

\author[Jing Niu et al.]{
Jing Niu,$^{1,2,3}$ 
Jie-Feng Chen,$^{4,5}$ 
Peng He,$^{6}$ 
Tong-Jie Zhang,$^{2,3}$\thanks{E-mail: tjzhang@bnu.edu.cn} 
Jie Zhang $^{1}$\thanks{E-mail: zhangjie@qlnu.edu.cn}
\\
$^{1}$School of arts and sciences, Shanghai Dianji University, Shanghai, 200240, China\\
$^{2}$Institute for Frontiers in Astronomy and Astrophysics, Beijing Normal University, Beijing 102206, China\\
$^{3}$School of Physics and Astronomy, Beijing Normal University, Beijing 100875, China\\
$^{4}$Institut de F\'{\i}sica d’Altes Energies (IFAE), The Barcelona Institute of Science and Technology, Campus UAB, 08193 Bellaterra (Barcelona), Spain\\
$^{5}$Port d'Informaci\'{o} Cient\'{i}fica (PIC), Campus UAB, C. Albareda s/n, 08193 Bellaterra (Barcelona), Spain\\
$^{6}$Burerau of Frontier Science and Education, Chinese Academy of Sciences, Beijing 100190, China
}

\pubyear{\the\year{}}

\begin{document}
\begin{CJK*}{UTF8}{gbsn}

\label{firstpage}
\pagerange{\pageref{firstpage}--\pageref{lastpage}}
\maketitle

\begin{abstract}
The Hubble constant ($H_0$) is essential for understanding the universe's evolution. Different methods, such as Affine Invariant Markov chain Monte Carlo Ensemble sampler (EMCEE), Gaussian Process (GP), and Masked Autoregressive Flow (MAF), are used to constrain $H_0$ using $H(z)$ data. However, these methods produce varying $H_0$ values when applied to the same dataset. To investigate these differences, we compare the methods based on their sensitivity to individual data points and their performance in constraining $H_0$. We apply Monte Carlo delete-$d$ jackknife (MCDJ) to assess their sensitivity to individual data points. Our findings reveal that GP is more sensitive to individual data points than both MAF and EMCEE, with MAF being more sensitive than EMCEE. Sensitivity also depends on redshift: EMCEE and GP are more sensitive to $H(z)$ at higher redshifts, while MAF is more sensitive at lower redshifts. In simulation-based performance tests, we generate an ensemble of mock CC datasets with a fixed input truth $H_{0,\mathrm{true}}$, apply each method to recover $H_0$ posteriors, and summarise performance by comparing the recovered posterior to $H_{0,\mathrm{true}}$: (i) posterior central value accuracy (bias and RMSE), (ii) credible-interval calibration (68\% and 95\% coverage), and (iii) overall posterior quality (log score), under two simulation prescriptions ($\Lambda$CDM-based and GP-based). Overall, EMCEE performs best, GP is intermediate, and MAF performs worst across the performance metrics.
\end{abstract}




\section{Introduction} \label{sec:intro}

Understanding the evolution of the universe has been a paramount goal in modern cosmology, primarily achieved through various observational techniques. The Hubble parameter, $H(z)$, indicates the rate of expansion of the universe at a given redshift $z$, while $H_0$ represents the current expansion rate. Accurate measurement of $H_0$ is crucial for understanding the universe's history. Currently, there are two important methods for measuring the Hubble constant. The Planck mission's measurements of the cosmic microwave background (CMB), under the base-$\Lambda$CDM cosmology, yield $H_0 = 67.4 \pm 0.5 \, \mathrm{km \, s^{-1} \, Mpc^{-1}}$ \citep{2020A&A...641A...6P}. In contrast, measurements using Cepheid variables in the host galaxies of 42 Type Ia supernovae via the Hubble Space Telescope give $H_0 = 73.04 \pm 1.04 \, \mathrm{km \, s^{-1} \, Mpc^{-1}}$ \citep{2022ApJ...934L...7R}. This discrepancy, known as the Hubble tension, is significant at a 5$\sigma$ level \citep{2019NatAs...3..891V,2020NatRP...2...10R}.

Constraining $H_0$ using the Hubble parameter $H(z)$, which measures the expansion rate of the universe at different redshifts, is an alternative approach. This method could potentially provide insights into resolving the Hubble tension. In this paper, we use the Cosmic Chronometers (CC) dataset for $H(z)$ because the CC method provides model-independent $H(z)$ measurements. Several methods can be employed to constrain $H_0$ using the $H(z)$ data, including Affine Invariant Markov chain Monte Carlo Ensemble sampler (EMCEE) \citep{2013PASP..125..306F}, Gaussian Process (GP) \citep{2006gpml.book.....R, 2012JCAP...06..036S}, and Masked Autoregressive Flow (MAF) \citep{2017arXiv170507057P}. EMCEE is an affine-invariant ensemble sampler for Markov Chain Monte Carlo (MCMC), GP is a non-parametric method that allows for model-independent reconstruction of the Hubble parameter, and MAF is a deep learning-based approach capable of modeling complex distributions with neural networks. Previous studies have applied these methods, or closely related implementations, to constrain $H_0$ from $H(z)$-based data. For instance, \cite{2021JCAP...08..051R} obtained $H_0 = 67.73^{+3.078}_{-3.077}\ \mathrm{km\,s^{-1}\,Mpc^{-1}}$ using MCMC implemented with EMCEE, \cite{2018JCAP...04..051G} found $H_0 = 67.42 \pm 4.75\ \mathrm{km\,s^{-1}\,Mpc^{-1}}$ using Gaussian Processes, and \cite{2021ApJS..254...43W} derived $H_0 = 68.68^{+5.12}_{-5.07}\ \mathrm{km\,s^{-1}\,Mpc^{-1}}$ using MAF. These results provide useful context for the present work and motivate a controlled comparison in which EMCEE, GP, and MAF are applied to the same CC dataset under identical data conditions.

We compare EMCEE, GP, and MAF along two complementary dimensions: sensitivity to individual $H(z)$ measurements and performance in controlled simulations with known ground truth. For the sensitivity analysis, we apply the Monte Carlo delete-$d$ jackknife (MCDJ) and further examine the dependence on redshift range. For the simulation-based performance tests, we generate an ensemble of mock CC datasets with a fixed input truth $H_{0,\mathrm{true}}$, apply each method to recover $H_0$ posteriors, and summarise performance by comparing the recovered posterior to $H_{0,\mathrm{true}}$: accuracy (bias and RMSE), credible-interval calibration (68\% and 95\% coverage), and overall posterior quality (log score). Two simulation approaches are employed to reduce dependence on any single mock-data prescription.

In Section \ref{sec:data}, we introduce the measurement of the CC dataset and compile it in Table \ref{tab:cc_data}. Section \ref{sec:Sensitivity Analysis} presents a comparison of the sensitivity to individual $H(z)$ data points using the three methods (EMCEE, GP, and MAF) through the MCDJ. Subsection \ref{subsec:CC constrains H0} introduces these methods, while Subsection \ref{subsec:Monte Carlo delete d jackknife} details the MCDJ and its application in comparing the methods. Subsection \ref{subsec:Seperate to different areas} examines the sensitivity of these methods to data points in different redshift regions. In Section \ref{sec:Simulation}, we evaluate method performance in the simulated $H_{\text{sim}}$ datasets with a fixed input truth $H_{0,\mathrm{true}}$, summarising results using accuracy, credible-interval calibration, and overall posterior quality. Finally, Section \ref{sec:discussions and conclusions} provides the conclusions and discussions of this study.

\section{data}\label{sec:data}

The Hubble parameter, $H(z)$, describes the rate of expansion of the universe and is defined by the equation:
\begin{equation}
H(z)=-\frac{1}{1+z}\frac{\mathrm{d}z}{\mathrm{d}t}.
\label{eq:1}
\end{equation}
One effective method for determining $H(z)$ is the differential age method, utilized in the cosmic chronometers approach. This method involves observing the ages of massive, passive galaxies to estimate the rate of change of redshift with time \citep{2002ApJ...573...37J, 2020ApJ...898...82M, 2023ApJS..265...48J}. The CC method is advantageous because it provides a model-independent way to determine the Hubble parameter, not relying on other cosmic probes. In this paper, we use the CC dataset, which offers independent $H(z)$ measurements. The CC $H(z)$ dataset used in this work is summarised in Table~\ref{tab:cc_data}, comprising $N=33$ measurements spanning $z \in [0.07,1.965]$. In the following, we represent the dataset as $\{z_i,\,H(z_i),\,\sigma_i\}$, with $H(z)$ in units of km\,s$^{-1}$\,Mpc$^{-1}$. In this work, we assume the CC measurements are independent and use only the published uncertainties $\sigma_i$. Potential correlated systematics in CC measurements are briefly discussed in Section~\ref{sec:discussions and conclusions}. In Subsection~\ref{subsec:CC constrains H0}, we describe how these $H(z)$ measurements are used to constrain $H_0$ with EMCEE, GP, and MAF.

\begin{table}
    \centering
    \caption{Compiled CC data.}
    \label{tab:cc_data}
    \begin{tabular}{ccc}
        \hline
        Redshift $z$ & $H(z)$ \textsuperscript{a} $\pm 1\sigma$ error & References \\
        \hline
        0.07 & $69 \pm 19.6$ & \cite{2014RAA....14.1221Z} \\
        0.09 & $69 \pm 12$ & \cite{2005PhRvD..71l3001S} \\
        0.12 & $68.6 \pm 26.2$ & \cite{2014RAA....14.1221Z} \\
        0.17 & $83 \pm 8$ & \cite{2005PhRvD..71l3001S} \\
        0.1791 & $75 \pm 4$ & \cite{2012JCAP...08..006M} \\
        0.1993 & $75 \pm 5$ & \cite{2012JCAP...08..006M} \\
        0.2 & $72.9 \pm 29.6$ & \cite{2014RAA....14.1221Z} \\
        0.27 & $77 \pm 14$ & \cite{2005PhRvD..71l3001S} \\
        0.28 & $88.8 \pm 36.6$ & \cite{2014RAA....14.1221Z} \\
        0.3519 & $83 \pm 14$ & \cite{2012JCAP...08..006M} \\
        0.382 & $83 \pm 13.5$ & \cite{2016JCAP...05..014M} \\
        0.4 & $95 \pm 17$ & \cite{2005PhRvD..71l3001S} \\
        0.4004 & $77 \pm 10.2$ & \cite{2016JCAP...05..014M} \\
        0.4247 & $87.1 \pm 11.2$ & \cite{2016JCAP...05..014M} \\
        0.4497 & $92.8 \pm 12.9$ & \cite{2016JCAP...05..014M} \\
        0.47 & $89 \pm 49.6$ & \cite{2017MNRAS.467.3239R} \\
        0.4783 & $80.9 \pm 9$ & \cite{2016JCAP...05..014M} \\
        0.48 & $97 \pm 62$ & \cite{2010JCAP...02..008S} \\
        0.5929 & $104 \pm 13$ & \cite{2012JCAP...08..006M} \\
        0.6797 & $92 \pm 8$ & \cite{2012JCAP...08..006M} \\
        0.7812 & $105 \pm 12$ & \cite{2012JCAP...08..006M} \\
        0.8 & $113.1 \pm 25.22$ & \cite{2023ApJS..265...48J} \\
        0.8754 & $125 \pm 17$ & \cite{2012JCAP...08..006M} \\
        0.88 & $90 \pm 40$ & \cite{2010JCAP...02..008S} \\
        0.9 & $117 \pm 23$ & \cite{2005PhRvD..71l3001S} \\
        1.037 & $154 \pm 20$ & \cite{2012JCAP...08..006M} \\
        1.26 & $135 \pm 65$ & \cite{2023AA...679A..96T} \\
        1.3 & $168 \pm 17$ & \cite{2005PhRvD..71l3001S} \\
        1.363 & $160 \pm 33.6$ & \cite{2015MNRAS.450L..16M} \\
        1.43 & $177 \pm 18$ & \cite{2005PhRvD..71l3001S} \\
        1.53 & $140 \pm 14$ & \cite{2005PhRvD..71l3001S} \\
        1.75 & $202 \pm 40$ & \cite{2005PhRvD..71l3001S} \\
        1.965 & $186.5 \pm 50.4$ & \cite{2015MNRAS.450L..16M} \\
        \hline
    \end{tabular}
    \vspace{3mm}
    \parbox{\linewidth}{\small
    \textsuperscript{a} $H(z)$ figures are in the unit of km s$^{-1}$ Mpc$^{-1}$.}
\end{table}

\section{Sensitivity Analysis} \label{sec:Sensitivity Analysis}

The Hubble constant $H_0$ is crucial for understanding the evolution of the universe. One approach to determine $H_0$ is to constrain it using the CC data compiled in Table \ref{tab:cc_data}. Several methods can be used to constrain $H(z)$, including EMCEE\citep{2013PASP..125..306F}, GP \citep{2012JCAP...06..036S} and MAF \citep{2017arXiv170507057P}, all of which are introduced in Subsection \ref{subsec:CC constrains H0}. The $H_0$ values derived from these three methods using the CC data vary. In Subsection \ref{subsec:Monte Carlo delete d jackknife}, we propose using MCDJ to compare the sensitivity of each method to individual $H(z)$ data points. We assess how $H(z)$ data points from different redshift regions affect the posterior summaries of $H_0$ across these methods. To do this, we split the redshift range into two regions, as outlined in Subsection \ref{subsec:Seperate to different areas}.

\subsection{CC constrains $H_0$} \label{subsec:CC constrains H0}

\subsubsection{EMCEE} \label{subsubsec:EMCEE}

To determine $H_0$ from CC data using \href{https://emcee.readthedocs.io/en/stable/index.html}{EMCEE}, it is essential to choose a cosmological model. In this paper, we adopt the flat $\Lambda$CDM cosmological model. The corresponding Friedmann equation is given by:
\begin{equation}
H(z)=H_0\sqrt{\Omega_M(1+z)^3
+(1-\Omega_M)}.
\label{eq:3}
\end{equation}
Where $H(z)$ represents the Hubble parameter, $H_0$ is the Hubble constant, $\Omega_M$ denotes the matter density, and $z$ indicates the redshift. The observed CC data, denoted as $\bm{H_{\mathrm{obs}}(z)}$, is presented in Table \ref{tab:cc_data}. It provides the values $\bm{H_{\mathrm{obs}}}=(H_{\mathrm{obs,1}},\cdot\cdot\cdot,H_{\mathrm{obs,N}})^T$, along with their corresponding errors denoted as $\bm{\sigma_{\mathrm{obs}}}=(\sigma_{\mathrm{obs,1}},\cdot\cdot\cdot,\sigma_{\mathrm{obs,N}})^\mathrm{T}$, at various redshifts $\bm{z}=(z_1,\cdot\cdot\cdot,z_\mathrm{N})^T$.

To constrain the parameters $H_0$ and $\Omega_M$, we use EMCEE, a method based on Bayes' theorem. The application of Bayes' theorem in this context is as follows:
\begin{equation}
        P(H_0,\Omega_M|\bm{H_{\mathrm{obs}}})=\frac{P(\bm{H_{\mathrm{obs}}}|H_0,\Omega_M)P(H_0,\Omega_M)}{P(\bm{H_{\mathrm{obs}}})}.
        \label{eq:4}
\end{equation}
We define the likelihood in Equation (\ref{eq:5}) and use EMCEE to sample the posterior $P(H_0,\Omega_M|\bm{H_{\mathrm{obs}}})$. The likelihood function, denoted by $\mathcal{L}$, is expressed as:
\begin{equation}
        \mathcal{L}(\bm{H_{\mathrm{obs}}}|H_0,\Omega_M)=\displaystyle\prod_{i=1}^{N}P(H_{\mathrm{obs,i}}|H_0,\Omega_M).
        \label{eq:5}
\end{equation}

Assuming that the measurement errors are Gaussian with standard deviations $\mathrm{\sigma_{obs,i}}$ and are independent of each other, Equation (\ref{eq:5}) can be rewritten as\citep{2011ApJ...730...74M,2021ApJS..254...43W}:
\begin{equation}
        \mathcal{L}(\bm{H_\mathrm{{obs}}}|H_0,\Omega_M)=\left(\overset{N}{\underset{i=1}{\prod}}\frac{1}{\sqrt{2\pi\sigma_i^2}}\right)\exp{\left(-\frac{\chi^2}{2}\right)},
        \label{eq:6}
\end{equation}
the $\chi^2$ is given by: 
\begin{equation}
       \chi^2=\underset{i}\sum\frac{[H_{\mathrm{model},i}-H_{\mathrm{obs},i}]^2}{\sigma_i^2}.
         \label{eq:7}
\end{equation}
We sample the posterior $P(H_0,\Omega_M|\bm{H_{\mathrm{obs}}})$ using an affine-invariant ensemble sampler with 33 walkers and 10000 production steps after an initial burn-in. Convergence is assessed using the acceptance fraction and the integrated autocorrelation time $\tau$, and we verify that the chain length satisfies $\mathrm{length}\gg\tau$ (approximately $300\,\tau$). We then summarise the marginal $H_0$ posterior by its sample median (central value) and by credible intervals obtained from posterior percentiles.

\subsubsection{GP} \label{subsubsec:GP}

$H_0$ can also be determined by reconstructing the CC data using the Gaussian process, as outlined by \citet{2006gpml.book.....R}. GP is a powerful tool that models the data by assuming that the reconstructed function values at different redshifts follow a joint Gaussian distribution. It estimates values at new points without requiring additional parameters. In this study, we use a Gaussian process to estimate the Hubble parameter, $H(z)$, as a function of redshift $z$. In practice, we take $H_0$ from the GP predictive distribution at $z=0$, summarised by its mean and standard deviation. This model-independent approach provides a major advantage over other methods that rely on cosmological assumptions.

The observed CC data ($z_i, H_i, \sigma_i$) can be represented as a function $\bm{y}$, assuming Gaussian measurement errors with standard deviations $\sigma_i$.
\begin{equation}
    \bm{y} \thicksim \mathcal{N} ( \bm{\mu}, \mathbfss{K}(\bm{Z},\bm{Z}) + \mathbfss{C} ),
    \label{eq:gp1}
\end{equation}
where $\mathcal{N}$ represents the evaluated Gaussian Process, and $\mathbfss{C}$ is the covariance matrix of the data. The mean and the covariance function between two data points, $y(z_i)$ and $y(z_j)$, are denoted by $\bm{\mu}$ and $\mathbfss{K}(\bm{Z},\bm{Z})$, respectively. There are various types of covariance functions available, as discussed by \citet{2023ApJS..266...27Z}. Their work compares the performance of different covariance functions in reconstructing cosmological data. In this study, we use the Squared Exponential covariance function, which is widely recognized and commonly applied in cosmology \citep{2012JCAP...06..036S,2021ApJS..254...43W,2021ApJ...915..123S}. Here, $[\mathbfss{K}(\bm{Z},\bm{Z})]_{ij} = k( z_i, z_j )$,
\begin{equation}
    k( z_i, z_j ) = \sigma^2_{f} \exp{\left( -\frac{( z_i, z_j )^2}{2l^2} \right)} .
    \label{eq:gp2}
\end{equation}
Here, $l$ represents the length scale, and $\sigma_f$ signifies the signal variance. Both are considered 'hyperparameters' in Equation (\ref{eq:gp2}).

Similarly, we can generate a Gaussian vector $\bm{f^{\ast}}$ at $\bm{Z^{\ast}}$:
\begin{equation}
    \bm{f^{\ast}} \thicksim \mathcal{N} ( \bm{\mu^{\ast}}, \mathbfss{K}(\bm{Z^{\ast}},\bm{Z^{\ast}}) ).
    \label{eq:gp3}
\end{equation}
The observed values of $\bm{y}$ can be obtained from the data. We can then reconstruct $\bm{f^{\ast}}$ ($\bm{\bar{f^{\ast}}}$ and cov($\bm{f^{\ast}}$)) by combining Equation (\ref{eq:gp1}) and (\ref{eq:gp3}) in the joint distribution \citep{2012JCAP...06..036S}:
\begin{equation}
    \begin{bmatrix}
        \bm{y}  \\
        \bm{f^{\ast}}  \\
    \end{bmatrix}
    \thicksim \mathcal{N}  
    \left( 
    \begin{bmatrix}
        \bm{\mu} \\
        \bm{\mu^{\ast}} \\
    \end{bmatrix}
    ,
    \begin{bmatrix}
        \mathbfss{K}(\bm{Z},\bm{Z}) + \mathbfss{C} & \mathbfss{K}(\bm{Z},\bm{Z^{\ast}}) \\
        \mathbfss{K}(\bm{Z^{\ast}},\bm{Z}) & \mathbfss{K}(\bm{Z^{\ast}},\bm{Z^{\ast}}) \\
    \end{bmatrix}    
    \right).
    \label{eq:gp4}
\end{equation}
From Equation (\ref{eq:gp4}), we can derive $\bm{\bar{f^{\ast}}}$ and cov($\bm{f^{\ast}}$) \citep{2006gpml.book.....R,2012JCAP...06..036S}:
\begin{equation}
    \bm{\bar{f^{\ast}}} = \bm{\mu^{\ast}} + \mathbfss{K}(\bm{Z^{\ast}},\bm{Z}) [\mathbfss{K}(\bm{Z},\bm{Z}) + \mathbfss{C}]^{-1} (\bm{y} - \bm{\mu}),
    \label{eq:gp5}
\end{equation}
\begin{align}
    \operatorname{cov}(\bm{f^{\ast}}) &= \mathbfss{K}(\bm{Z^{\ast}}, \bm{Z})
    - \mathbfss{K}(\bm{Z^{\ast}}, \bm{Z}) [\mathbfss{K}(\bm{Z}, \bm{Z}) + \mathbfss{C}]^{-1} \notag \\
    &\quad - \mathbfss{K}(\bm{Z}, \bm{Z^{\ast}}).
    \label{eq:gp6}
\end{align}
To reconstruct the functions in Equations (\ref{eq:gp5}) and (\ref{eq:gp6}), we need the hyperparameters values, $l$ and $\sigma_f$, from Equation (\ref{eq:gp2}). These values are determined by maximizing the log marginal likelihood:
\begin{align}
    \ln \mathcal{L} &= - \frac{1}{2} ( \bm{y} - \bm{\mu} )^{T} [\mathbfss{K}(\bm{Z}, \bm{Z}) + \mathbfss{C}]^{-1} ( \bm{y} - \bm{\mu} ) \notag \\
    &\quad - \frac{1}{2} \ln |\mathbfss{K}(\bm{Z}, \bm{Z}) + \mathbfss{C}| - \frac{n}{2} \ln 2\pi.
    \label{eq:gp7}
\end{align}
Following the steps outlined above, we can reconstruct the $H(z)$ function using CC data and then derive the Hubble constant $H_0$. We used the Gaussian process algorithm GAPP (Gaussian Processes in Python), as proposed by \citet{2012JCAP...06..036S}.

\subsubsection{MAF} \label{subsubsec:MAF}

In addition to EMCEE and GP, we employ Masked Autoregressive Flows (MAF) to constrain $H_0$ from CC data. MAF can represent non-Gaussian posteriors for $H_0$ and other cosmological parameters, while remaining fully normalized and straightforward to sample. In this paper we first train the MAF on simulated $H(z)$ data. After training, we input the CC dataset in Table~\ref{tab:cc_data} into the network to obtain samples from the learned conditional posterior $P(\bm{\theta} \mid \bm{H}_{\rm CC})$, and we then infer the posterior summary of $H_0$. We now briefly review the MAF method and then describe how we train and apply it in this work.

MAF encompasses both autoregressive models and normalizing flows, which are widely used for neural density estimation and offer a good balance between flexibility and tractability. Concretely, it stacks multiple Masked Autoencoders for Distribution Estimation (MADE) \citep{2015arXiv150203509G} in a normalizing flow to estimate parameters. Compared to the original MADE, MAF provides greater flexibility while maintaining tractability \citep{2017arXiv170507057P}.

MADE is a type of autoregressive model. To estimate the joint density $p(\bm{x})$, where $\bm{x}$ is a D-dimensional vector, $p(\bm{x})$ can be decomposed using the chain rule of probability:
\begin{equation}
    p(\bm{x}) = \prod_{i}^{D} p(x_i | \bm{x}_{1:i-1}),
    \label{eq:maf1}
\end{equation}
where $\bm{x}_{1:i-1} = ( x_1, x_2, ..., x_{i-1})^T$. The autoregressive density estimator models each density $p(x_i | \bm{x}_{1:i-1})$ \citep{2016arXiv160502226U}. By multiplying all of them according to Equation (\ref{eq:maf1}), we can derive the joint density $p(\bm{x})$. MADE has an advantage over straightforward recurrent autoregressive models: it can compute on parallel architectures by using binary masks to drop connections \citep{2015arXiv150203509G}. 

The normalizing flow \citep{2015arXiv150505770J} computes the density $p(\bm{x})$ as:
\begin{equation}
    p(\bm{x}) = \pi_u (f^{-1}(\bm{x})) \left| \det{ \left( \frac{\partial f^{-1}}{\partial \bm{x}} \right)} \right|. 
    \label{eq:maf3}
\end{equation}
Here, $\bm{x} = f(\bm{u})$, and $\bm{u} \sim \pi_u(\bm{u})$, where the base density $\pi_u(\bm{u})$ is typically chosen as a standard Gaussian distribution. The functions $f$ and $\pi_u(\bm{u})$ are inverses of each other. To enhance the transformation of $f$, multiple instances can be composed, such as $f_1 \circ f_2$. This composition still forms a valid normalizing flow \citep{2017arXiv170507057P}.

Since the conditionals of MADE are parameterized as single Gaussians, the $i^{th}$ conditional is given by:
\begin{equation}
    p(x_i | \bm{x}_{1:i-1}) = \mathcal{N} (x_1 | \mu_i , (\exp{\alpha_i})^2). 
    \label{eq:maf4}
\end{equation}
Where $\mu_i = f_{\mu_{i}} (\bm{x}_{1:i-1})$ and $ \alpha_i = f_{\alpha_{i}} (\bm{x}_{1:i-1})$, they represent the mean and log standard deviation of $i^{th}$ conditional, respectively. Data can be generated as follows:
\begin{equation}
    x_i = u_i \exp{\alpha_i} + \mu_i
    \label{eq:maf5},
\end{equation}
$u_i \sim \mathcal{N} (0,1)$. Equation (\ref{eq:maf5}) can be represented as $\bm{x} = f(\bm{u})$. Since $f$ is easily invertible, we can transform $\{ \bm{x}_n\}$ into $\{\bm{u}_n\}$. At this point, we have stacked multiple MADE models into a deeper flow, known as MAF \citep{2017arXiv170507057P}. 

In our application, the parameter vector is $\boldsymbol{\theta} = (H_0, \Omega_{\mathrm{m}}, \Omega_{\Lambda})$, and the conditioning input is the CC dataset in Table~\ref{tab:cc_data}. In the MAF, the quantities $\mu_i$ and $\alpha_i$ in the Gaussian conditional of Equation~(\ref{eq:maf4}) depend on both the previous components of $\boldsymbol{\theta}$ and on the CC dataset, so that the model approximates the conditional density $P(\boldsymbol{\theta} \mid \bm{H}_{\rm CC})$. Given a set of training pairs $\{(\boldsymbol{\theta}_n, \bm{H}_n)\}$ \citep{2021ApJS..254...43W, 2023PhRvD.107f3517C}, the parameters of the flow are determined by maximizing the conditional likelihood, or equivalently by minimizing the negative log-probability
\begin{equation}
\mathcal{L} = -\sum_n \ln P(\boldsymbol{\theta}_n \mid \bm{H}_n).
\label{eq:maf6}
\end{equation}
After the training has converged, we feed the observed CC dataset in Table~\ref{tab:cc_data} into the network, draw posterior samples of $\boldsymbol{\theta}$ from $P(\boldsymbol{\theta} \mid \bm{H}_{\rm CC})$, and obtain the posterior summary of $H_0$.

To train the MAF, we construct simulated parameter--data pairs 
$(\boldsymbol{\theta}_n, \bm{H}_n)$ that follow the same statistical model. We adopt a flat $\Lambda$CDM model for the expansion rate
\begin{equation}
H(z; \boldsymbol{\theta}) = H_0 \sqrt{ \Omega_{\rm m} (1+z)^3 + \Omega_{\Lambda} },
\label{eq:maf7}
\end{equation}
where $\boldsymbol{\theta} = (H_0, \Omega_{\rm m}, \Omega_{\Lambda})$ denotes the set of cosmological parameters. The index $n$ labels different simulated training examples, and the index $i$ introduced below labels the individual redshift points within one training example. Each CC data point in Table~\ref{tab:cc_data} contains a redshift $z_i$, an observed value $H_i$ and a $1\sigma$ credible interval $\sigma_i$. For a given CC dataset, we proceed as follows.

\begin{itemize}
\item Draw cosmological parameters. For each training example $n$ we draw $H_{0,n}$, $\Omega_{{\rm m},n}$ and $\Omega_{\Lambda,n}$ from broad uniform priors (in this paper, $H_{0,n} \sim \mathcal{U}(40,100)~{\rm km\,s^{-1}Mpc^{-1}}$, $\Omega_{{\rm m},n} \sim \mathcal{U}(0,0.7)$ and $\Omega_{\Lambda,n} \sim \mathcal{U}(0.3,1.0)$). These prior ranges are chosen to be wide enough to cover the plausible cosmological parameter region, similar in spirit to previous works using MAF \citep{2021ApJS..254...43W, 2023PhRvD.107f3517C}.

\item Compute noise-free model values. For the drawn $\boldsymbol{\theta}_n = (H_{0,n}, \Omega_{{\rm m},n} , \Omega_{\Lambda,n})$ we compute the fiducial expansion rate at all CC redshifts using Equation (\ref{eq:maf7}), $H_{\rm fid}(z_i; \boldsymbol{\theta}_n) = H(z_i; \boldsymbol{\theta}_n)$.

\item Add Gaussian observational errors. We then add Gaussian noise with standard deviation tied to the CC uncertainties at each redshift,
\begin{equation}
H_n(z_i) = H_{\rm fid}(z_i; \boldsymbol{\theta}_n) + \Delta H_{i,n},
\label{eq:maf8}
\end{equation}
where $\Delta H_{i,n} \sim \mathcal{N}(0, \sigma_i^2)$, so that the simulated data have the same redshift sampling and typical noise level as the real CC data.

\item Form training pairs. The simulated data vector for example $n$ is $\bm{H}_n = (H_n(z_1), \ldots, H_n(z_{N_z}))$; the associated label is $\boldsymbol{\theta}_n$. In our implementation we produce $N_{\mathrm{train}} = 8000$ simulated parameter-data pairs $\{(\boldsymbol{\theta}_n, \bm{H}_n)\}$.

\end{itemize}

The conditional MAF takes $\bm{H}_n$ as input and is trained to approximate the posterior density $P(\boldsymbol{\theta} \mid \bm{H}_n)$ by minimizing the loss $\mathcal{L}$ in equation~(\ref{eq:maf6}) on this simulated training set. After the MAF has been trained on simulated parameter-data pairs $(\boldsymbol{\theta}_n, \bm{H}_n)$, we keep the network weights fixed and use it to analyse the real CC dataset in Table~\ref{tab:cc_data}. We feed the CC $H(z)$ data into the trained MAF, which returns an approximation to the posterior distribution of the cosmological parameters $\boldsymbol{\theta} = (H_0, \Omega_{\mathrm{m}}, \Omega_{\Lambda})$. We then take the $H_0$ component of these samples as the marginal posterior samples for $H_0$.

Given a dataset, each method constrains $H_0$ in the form of a posterior distribution. We report a posterior summary of $H_0$ consisting of a central value and associated $1\sigma$ and $2\sigma$ credible intervals. Throughout the paper we use the shorthand 1$\sigma$ and 2$\sigma$ to denote the central 68$\%$ and 95.45$\%$ credible intervals, respectively (equal to $\pm 1\sigma$ and $\pm 2\sigma$ for a Gaussian posterior). For EMCEE and MAF, the $H_0$ posterior is represented by samples. We take the central value as the posterior median. Denoting by $q_p$ the $p$-th percentile of the sampled $H_0$ values, the $1\sigma$ credible interval is the central interval $[q_{16},\,q_{84}]$, and the $2\sigma$ credible interval is $[q_{2.275},\,q_{97.725}]$. For GP, $H_0$ is obtained from the predictive distribution at $z=0$. We take the central value as the predictive mean (which equals the median for a Gaussian) and use the predictive standard deviation as the corresponding $1\sigma$ credible interval. The associated $1\sigma$ and $2\sigma$ intervals follow from the Gaussian predictive distribution at $z=0$.

\subsection{Monte Carlo delete-$d$ jackknife} \label{subsec:Monte Carlo delete d jackknife}

\begin{figure}
    \centering
    \includegraphics[width=8cm]{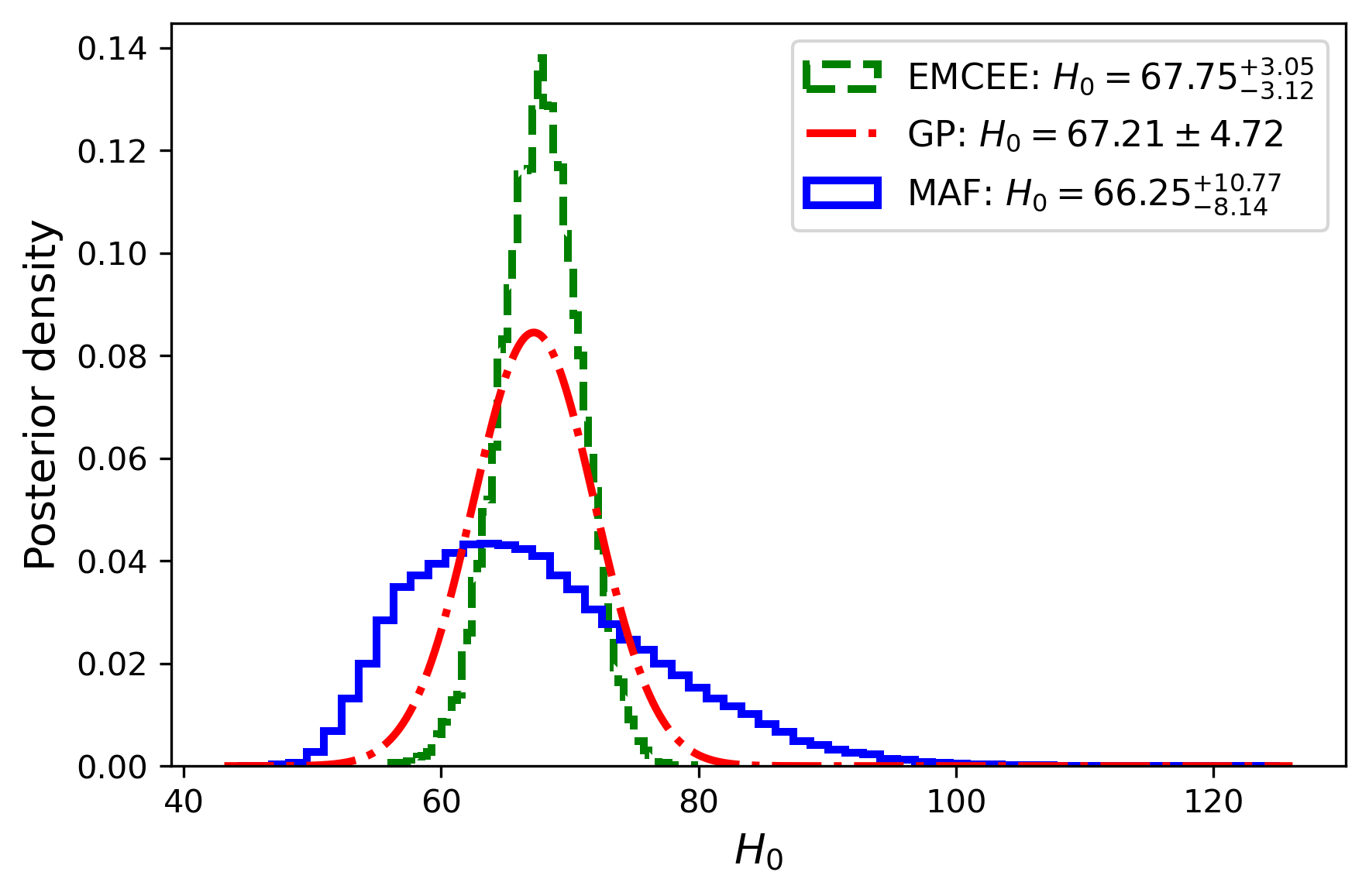}
    \caption{Posterior distributions on $H_0$ from EMCEE, GP, and MAF using the full CC dataset. The EMCEE and MAF curves are posterior densities estimated from samples, and the quoted values are the posterior median with the 16th and 84th percentiles. The GP curve is the Gaussian predictive distribution at $z=0$, and the quoted value is the predictive mean with its standard deviation.}
    \label{fig:3_H0_posteriors}
\end{figure}

\begin{figure*}
    \centering
    \includegraphics[width=0.8\linewidth]{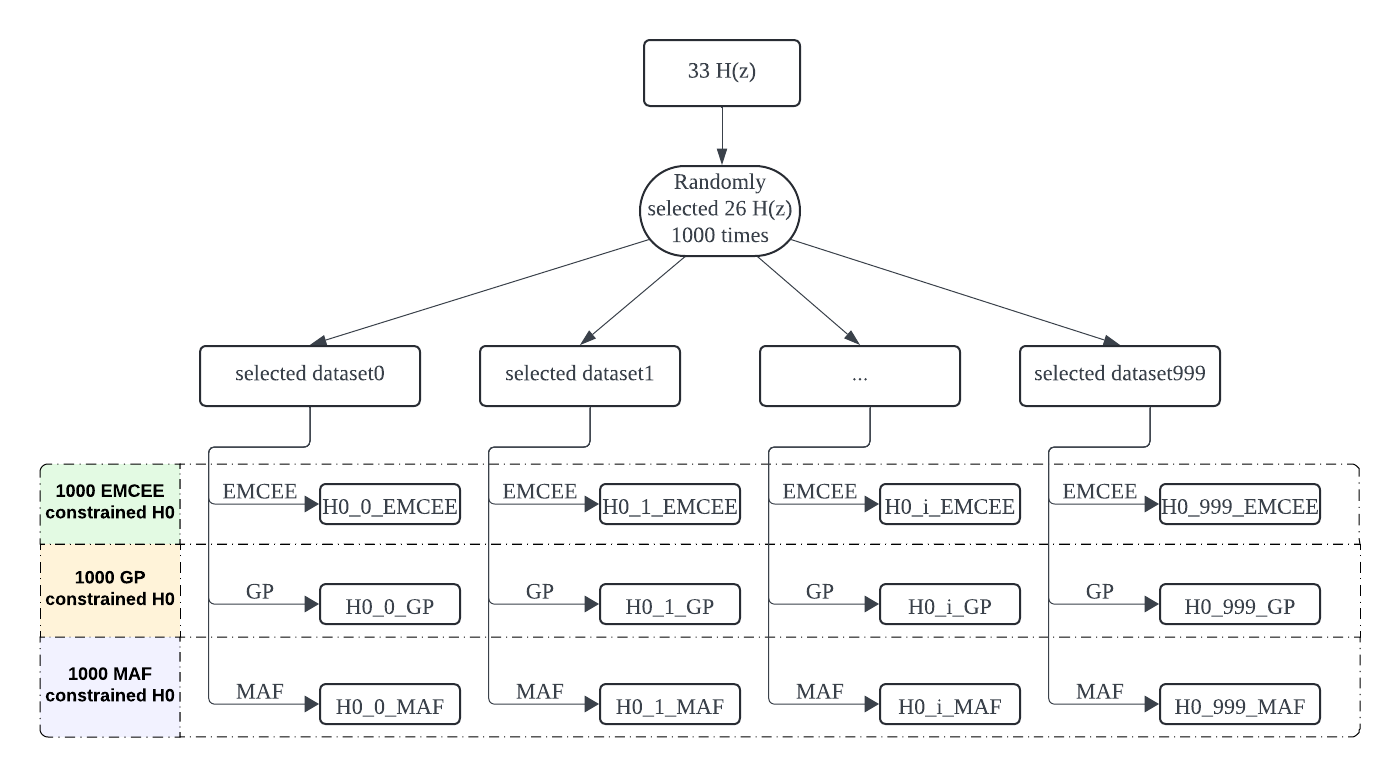} 
    \caption{Diagram of MCDJ procedure. At the top, 26 out of 33 $H(z)$ data points from the CC dataset are randomly selected. This is repeated 1000 times to create 1000 different datasets, shown in the middle. At the bottom, each dataset is used to constrain $H_0$ using the EMCEE, GP, and MAF methods, resulting in 1000 $H_0$ values for each method.}
    \label{fig:ProcedureMCDJ}
\end{figure*}

\begin{figure*}
    \centering
    \includegraphics[width=1\linewidth]{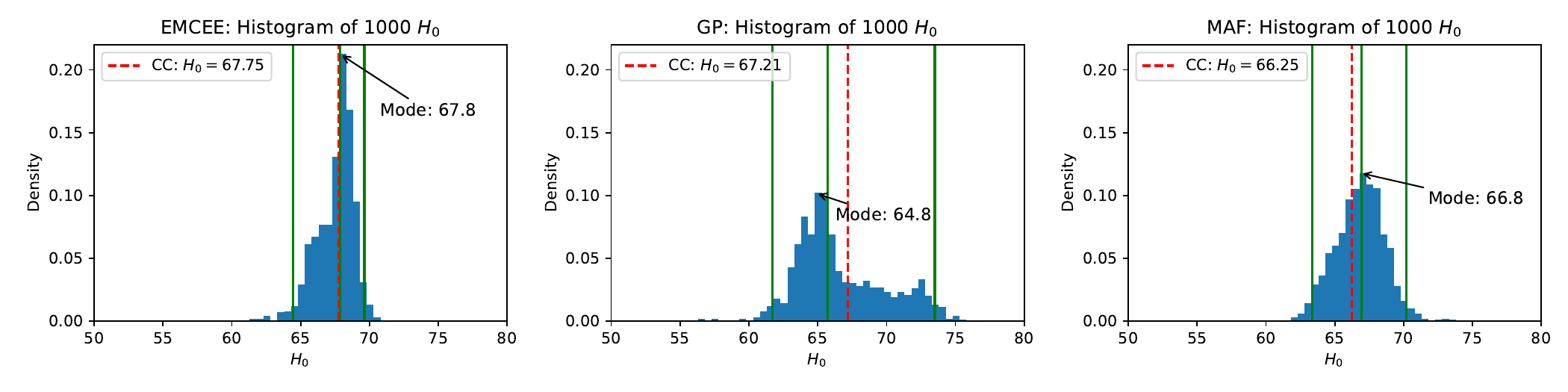} 
    \caption{The histogram of 1000 posterior central values of $H_0$ obtained from 1000 selected $H(z)$ datasets using MCDJ with EMCEE (left), GP (middle), and MAF (right). The red dashed line represents the posterior central value of $H_0$ using 33 CC data compiled in Table \ref{tab:cc_data}, while the green lines denote the $2\sigma$ range and the median of the distribution (at $2.5\%$, $50\%$, and $97.5\%$ percentile). And the mode is marked in each histogram.}
    \label{fig:3Methods_H0_Histogram}
\end{figure*}

In Subsection \ref{subsec:CC constrains H0}, three methods are introduced. Although these methods use the same CC dataset compiled in Table \ref{tab:cc_data}, they yield the different posterior summaries for $H_0$: $H_0=67.75^{+3.05}_{-3.12}\,\mathrm{km\,s^{-1}\,Mpc^{-1}}$ for EMCEE, $H_0 = 67.21 \pm 4.72\,\mathrm{km\,s^{-1}\,Mpc^{-1}}$ for GP, and $H_0~=~ 66.25^{+10.77}_{-8.14}\,\mathrm{km\,s^{-1}\,Mpc^{-1}}$ for MAF. Fig.~\ref{fig:3_H0_posteriors} compares the corresponding posterior distributions on $H_0$ from the three methods using the full CC dataset.
To assess the sensitivity of each method to individual CC data points, we apply a Monte Carlo delete-$d$ jackknife (MCDJ) scheme. In the classical delete-$d$ jackknife \citep{ShaoTu1995}, one forms all possible subsets of size $n-d$ from an $n$-point dataset, refits the model on each subset, and studies the variability of the resulting estimator; this quantifies how sensitive the estimator is to removing data points. For our CC sample with $n=33$ and $d=7$, the number of subsets, $\binom{33}{26} \sim 4\times 10^6$, is far too large to enumerate. We therefore adopt a Monte Carlo version of the delete-$d$ jackknife: instead of using all subsets, we randomly select $N$ subsets of size 26, fit each subset with EMCEE, GP, or MAF, and analyse the distribution of the resulting $H_0$ values. This directly measures the sensitivity of the inferred $H_0$ to the deletion of individual CC data points. The procedure of MCDJ is illustrated in Fig.~\ref{fig:ProcedureMCDJ},
and the details are outlined as follows:
\begin{itemize}
  \item Step 1: We randomly select a subset of data points from the CC dataset. In this paper, we randomly select 26 out of 33 $H(z)$ data points, corresponding to a $79 \%$ selection rate. This corresponds to a delete-$d$ jackknife with $d=7$. We then constrain $H_0$ using these selected 26 data points, denoted as $H_{0\text{-}i\text{-}EMCEE}$, indicating the $H_0$ constrained by EMCEE using the $i_{th}$ selected dataset.
  \item Step 2: We repeat Step 1 multiple times. In this paper, Step 1 is repeated $N$ times, with $N$ set to 1000. This results in the creation of 1000 randomly selected datasets, labeled with an index $i$ ranging from 0 to 999.
  \item Step 3: Generate 1000 $H_0$ posterior summaries with EMCEE, GP, and MAF respectively using the 1000 randomly selected $H(z)$ datasets from Step 2. Each method generates 1000 values of $H_0$, and these results are shown in Fig.~\ref{fig:3Methods_H0_Histogram}.
\end{itemize}

To evaluate the sensitivity of $H_0$ posterior summaries from different methods to individual $H(z)$ data points, we propose MCDJ. The distribution of 1000 posterior central values of $H_0$ from selected $H(z)$ datasets is displayed in Fig.~\ref{fig:3Methods_H0_Histogram}. We compare three metrics: (1) the absolute difference between the mode of the $H_0$ distribution and the $H_0$ value constrained from the full 33 CC dataset, denoted as $\Delta H_\mathrm{0,mode-CC} = |H_\mathrm{0, mode} - H_\mathrm{0,CC}|$, (2) the absolute difference between the median of the $H_0$ distribution and $H_\mathrm{0,CC}$, denoted as $\Delta H_\mathrm{0,median-CC} = |H_\mathrm{0, median} - H_\mathrm{0,CC}|$, and (3) the range spanned by $2\sigma$ in the distribution, denoted as $\Delta H_{0,2\sigma}$. We summarize these results in Table \ref{tab:MCDJ_33}. Here, $H_\mathrm{0, mode}$ represents the mode of the distribution of 1000 posterior central values of $H_0$ values shown in Fig.~\ref{fig:3Methods_H0_Histogram}. The $H_0$ value at the median density, $H_\mathrm{0, median}$, and the $2\sigma$ boundaries are indicated by green lines. Additionally, the red dashed line represents the Hubble constant constrained from the full 33 CC dataset, denoted as $H_\mathrm{0,CC}$.

We analyze the MCDJ results in Fig.~\ref{fig:3Methods_H0_Histogram} from two perspectives to understand how different methods affect the sensitivity of posterior summary of $H_0$ to individual $H(z)$ data points. First, we compare the distribution of 1000 posterior central values of $H_0$ from MCDJ with $H_\mathrm{0,CC}$, which acts as the standard for $H_0$. Each of the three methods also produces its own $H_\mathrm{0,CC}$ values, as shown in Fig.~\ref{fig:3Methods_H0_Histogram}. Based on this analysis, we calculate the $\Delta H_\mathrm{0,mode-CC}$ and $\Delta H_\mathrm{0,median-CC}$ values, summarized in Table \ref{tab:MCDJ_33}. The results show that, for both measures, GP values are higher than MAF, and MAF values are higher than EMCEE. This indicates that, when constraining $H_0$, GP is more sensitive to individual $H(z)$ values than MAF, and MAF is more sensitive than EMCEE. The second perspective is the degree of dispersion in the distribution of 1000 posterior central values of $H_0$ values obtained through MCDJ. Greater dispersion suggests higher sensitivity to individual $H(z)$ values in constraining $H_0$. We calculate the range of the distribution within the $2\sigma$ interval. The $\Delta H_{0,2\sigma}$ results are summarized in Table \ref{tab:MCDJ_33}. These results show that GP is more sensitive to individual $H(z)$ data points than MAF, and MAF is more sensitive than EMCEE. Both perspectives confirm this finding. From this result, we infer that as more CC data are observed, the posterior central value of $H_0$ will vary more with GP than with MAF, and more with MAF than with EMCEE.

GP is more sensitive to individual $H(z)$ points when constraining $H_0$ from the CC dataset compared to MAF and EMCEE. This increased sensitivity is likely due to the use of the $\Lambda$CDM model within EMCEE and MAF, as discussed in Subsection \ref{subsec:CC constrains H0}. However, this explanation is speculative, and further investigation is needed. The results in Table \ref{tab:MCDJ_33} on sensitivity to individual $H(z)$ points using three different methods help clarify how $H_0$ posterior central values differ among these methods when using CC data. This information will help us choose the most suitable method for constraining $H_0$ based on our research objectives. To check that these conclusions do not depend on the particular choice of retaining 26 out of 33 CC points in the MCDJ resampling, we repeated the analysis with a smaller subset of 22/33 points, using 500 realizations. The corresponding sensitivity measures are summarised in Table~\ref{tab:MCDJ_test_22_33} in Appendix~\ref{app:MCDJ_22}. While the numerical values of $\Delta H_{0,\mathrm{mode-CC}}$, $\Delta H_{0,\mathrm{median-CC}}$, and $\Delta H_{0,2\sigma}$ change moderately, the qualitative ordering of the methods is unchanged: GP remains the most sensitive to the deletion of CC points, MAF intermediate, and EMCEE the least sensitive. Our conclusions from Section~\ref{subsec:Monte Carlo delete d jackknife} are therefore insensitive to the exact number of CC data points retained in the MCDJ resampling.

\begin{table}
    \centering
    \caption{Comparing results using MCDJ. $\Delta H_{0,\mathrm{mode-CC}}$ ($\Delta H_{0,\mathrm{median-CC}}$) denotes the absolute difference between the mode (median) of the MCDJ $H_0$ distribution and the full-sample CC result. $\Delta H_{0,2\sigma}$ is the span of the central $2\sigma$ interval of the MCDJ $H_0$ distribution. Values are in km s$^{-1}$ Mpc$^{-1}$.}
    \label{tab:MCDJ_33}
    \begin{tabular}{cccc}
        \hline
        Methods & $\Delta H_\mathrm{0,mode-CC}$ & $\Delta H_\mathrm{0,median-CC}$ & $\Delta H_{0,2\sigma}$ \\
        \hline
        EMCEE & 0.05 & 0.11 & 5.18\\
        GP & 2.41 & 1.51 & 11.77\\
        MAF & 0.55 & 0.69 & 6.84\\
        \hline
    \end{tabular}
\end{table}


\begin{table*}
\centering
\caption{Comparison of $1\sigma$ uncertainties on $H_0$ from the full 33-point CC dataset and from the Monte Carlo delete-$d$ jackknife resampling (26/33, 1000 realizations). For EMCEE and MAF, $\sigma_{H_0}$ is defined as half the width of the central 68\% credible interval of the posterior samples. For GP, $\sigma_{H_0}$ denotes the standard deviation of the Gaussian predictive distribution. All values are in $\mathrm{km\,s^{-1}\,Mpc^{-1}}$.}
\label{tab:MCDJ_sigma}
\begin{tabular}{lcccc}
\hline
Method &
$\sigma_{H_0,\mathrm{CC}}$ &
$\langle \sigma_{H_0} \rangle_{\mathrm{MCDJ}}$ &
$\mathrm{median}(\sigma_{H_0})_{\mathrm{MCDJ}}$ &
$[\sigma_{H_0}^{16\%},\,\sigma_{H_0}^{84\%}]_{\mathrm{MCDJ}}$ \\
\hline
EMCEE & 3.08 & 3.56 & 3.38 & [3.16, 4.00] \\
GP    & 4.72 & 5.22 & 4.99 & [4.31, 6.10] \\
MAF   & 9.46 & 9.98 & 9.93 & [9.41, 10.59] \\
\hline
\end{tabular}
\end{table*}

To quantify the impact of individual CC data points we have so far examined how the posterior central value of $H_0$ varies across the MCDJ realizations. As a complementary check, we now study the behaviour of the $1\sigma$ posterior credible intervals under the same resampling. For each MCDJ realization $i$ and each method $m$ we record the corresponding posterior width $\sigma_{H_0,m,i}$ together with $H_{0,m,i}$, and from the 1000 values of $\sigma_{H_0,m,i}$ we compute summary statistics that can be compared to the full 33-point result. These are summarized in Table~\ref{tab:MCDJ_sigma}, which contrasts the $1\sigma$ uncertainties on $H_0$ from the full CC dataset with those from the MCDJ resampling (26/33, 1000 realizations). In all three methods the MCDJ median uncertainties are only slightly larger (by about 5--10\%) than the full-data values, as expected when some data points are removed. Apart from this mild inflation, the behaviour of the uncertainties is very similar: the ordering $\sigma_{H_0}^{\mathrm{EMCEE}} < \sigma_{H_0}^{\mathrm{GP}} < \sigma_{H_0}^{\mathrm{MAF}}$ is unchanged, and the 16th--84th percentile ranges indicate that most resampled $\sigma(H_0)$ values stay close to the corresponding full-data $\sigma_{H_0,\mathrm{CC}}$. This shows that deleting seven CC points increases the error bars modestly but does not significantly alter the relative performance of the three methods or the stability of their posterior widths.
We have repeated the same analysis of $\sigma_{H_0}$ for the alternative MCDJ setup with 22 out of 33 CC points and obtain very similar median values and percentile ranges, with the ordering $\sigma_{H_0}^{\mathrm{EMCEE}} < \sigma_{H_0}^{\mathrm{GP}} < \sigma_{H_0}^{\mathrm{MAF}}$ unchanged. This confirms that the behaviour of the posterior credible intervals is insensitive to the exact subset size adopted in the MCDJ resampling.

\subsection{Seperate to two redshift regions} \label{subsec:Seperate to different areas}

\begin{figure*}
    \centering
    \includegraphics[width=1\linewidth]{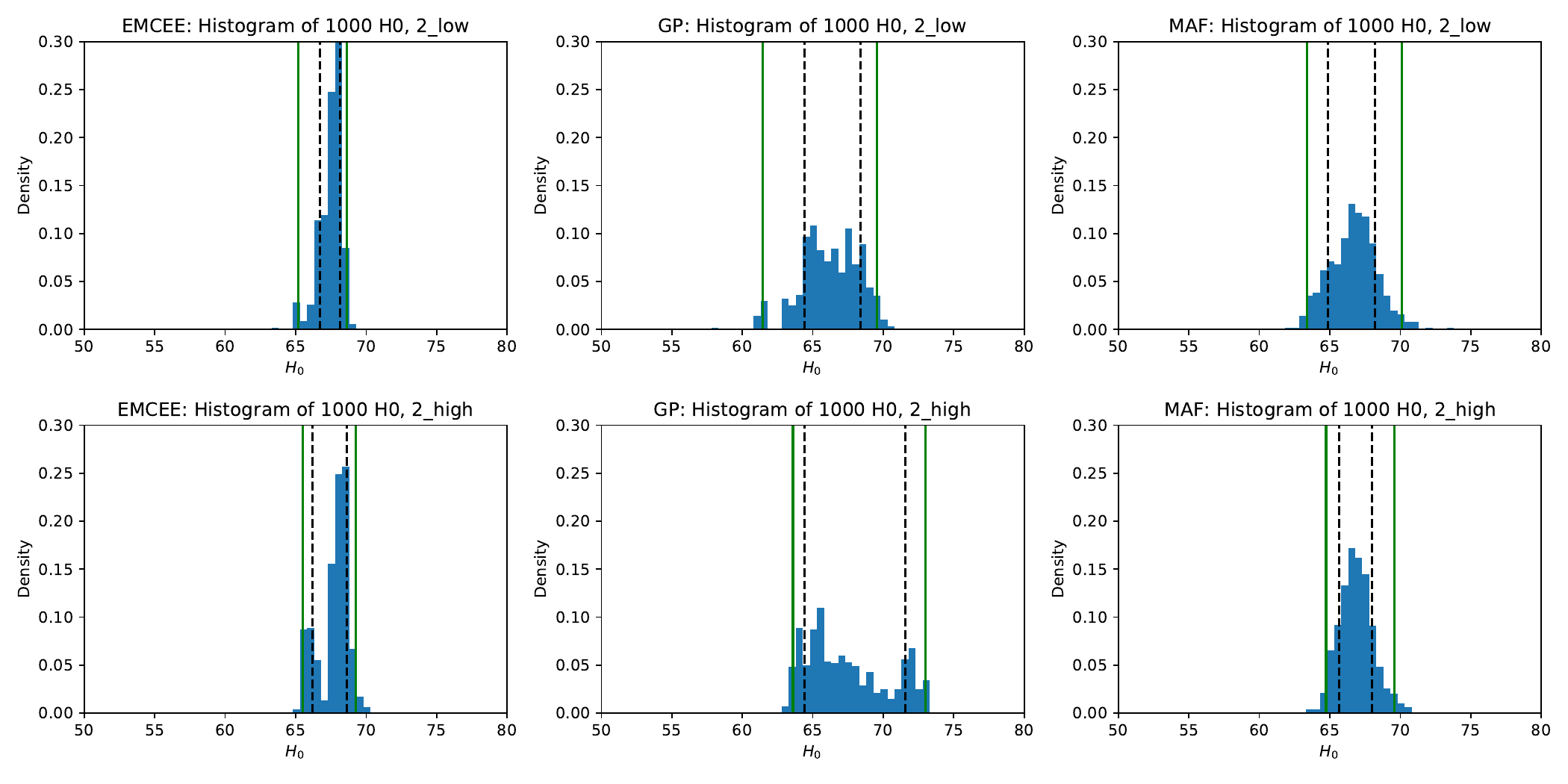} 
    \caption{The histograms of the distribution of 1000 posterior central values of $H_0$ using the selected datasets from the CC dataset. In the first and second rows, the histograms illustrate the distribution of 1000 posterior central values of $H_0$ values using datasets dataset$2_{L-i}$ and dataset$2_{H-i}$, respectively. These datasets are generated in Subsection \ref{subsec:Seperate to different areas}. In each row, the histograms are plotted using three different methods: EMCEE (left), GP (middle), and MAF (right). Solid green lines indicate the $2\sigma$ range of the distribution (at $2.5\%$ and $97.5\%$ percentiles), while dashed black lines represent the $1\sigma$ range (at $16\%$ and $84\%$ percentiles).}
    \label{fig:Separate2_H0_Histogram}
\end{figure*}

In this subsection, we explore whether the posterior central value of $H_0$ shows different sensitivities to $H(z)$ across various redshift regions using three methods: EMCEE, GP, and MAF. We sort the 33 CC $H(z)$ datasets by redshift $z$ and divide them into two groups with as equal a number of data points as possible, using $z_{\rm split}=0.48$ as the boundary. Group 1 consists of 17 $H(z)$ data points from the low-redshift region ($z < 0.48$), while Group 2 contains 16 $H(z)$ data points from the high-redshift region ($z \geq 0.48$). For instance, to evaluate the sensitivity of posterior central value of $H_0$ to $H(z)$ in the low redshift region, we apply MCDJ to this group:
\begin{itemize}
  \item Step 1: We randomly selected 13 $H(z)$ data points from the 17 data points in Group 1, with a selection rate of approximately $76 \%$. In addition, all 16 $H(z)$ data points from Group 2, representing the high-redshift region, are included. This results in a combined dataset of 29 $H(z)$ points: 13 from Group 1 and 16 from Group 2. To isolate the effect of the low-redshift data on $H_0$ sensitivity, we retain all high-redshift points to prevent them from influencing the results.
  \item Step 2: Repeat Step 1 one thousand times to create 1000 different datasets, labeled as dataset$2_{L-i}$. In this label, "dataset$2$" indicates that the CC data has been divided into two groups. "$L$" refers to the selected $H(z)$ data from the low-redshift region, while all high-redshift data is kept intact. The variable "$i$" represents the dataset number, ranging from 0 to 999.
  \item Step 3: Use EMCEE, GP, and MAF to constrain $H_0$ for the 1000 generated dataset$2_{L-i}$. The results are shown in Fig.~\ref{fig:Separate2_H0_Histogram}.
\end{itemize}
The dataset$2_{H-i}$ is created in a similar way, by randomly selecting 12 $H(z)$ data points from the 16 data points in Group 2, and including all $H(z)$ data points from Group 1. The results are shown in Fig.~\ref{fig:Separate2_H0_Histogram}.

By analyzing Fig.~\ref{fig:Separate2_H0_Histogram}, we can assess whether $H(z)$ values from different redshift regions have varying impacts on the sensitivity of the posterior central value of $H_0$. The wider range of $H_0$ within the $1\sigma$ ($\Delta H_{0,1\sigma}$) and $2\sigma$ ($\Delta H_{0,2\sigma}$) intervals indicates greater sensitivity of posterior central value of $H_0$ to individual $H(z)$ data points in the corresponding redshift region for that method. To compare the sensitivity of posterior central value of $H_0$ to individual $H(z)$ data points across different redshift regions, we divide the redshift into two regions and calculate
\begin{equation}
    \Delta H_{0,1\sigma,\mathrm{L-H}} = \Delta H_{0,1\sigma,\mathrm{Low}} - \Delta H_{0,1\sigma,\mathrm{High}},
    \label{eq:separate1}
\end{equation}
\begin{equation}
    \Delta H_{0,2\sigma,\mathrm{L-H}} = \Delta H_{0,2\sigma,\mathrm{Low}} - \Delta H_{0,2\sigma,\mathrm{High}}.
    \label{eq:separate2}
\end{equation}
where $\Delta H_{0,1\sigma,\mathrm{Low}}$ refers to the $H_0$ range within the $1\sigma$ area in the first row of Fig.~\ref{fig:Separate2_H0_Histogram}, while $\Delta H_{0,1\sigma,\mathrm{High}}$ refers to the $1\sigma$ area in the second row. The same applies for $\Delta H_{0,2\sigma,\mathrm{Low}}$ and $\Delta H_{0,2\sigma,\mathrm{High}}$. The results of $\Delta H_{0,1\sigma,\mathrm{L-H}}$ and $\Delta H_{0,2\sigma,\mathrm{L-H}}$ for three different methods in Fig.~\ref{fig:Separate2_H0_Histogram} are summarized in Table \ref{tab:2_separate}.
\begin{table}
    \centering
    \caption{Comparison of $H_0$ sensitivity between two redshift regions. $\Delta H_{0,1\sigma,\mathrm{L-H}}$ and $\Delta H_{0,2\sigma,\mathrm{L-H}}$ are defined as the difference between the low-redshift and high-redshift sensitivity ranges (low minus high), computed from the $1\sigma$ and $2\sigma$ spans of the MCDJ $H_0$ distributions. Positive (negative) values indicate stronger sensitivity to low-$z$ (high-$z$) points. Values are in km~s$^{-1}$~Mpc$^{-1}$.}
    \label{tab:2_separate}
    \begin{tabular}{ccc}
        \hline
        Methods & $\Delta H_{0,1\sigma,\mathrm{L-H}}$ & $\Delta H_{0,2\sigma,\mathrm{L-H}}$ \\
        \hline
        EMCEE & -0.95 & -0.30 \\
        GP & -3.16 & -1.33 \\
        MAF & 1.05 & 1.90 \\
        \hline
    \end{tabular}
\end{table}


In Equation (\ref{eq:separate1}), if $\Delta H_{0,1\sigma,\mathrm{L-H}}>0$, it means that $\Delta H_{0,1\sigma,\mathrm{Low}} $ is greater than $\Delta H_{0,1\sigma,\mathrm{High}}$, indicating that the posterior central value of $H_0$ is more responsive to the $H(z)$ in the low redshift region than in the high redshift region. On the other hand, if $\Delta H_{0,1\sigma,\mathrm{L-H}}<0$, it shows that the posterior central value of $H_0$ is less sensitive to the $H(z)$ in the low redshift region than in the high redshift region. Table \ref{tab:2_separate} shows that for the EMCEE and GP methods, the posterior central value of $H_0$ is more sensitive to $H(z)$ data in the high redshift region than in the low redshift region, as indicated by $\Delta H_{0,1\sigma,\mathrm{L-H}}$ and $\Delta H_{0,2\sigma,\mathrm{L-H}}$ being less than 0. In contrast, the MAF method shows that the posterior central value of $H_0$ is more sensitive to $H(z)$ data in the low redshift region than in the high redshift region.

All three methods (EMCEE, GP, and MAF) are sensitive to individual data points in both low and high redshift regions. For MAF, we find stronger sensitivity to low-$z$ points, which is consistent with the common heuristic that low-$z$ measurements more directly anchor the local normalisation relevant to the posterior central value of $H_0$. However, our results show this heuristic is not universal across methods. For EMCEE and GP, the posterior central value of $H_0$ is more sensitive to $H(z)$ data in higher redshift regions. A plausible interpretation is that the three approaches couple $H_0$ to the CC measurements in different ways, which can lead to different redshift sensitivities. In EMCEE, $H_0$ is inferred jointly with parameters governing the global expansion history, so high-redshift points can influence $H_0$ through their leverage on the overall shape of $H(z)$. In GP, $H_0$ is obtained from a reconstruction (and extrapolation) to $z=0$, and the relative impact of different redshifts is mediated by the fitted GP covariance, so higher-$z$ points may still propagate to $H_0$. In contrast, MAF is a simulation-trained conditional density estimator, so its sensitivity can concentrate on the redshift range that is most informative for $H_0$ within the learned mapping. These points provide a qualitative interpretation of the observed trend.

To test whether this redshift dependence is sensitive to the deletion fraction, we repeated the analysis using a more aggressive deletion in each region. In this additional MCDJ run, we select 11 of the 17 low-$z$ points (and keep all high-$z$ points) and, in a symmetric test, select 11 of the 16 high-$z$ points while keeping all low-$z$ points, with 500 realizations in each case. The corresponding values of $\Delta H_{0,1\sigma,\mathrm{L-H}}$ and $\Delta H_{0,2\sigma,\mathrm{L-H}}$ are reported in Appendix \ref{app:2redshift_27} (Table~\ref{tab:2_separate_27}). The 1$\sigma$ indicators still show EMCEE and GP to be more sensitive to the high-redshift region and MAF to be more sensitive to the low-redshift region, although the numerical differences move closer to zero and some 2$\sigma$ values become nearly symmetric. This shows that the qualitative redshift dependence of the three methods is unchanged and is insensitive to the deletion fraction used in the subsampling test (29/33 versus 27/33), with EMCEE and GP remaining more sensitive to the high-redshift subset and MAF remaining more sensitive to the low-redshift subset (Appendix \ref{app:2redshift_27}).


\section{Simulation} \label{sec:Simulation}

The $H_0$ posterior summary differs when using the CC dataset with three different methods: EMCEE, GP, and MAF. In Subsection \ref{subsec:Monte Carlo delete d jackknife}, we compare their sensitivity in posterior summary of $H_0$ to individual $H(z)$ values using MCDJ. In this section we instead assess their overall performance in a controlled setting by validating each method on ensembles of mock CC datasets with known ground truth. For each simulation realisation, we generate a CC-like dataset $\{z_i,\,H_{\rm sim}(z_i),\,\sigma_{\rm sim}(z_i)\}$ and apply EMCEE, GP, and MAF to infer the $H_0$ posterior for each method. We then compare the resulting posterior summaries to the fixed input value $H_{0,{\rm true}}$. To reduce the impact of any single noise realisation and to enable statistically meaningful comparisons, we generate $R=100$ independent mocks for each simulation design. We consider two complementary mock-generation schemes: (i) a $\Lambda$CDM-based simulation (Section~\ref{subsec:Simulation based on LCDM}), and (ii) a GP-based simulation (Section~\ref{subsec:Simulation based on Gaussian Process}). Method performance is summarised in terms of central value accuracy, credible interval calibration, and overall posterior quality, quantified by the bias, RMSE, coverage, and log score defined below.

\subsection{Simulation based on $\Lambda$CDM}\label{subsec:Simulation based on LCDM}

In this section we perform a controlled $\Lambda$CDM simulation to validate the three constraining $H_0$ methods considered in this work. We adopt a fixed input truth, $H_{0,\mathrm{true}}=67.75\ \mathrm{km\,s^{-1}\,Mpc^{-1}}$, so that every mock dataset shares the same well-defined ground truth and the accuracy of the recovered posteriors can be assessed directly. We then generate an ensemble of $R=100$ independent mock CC datasets to reduce the impact of any single noise realization and to enable robust, statistically meaningful performance comparisons. Section~\ref{subsubsec:LCDM-based mock-data simulation} describes the mock-data simulation procedure, and Section~\ref{subsubsec:Method validation on LCDM based simulations} applies EMCEE, GP, and MAF to the mocks and summarises their performance using accuracy, credible interval calibration, and overall posterior-quality.

\subsubsection{$\Lambda$CDM-based mock-data simulation}
\label{subsubsec:LCDM-based mock-data simulation}

To test the performance of EMCEE, GP, and MAF under controlled conditions, we generate mock datasets from a flat $\Lambda$CDM model with a fixed input truth of $H_{0,\mathrm{true}}$. We adopt a single fiducial cosmology for all simulated realizations, $H_{0,\mathrm{true}}=67.75\ \mathrm{km\,s^{-1}\,Mpc^{-1}}$ and $\Omega_{M,\mathrm{true}}=0.328$. Fixing the cosmological parameters ensures that every realization has a well-defined and identical ground truth against which the inferred posteriors can be evaluated.

First, for each realization, to ensure that the simulated dataset closely matches the observed CC dataset, we generate 33 values for $\{z_i^{\mathrm{sim}}\}$ within the range $[0,2.0]$, which covers the full observed CC redshift range $0.07 \leq z \leq 1.965$. Our aim is to make the simulated redshifts resemble the observed CC sample as closely as possible in their overall redshift distribution, while still retaining randomness in each realization. To do this, we divide the interval into 10 equal-width bins and count the number of observed CC data points in each bin. We then generate the same number of simulated redshift values within each bin by randomly assigning them between the lower and upper boundaries of that bin. In this way, the simulated redshifts preserve the overall distribution pattern of the observed CC data in redshift, while allowing the exact $z_i^{\mathrm{sim}}$ values to vary from one realization to another. Using the fixed input parameters, we compute the fiducial $\Lambda$CDM prediction at each $z_i^{\mathrm{sim}}$,
\begin{equation}
H_{\mathrm{fid}}(z_i^{\mathrm{sim}})=H_{0,\mathrm{true}}
\sqrt{\Omega_{M,\mathrm{true}}(1+z_i^{\mathrm{sim}})^3+\left(1-\Omega_{M,\mathrm{true}}\right)}.
\label{eq:Hfid_LCDM}
\end{equation}

\begin{figure}
    \centering
    \includegraphics[width=8cm]{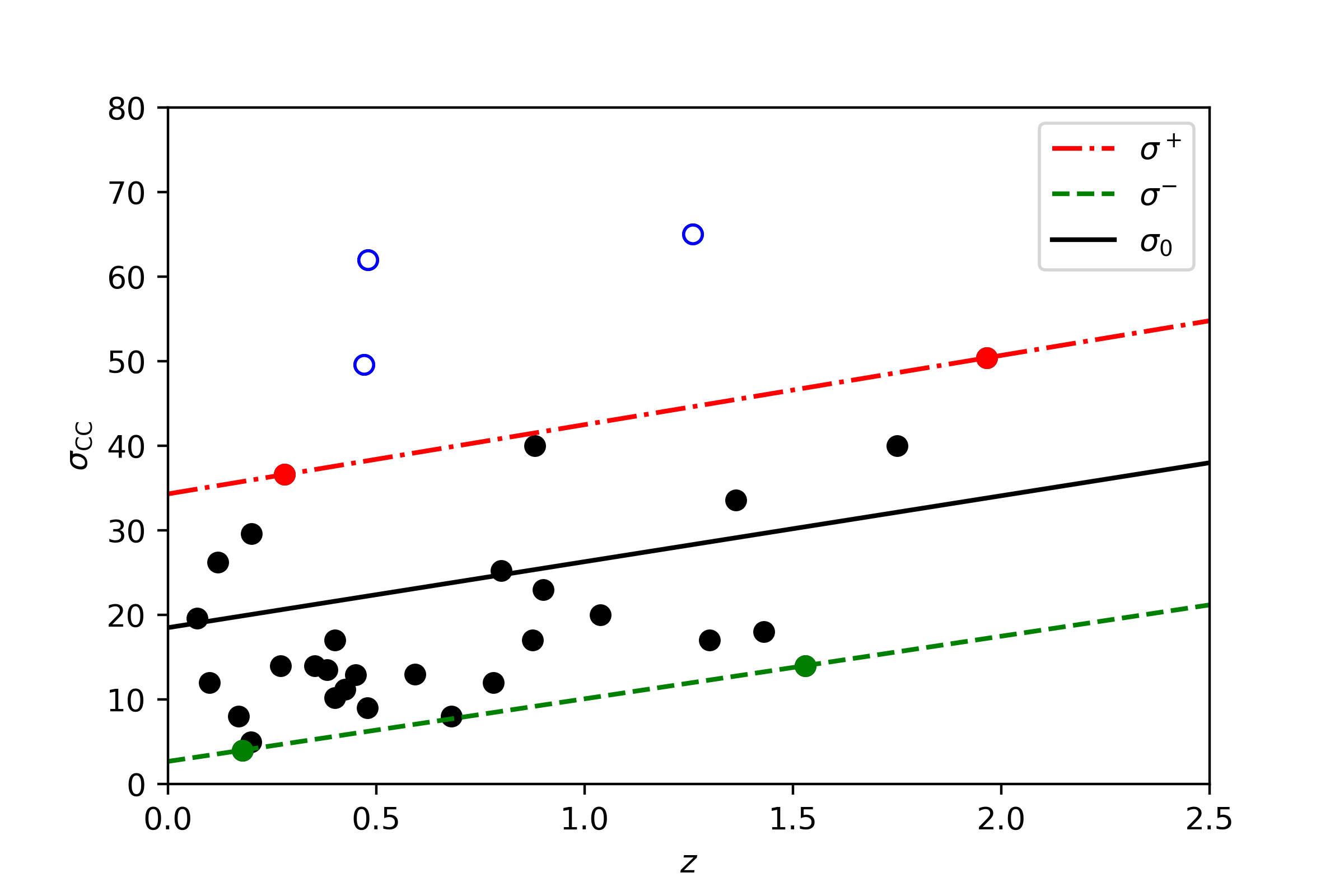}
    \caption{The 1$\sigma$ uncertainty of CC data. Black dots represent non-outliers, while blue circles indicate outliers. The red dash-dotted line and the green dashed line mark the bounds of the non-outliers, denoted as $\sigma^+$ and $\sigma^-$, respectively. The solid black line represents the mean uncertainty of the non-outliers, denoted as $\sigma_0$.}
    \label{fig:cc_sigma_line}
\end{figure}

Next, to mimic realistic CC measurement uncertainties, we assign each simulated point an error bar
$\sigma_{\mathrm{sim}}(z_i^{\mathrm{sim}})$. Based on the observed CC uncertainties $\sigma_{\mathrm{CC}}$, we model
the mean trend as $\sigma_0(z)$ and the upper and lower envelopes as $\sigma_{+}(z)$ and $\sigma_{-}(z)$,
respectively, after excluding outliers. Each of these functions is taken to be linear in $z$, as
illustrated in Fig.~\ref{fig:cc_sigma_line}.
We assume that this bounded region contains 95$\%$ of the $\sigma_{\mathrm{CC}}$ values, and we model the scatter about the mean trend $\sigma_0(z)$ with a Gaussian approximation when generating mock uncertainties. At each simulated redshift, we draw
\begin{equation}
\sigma_{\mathrm{sim}}(z_i^{\mathrm{sim}})\sim
\mathcal{N}\!\left(\sigma_0(z_i^{\mathrm{sim}}),\ \left[\frac{\sigma_{+}(z_i^{\mathrm{sim}})-\sigma_{-}(z_i^{\mathrm{sim}})}{4}\right]^2\right).
\label{eq:sigmasim}
\end{equation}
In practice, after drawing $\sigma_{\mathrm{sim}}(z_i^{\mathrm{sim}})$ from Equation~(\ref{eq:sigmasim}), we take its absolute value to ensure that all simulated uncertainties are positive.

Given $H_{\mathrm{fid}}(z_i^{\mathrm{sim}})$ and $\sigma_{\mathrm{sim}}(z_i^{\mathrm{sim}})$, the simulated Hubble-parameter measurement is generated by adding a Gaussian noise term $\epsilon_i$,
\begin{equation}
H_{\mathrm{sim}}(z_i^{\mathrm{sim}})=H_{\mathrm{fid}}(z_i^{\mathrm{sim}})+\epsilon_i,
\label{eq:Hsim}
\end{equation}
with
\begin{equation}
\epsilon_i\sim\mathcal{N}\!\left(0,\ (f\,\sigma_{\mathrm{sim}}(z_i^{\mathrm{sim}}))^2\right).
\label{eq:eps}
\end{equation}
Here $\sigma_{\mathrm{sim}}$ defines the quoted uncertainty of each simulated point, while $\epsilon_i$ generates the scatter of $H_{\mathrm{sim}}$ around the fiducial $\Lambda$CDM curve. The factor $f$ controls the scatter amplitude relative to the assigned uncertainties.

\begin{figure}
    \centering
    \includegraphics[width=8cm]{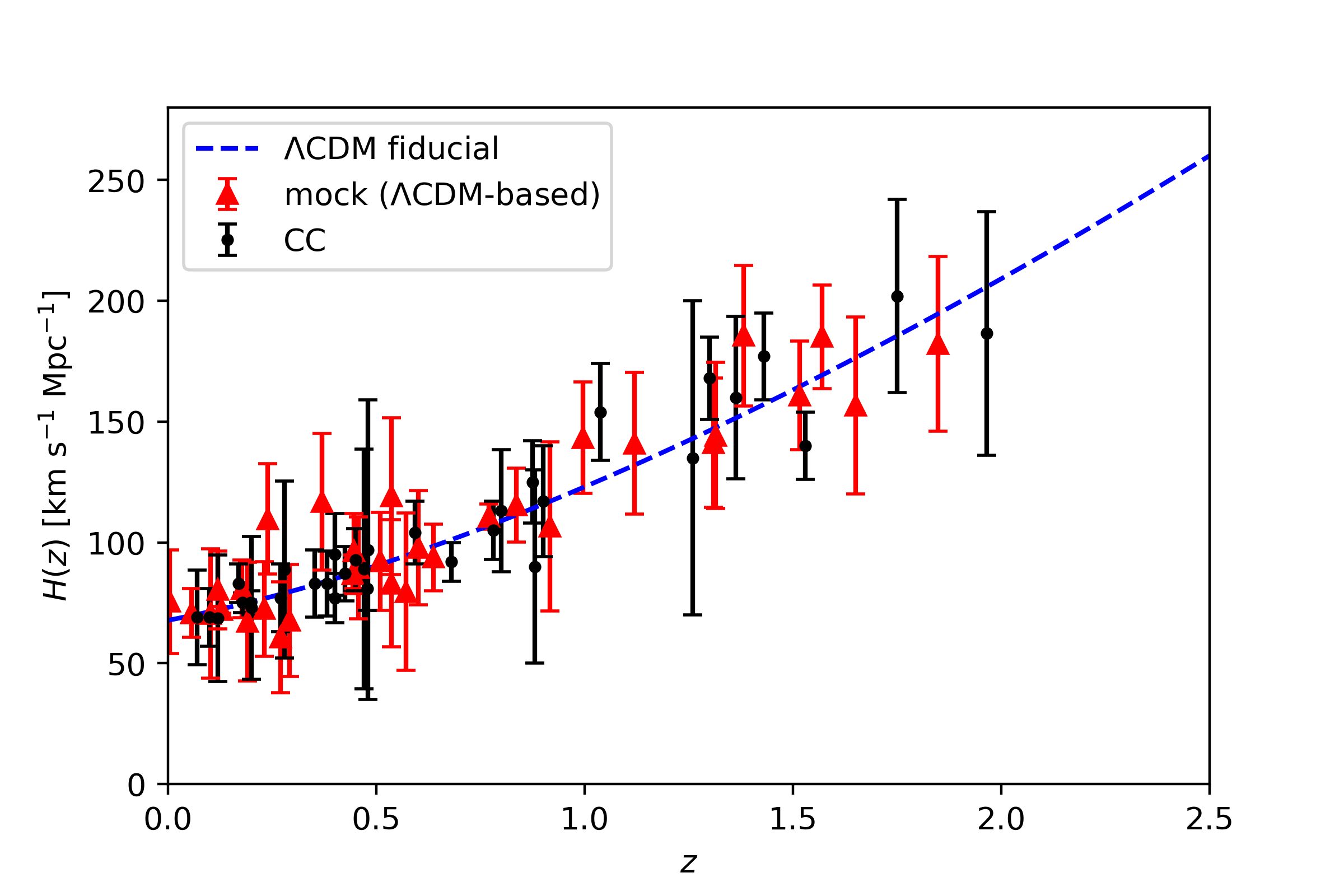}
    \caption{An example simulated $H_{\mathrm{sim}}$ dataset based on the $\Lambda$CDM model using CC data. The simulated mock-data points, $H_{\mathrm{sim}}(z_i^{\mathrm{sim}})$, are shown as red triangles with uncertainties $\sigma_{\mathrm{sim}}(z_i^{\mathrm{sim}})$. For comparison, the observational CC data are represented by black dots with error bars. The blue dashed curve shows the fiducial model $H_{\mathrm{fid}}(z)$ computed at the fixed input parameters $(H_{0,\mathrm{true}}, \Omega_{M,\mathrm{true}})$.}
    \label{fig:LCDM_simulated_dataset}
\end{figure}

We determine $f$ from the observed CC dataset so that the mock datasets reproduce the standardized residual scatter seen in the data. Using the observed CC redshifts $\{z_i^{\mathrm{CC}}\}$, measurements $H_{\mathrm{CC}}(z_i^{\mathrm{CC}})$, and uncertainties $\sigma_{\mathrm{CC}}(z_i^{\mathrm{CC}})$, we compute standardized residuals relative to the same fiducial curve,
\begin{equation}
r_i=\frac{H_{\mathrm{CC}}(z_i^{\mathrm{CC}})-H_{\mathrm{fid}}(z_i^{\mathrm{CC}})}{\sigma_{\mathrm{CC}}(z_i^{\mathrm{CC}})},
\qquad
f=\mathrm{Std}\!\left(\{r_i\}\right),
\label{eq:fcal}
\end{equation}
so that the typical scatter of the simulated measurements in $\sigma$-units matches that of the observed CC compilation. An example realization of the simulated CC dataset is shown in Fig.~\ref{fig:LCDM_simulated_dataset}.

Finally, we generate an ensemble of $R=100$ independent realizations by repeating the above procedure with new random draws of $\{z_i^{\mathrm{sim}}\}$, $\{\sigma_{\mathrm{sim}}(z_i^{\mathrm{sim}})\}$, and $\{\epsilon_i\}$ for each realization, while keeping $H_{0,\mathrm{true}}$ and $\Omega_{M,\mathrm{true}}$ fixed. Each realization therefore consists of the triplets $\{z_i^{\mathrm{sim}},\,H_{\mathrm{sim}}(z_i^{\mathrm{sim}}),\,\sigma_{\mathrm{sim}}(z_i^{\mathrm{sim}})\}_{i=1}^{33}$, which are analysed by EMCEE, GP, and MAF to obtain $H_0$ posterior summaries validated in Section~\ref{subsubsec:Method validation on LCDM based simulations}.

\subsubsection{Method validation on $\Lambda$CDM-based simulations}
\label{subsubsec:Method validation on LCDM based simulations}

In this section we quantify how well EMCEE, GP, and MAF recover a known input value $H_{0,\mathrm{true}}=67.75\ \mathrm{km\,s^{-1}\,Mpc^{-1}}$ from $R=100$ $\Lambda$CDM mock CC datasets. For each realization $r$, each method returns a posterior distribution for $H_0$, which we summarise by a posterior central value $\hat{H}_{0,r}$ and by posterior credible intervals derived from the posterior. We evaluate method performance using three complementary criteria: (i) posterior central value accuracy (Bias and root-mean-square error), (ii) credible interval calibration (coverage of credible intervals), and (iii) overall posterior quality (log score at the truth). The results for all three methods are summarised in Table~\ref{tab:sim_metrics_LCDM}.


\begin{table}
\centering
\caption{Performance metrics for $H_0$ recovery from $R=100$ $\Lambda$CDM-based mock CC datasets.}
\label{tab:sim_metrics_LCDM}
\begin{tabular}{lccccc}
\hline
Method & Bias & RMSE & Cov68 & Cov95 & LogScore \\
\hline
EMCEE & $-0.83$ & $3.15$ & $0.85$ & $1.00$ & $-2.53$ \\
GP    & $-2.25$ & $5.33$ & $0.64$ & $0.99$ & $-2.95$ \\
MAF   & $-0.01$ & $7.22$ & $0.79$ & $0.91$ & $-4.25$ \\
\hline
\end{tabular}
\end{table}


We assess posterior central value accuracy using the bias and the root-mean-square error (RMSE) relative to the truth. Bias diagnoses systematic offset, whereas RMSE captures the combined effect of bias and realization-to-realization variability. The bias is
\begin{equation}
\mathrm{Bias}=\frac{1}{R}\sum_{r=1}^{R}\left(\hat{H}_{0,r}-H_{0,\mathrm{true}}\right),
\label{eq:bias_H0}
\end{equation}
and the RMSE is
\begin{equation}
\mathrm{RMSE}=\left[\frac{1}{R}\sum_{r=1}^{R}\left(\hat{H}_{0,r}-H_{0,\mathrm{true}}\right)^2\right]^{1/2}.
\label{eq:rmse_H0}
\end{equation}
EMCEE achieves the smallest RMSE, indicating the most accurate recovery of $H_{0,\mathrm{true}}$ in typical realizations, whereas GP shows a substantially larger negative bias and larger RMSE. MAF is nearly unbiased on average but exhibits the largest RMSE, implying considerable realization-to-realization variability.

To evaluate whether the quoted posterior credible intervals are calibrated, we compute empirical coverage. For each realisation $r$, we construct the central $68\%$ and $95\%$ credible intervals for $H_0$, and record the fraction of realisations whose interval contains the ground-truth value $H_{0,\mathrm{true}}$. For a well-calibrated method, the empirical coverages should be close to their nominal targets ($\mathrm{Cov}_{68}\simeq 0.68$ and $\mathrm{Cov}_{95}\simeq 0.95$). We emphasize that coverage assesses credible-interval calibration rather than posterior central-value accuracy: values above (below) the nominal targets indicate conservative (overconfident) credible intervals. The $68\%$ and $95\%$ coverages are
\begin{equation}
\mathrm{Cov}_{68}=\frac{1}{R}\sum_{r=1}^{R}\mathbf{1}\!\left(
H_{0,\mathrm{true}}\in \left[H^{\mathrm{lo}}_{68,r},\,H^{\mathrm{hi}}_{68,r}\right]
\right),
\label{eq:cov68_H0}
\end{equation}
\begin{equation}
\mathrm{Cov}_{95}=\frac{1}{R}\sum_{r=1}^{R}\mathbf{1}\!\left(
H_{0,\mathrm{true}}\in \left[H^{\mathrm{lo}}_{95,r},\,H^{\mathrm{hi}}_{95,r}\right]
\right),
\label{eq:cov95_H0}
\end{equation}
where $\left[H^{\mathrm{lo}}_{68,r},H^{\mathrm{hi}}_{68,r}\right]$ and $\left[H^{\mathrm{lo}}_{95,r},H^{\mathrm{hi}}_{95,r}\right]$ denote the central $68\%$ and $95\%$ posterior credible intervals of $H_0$ for realisation $r$. For EMCEE and MAF, these are taken from posterior sample percentiles $\left[q_{16,r},\,q_{84,r}\right]$ and $\left[q_{2.5,r},\,q_{97.5,r}\right]$, while for GP (Gaussian predictive at $z=0$) they correspond to $\left[\mu_r-\sigma_r,\,\mu_r+\sigma_r\right]$ and $\left[\mu_r-1.96\sigma_r,\,\mu_r+1.96\sigma_r\right]$. Here $\mathbf{1}(\cdot)$ denotes the indicator function, equal to $1$ if the condition is satisfied and $0$ otherwise. At the $68\%$ level, GP is closest to the nominal target, whereas EMCEE and MAF show over-coverage, indicating conservative (over-wide) intervals. At the $95\%$ level, EMCEE and GP over-cover, while MAF under-covers, indicating comparatively tighter or slightly overconfident $95\%$ intervals under this simulation setup.

Finally, we summarise posterior quality using the average log score evaluated at the truth. 
For each realisation $r$, each method yields a posterior density $p_r(H_0)$ for $H_0$. 
We define the per-realisation log score as the log posterior density at the known input value,
\begin{equation}
s_r =\log p_r\!\left(H_{0,\mathrm{true}}\right),
\label{eq:logscore_single_H0}
\end{equation}
and average over $R$ realisations,
\begin{equation}
\mathrm{LogScore}=\frac{1}{R}\sum_{r=1}^{R}s_r.
\label{eq:logscore_H0}
\end{equation}
In practice, $p_r(H_0)$ is evaluated from posterior samples for EMCEE and MAF, for GP we use the analytic Gaussian predictive density at $z=0$. EMCEE attains the highest (least negative) log score, indicating the best overall posterior quality under this simulation, while MAF yields the lowest log score and GP lies in between.

Taken together, the metrics in Table~\ref{tab:sim_metrics_LCDM} indicate that EMCEE provides the best overall performance under the present $\Lambda$CDM simulation: it achieves the smallest RMSE and the highest log score, with mildly conservative coverage. GP is intermediate, showing near-nominal coverage overall but a substantial negative bias and a larger RMSE. MAF performs worst overall: although its mean bias is negligible, its RMSE is the largest and its log score is lowest, indicating large realization-to-realization variability and reduced posterior quality.

Since the GP shows the most substantial negative bias in Table~\ref{tab:sim_metrics_LCDM}, we performed an additional sanity check to test whether it could simply arise from a systematic downward shift of the simulated mock-data points $H_{\mathrm{sim}}(z_i^{\mathrm{sim}})$ relative to the fiducial curve $H_{\mathrm{fid}}(z)$. For this check, we define the normalized residual of each simulated mock-data point as 
\begin{equation}
r_i = \frac{ [H_{\mathrm{sim}}(z_i^{\mathrm{sim}})-H_{\mathrm{fid}}(z_i^{\mathrm{sim}})]}{\sigma_{\mathrm{sim}}(z_i^{\mathrm{sim}})}.
\label{eq:normalized_residual}
\end{equation}
We found that these normalized residuals, including those of a low-$z$ subset defined in each realization as the five smallest-redshift points, remain centered close to zero. This indicates that the negative bias seen in the GP posterior central values of $H_0$ cannot be explained simply by a systematic input-side offset in the simulated mock-data points.

\subsection{Simulation based on Gaussian Process}\label{subsec:Simulation based on Gaussian Process}

In this section we perform a complementary simulation based on a Gaussian Process reconstruction of the observed CC compilation to further validate the three constraining $H_0$ methods considered in this work. We adopt a fixed input truth, $H_{0,\mathrm{true}} = 67.21~\mathrm{km\,s^{-1}\,Mpc^{-1}}$, and generate an ensemble of $R=100$ independent GP-based mock CC datasets, so that method performance can be assessed by comparing recovered posteriors to a single well-defined ground truth. Section~\ref{subsubsec:GP-based mock-data simulaion} describes the GP-based mock-data simulation procedure, and Section~\ref{subsubsec:Method validation on GP-based simulations} applies EMCEE, GP, and MAF to the mocks and summarises their performance using the same accuracy, credible interval calibration, and posterior-quality metrics as in Section~\ref{subsubsec:Method validation on LCDM based simulations}.

\subsubsection{GP-based mock-data simulation}\label{subsubsec:GP-based mock-data simulaion}

In Subsection \ref{subsec:Simulation based on LCDM}, we compare the performance of three different methods in constraining $H_0$ using $H(z)$ datasets simulated based on $\Lambda$CDM model. To examine the influence of different simulated model on this study, we simulated $H(z)$ dataset in this subsection based on the Gaussian Process. This GP-based simulation provides a smooth, data-driven fiducial curve and serves as a robustness test of EMCEE, GP, and MAF under a non-parametric $H(z)$ model.

\begin{figure}
    \centering
    \includegraphics[width=8cm]{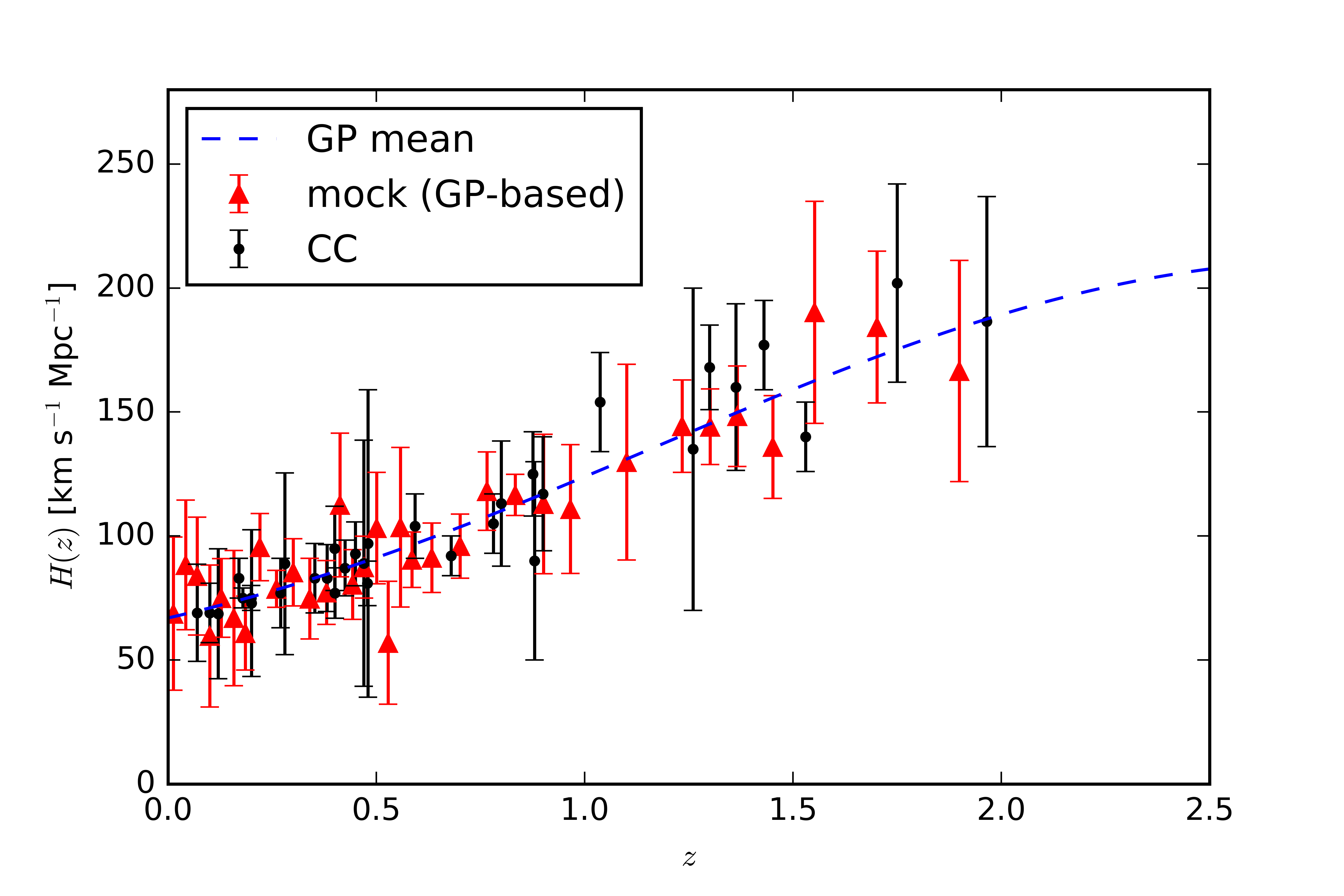}
    \caption{An example simulated $H_{\rm sim}$ dataset based on the GP reconstruction using CC data. The simulated mock-data points, $H_{\rm sim}(z_i^{\rm sim})$, are shown as red triangles with quoted uncertainties $\sigma_{\rm sim}(z_i^{\rm sim})$. For comparison, the observational CC data are shown as black points with error bars. The blue dashed curve shows the GP posterior mean $H_{\rm GP,mean}(z)$ used as the fiducial model in the simulation.}
    \label{fig:simulate_via_GassianProcess0}
\end{figure}

First, we select a fixed input truth at $z=0$. Following our original GP-based setup, we reconstruct the 33 observed CC data points in Table~\ref{tab:cc_data} with a GP and adopt the reconstructed value at $z=0$ as the input truth, $H_{0,\mathrm{true}}=67.21~\mathrm{km\,s^{-1}\,Mpc^{-1}}$ with an associated uncertainty $\sigma_{H_0}=4.72~\mathrm{km \, s^{-1} \, Mpc^{-1}}$. We then add this $z=0$ point to the 33 observed CC points (forming 34 points) and perform a GP reconstruction on the augmented dataset, yielding the GP posterior mean curve $H_{\mathrm{GP,mean}}$ (the blue dashed line in Fig.~\ref{fig:simulate_via_GassianProcess0}). This curve is taken as the fiducial model for generating mock $H(z)$ measurements.

For each realization, the simulated redshifts $z_i^{\mathrm{sim}}$ are generated following the same procedure as in Section~\ref{subsubsec:LCDM-based mock-data simulation}, so that their overall distribution in redshift remains close to that of the observed CC dataset in Table~\ref{tab:cc_data} while still retaining randomness from one realization to another. Each simulated point is assigned a quoted uncertainty $\sigma_{\mathrm{sim}}(z_i^{\mathrm{sim}})$ using the same uncertainty model as Equation~(\ref{eq:sigmasim}) in Section~\ref{subsubsec:LCDM-based mock-data simulation}. The simulated Hubble-parameter measurements are generated by perturbing the GP fiducial curve with Gaussian noise:
\begin{equation}
H_{\mathrm{sim}}(z_i^{\mathrm{sim}})=H_{\mathrm{GP,mean}}(z_i^{\mathrm{sim}})+\epsilon_i,
\label{eq:Hsim_gp}
\end{equation}
\begin{equation}
\epsilon_i \sim \mathcal{N}\!\left(0,\,(f\,\sigma_{\mathrm{sim}}(z_i^{\mathrm{sim}}))^2\right).
\label{eq:epsilon_gp}
\end{equation}
Thus each mock dataset has the CC-like triplet form $\{z_i^{\mathrm{sim}},\,H_{\mathrm{sim}}(z_i^{\mathrm{sim}}),\,\sigma_{\mathrm{sim}}(z_i^{\mathrm{sim}})\}$. The scale factor $f$ is calibrated following the same standardized-residual prescription as Eq.~(\ref{eq:fcal}) in Section~\ref{subsubsec:LCDM-based mock-data simulation}, with the fiducial curve set to $H_{\rm fid}(z)\equiv H_{\rm GP,mean}(z)$ when computing the residuals. Finally, we generate an ensemble of $R=100$ independent GP-based mock datasets by repeating the above procedure with new noise realisations.

\subsubsection{Method validation on GP-based simulations}
\label{subsubsec:Method validation on GP-based simulations}

We analyse the $R=100$ GP-based mock datasets with EMCEE, GP, and MAF using the same inference pipeline as in Section~\ref{subsubsec:Method validation on LCDM based simulations}. Method performance is quantified using the same validation metrics defined in Section~\ref{subsubsec:Method validation on LCDM based simulations} (Bias, RMSE, Cov68, Cov95, and LogScore), evaluated against the fixed input truth $H_{0,\mathrm{true}}=67.21~\mathrm{km\,s^{-1}\,Mpc^{-1}}$. The resulting performance statistics are summarised in Table~\ref{tab:gp_metrics}.
\begin{table}
\centering
\caption{Performance metrics for $H_0$ recovery from $R=100$ GP-based mock CC datasets.}
\label{tab:gp_metrics}
\begin{tabular}{lccccc}
\hline
Method & Bias & RMSE & Cov68 & Cov95 & LogScore \\
\hline
EMCEE & 0.09  & 3.36 & 0.82 & 0.97 & -2.61 \\
GP    & -3.14 & 5.41 & 0.64 & 0.95 & -3.01 \\
MAF   & -0.36 & 8.33 & 0.66 & 0.88 & -4.46 \\
\hline
\end{tabular}
\end{table}

Taken together, the GP-based simulation results indicate that EMCEE provides the best overall performance: it achieves the smallest RMSE and the highest (least negative) LogScore, with mildly conservative coverage. GP is intermediate, exhibiting a substantial negative bias and a larger RMSE, while its 95\% coverage remains close to nominal. MAF performs worst overall: although its mean bias is small, it yields the largest RMSE and the lowest LogScore, with clear under-coverage at the 95\% level, indicating larger realization-to-realization variability and reduced posterior quality under the GP-based mocks. These conclusions are consistent with the $\Lambda$CDM-based validation in Section~\ref{subsubsec:Method validation on LCDM based simulations}, suggesting that the qualitative method ranking is robust to the assumed mock-data generative model.

Since the GP again shows the most substantial negative bias in Table~\ref{tab:gp_metrics}, we performed an additional sanity check for the GP-based mocks, as in Section~\ref{subsubsec:LCDM-based mock-data simulation}. We again found that the residuals of the simulated mock-data points relative to the corresponding fiducial curve remain centered close to zero. This indicates that the negative bias seen in the GP posterior central values of $H_0$ in the GP-based mocks also cannot be explained simply by a systematic input-side offset in the simulated mock-data points.


\section{discussions and conclusions}\label{sec:discussions and conclusions}

The Hubble constant ($H_0$) is crucial for understanding the evolution of the universe. However, a significant 5$\sigma$ discrepancy exists between the two main methods for measuring $H_0$ \citep{2020A&A...641A...6P,2022ApJ...934L...7R}, known as the Hubble tension. Another approach to determining $H_0$ is using Cosmic Chronometers, a model-independent method for measuring $H(z)$. The CC dataset is presented in Table \ref{tab:cc_data}. CC provide an independent route to $H(z)$ and hence to $H_0$. Various methods can be used to constrain $H_0$, including EMCEE, GP, and MAF, as introduced in Subsection \ref{subsec:CC constrains H0}. Our aim is to examine how method choice affects the inferred $H_0$ posterior and to determine which approach is most appropriate for CC-only $H_0$ analyses given different analysis priorities.

Because EMCEE, GP, and MAF are based on different modelling assumptions and levels of flexibility, their CC-based $H_0$ posteriors can differ even when applied to the same dataset. In Subsection~\ref{subsec:Monte Carlo delete d jackknife}, MCDJ reveals clear differences in sensitivity to individual CC measurements: GP is the most sensitive to removing individual $H(z)$ points, whereas EMCEE is the least sensitive (MAF is intermediate). In terms of posterior credible-interval width (constraining power), EMCEE produces the tightest $H_0$ credible intervals, GP is intermediate, and MAF yields the widest intervals. Applying the MCDJ test separately to low- and high-redshift subsets in Subsection~\ref{subsec:Seperate to different areas} shows that all three methods are sensitive in both regimes, but with different patterns: for EMCEE and GP, the posterior central value of $H_0$ is more sensitive to high-$z$ $H(z)$ points, whereas for MAF it is more sensitive to low-$z$ points. This behaviour may reflect methodological differences, but it may also be driven by the limited size of the current CC sample, motivating further tests with larger datasets. In Section~\ref{sec:Simulation}, simulations further show that these differences should be assessed using multiple validation metrics. When evaluated using simulation-based performance measures, central-value accuracy (bias/RMSE), credible-interval calibration (empirical 68\% and 95\% coverage), and overall posterior quality (log score at $H_{0,\mathrm{true}}$), EMCEE performs best overall, GP is intermediate, and MAF performs worst overall.

The three methods differ in model assumptions and effective flexibility: EMCEE performs parametric $\Lambda$CDM inference, GP provides a non-parametric and model-independent reconstruction, and MAF is a flexible density estimator trained on $\Lambda$CDM-simulated parameter-data pairs. These differences naturally lead to a bias-variance tradeoff, so performance must be judged not only by posterior tightness but also by stability to data perturbations and accuracy in controlled tests. In our point-deletion diagnostics (MCDJ, Section~\ref{subsec:Monte Carlo delete d jackknife} and \ref{subsec:Seperate to different areas}), GP is the most sensitive to removing CC points (largest instability indicators; Tables~\ref{tab:MCDJ_33} and \ref{tab:2_separate}), whereas EMCEE is the least sensitive. In terms of posterior credible interval (constraining power), MAF yields the largest $\sigma_{H_0}$, EMCEE the smallest, and GP is intermediate (Table~\ref{tab:MCDJ_sigma}). Systematic bias is assessed separately using controlled mock simulations: in both $\Lambda$CDM-based and GP-based simulations, GP shows the largest bias among the three methods (Tables~\ref{tab:sim_metrics_LCDM} and \ref{tab:gp_metrics}), while the full comparison is based on multiple metrics in Section~\ref{sec:Simulation}. This knowledge helps identify the most suitable method for different research contexts.

If adopting an explicit parametric cosmological model for $H(z)$ is acceptable (we adopt $\Lambda$CDM in this paper), EMCEE is the recommended default. It yields the tightest $H_0$ posterior credible intervals, is the least sensitive to individual CC points in the MCDJ, and performs best overall in controlled simulations when judged jointly by accuracy, interval calibration, and posterior quality. If the priority is model independence, GP is the natural choice for reconstructing $H(z)$, but in our tests it is the most point-sensitive and shows the largest systematic bias in controlled simulations, so it is best used as a complementary diagnostic rather than the default precision $H_0$ method. In our CC-only setup, MAF performs worst overall under the simulation-based validation metrics (bias/RMSE/coverage/log-score), so we do not recommend it as the default. Its main advantage is speed after training when many posteriors must be evaluated, but this does not translate into improved $H_0$ performance for the present CC-only dataset.

An important implication of our validation tests is that the GP reconstruction shows a non-negligible negative bias in the posterior central values of $H_0$. To examine whether this could be caused by an obvious asymmetry in the simulated mock-data points themselves, we carried out additional sanity checks. For both the $\Lambda$CDM-based and GP-based mock datasets, the simulated mock-data points are centered close to the corresponding fiducial curves, both for the full set of simulated points and for a low-redshift subset defined in each realization as the five smallest-redshift points. This indicates that the mock generation itself does not introduce an obvious systematic downward shift near the extrapolation end. However, the negative GP bias persists in both simulation frameworks. We further analyze this point in Appendix~\ref{app:SanityChecks}, where we compare the GP bias with realization-by-realization low-$z$ fluctuations in the simulated mock-data points. The appendix results show that, in the $\Lambda$CDM-based mocks, the GP bias is positively correlated with realization-by-realization low-$z$ fluctuations, as shown in Appendix Fig.~\ref{fig:LCDMbased_sanitycheck_A2_GP_bias_vs_lowz_r}, whereas in the GP-based mocks the negative bias persists without an equally strong low-$z$ correlation, as shown in Appendix Fig.~\ref{fig:GPbased_sanitycheck_A2_GP_bias_vs_lowz_r}. This shows that the GP bias is robust across the two simulation frameworks, but it does not manifest in the same way in the $\Lambda$CDM-based and GP-based simulations. For this reason, we do not apply a simple additive correction to the GP result obtained from the observed CC data. The inferred offset is not a universal calibration constant, but depends on the simulation framework and, in some cases, on realization-specific low-$z$ fluctuations. In this sense, the GP result is best interpreted as a complementary model-independent reconstruction rather than the default precision estimator of $H_0$ in the present CC-only analysis.

The method-selection conclusions above follow from the different assumptions built into the three approaches. EMCEE performs inference within an explicit parametric cosmological model, which ties all CC measurements together through a small set of shared parameters and leads to stable $H_0$ posteriors under data perturbations. GP allows a model-independent, non-parametric reconstruction, which provides greater functional freedom and is consistent with the stronger point sensitivity seen in MCDJ and the larger bias observed in our controlled simulations. MAF is a simulation-trained density estimator; in our CC-only setup it shows weaker posterior quality in the simulation-based validation metrics, consistent with its overall poorer performance relative to EMCEE and GP.

A practical implication of this study is that CC-based $H_0$ results can depend materially on the inference pipeline, even for the same dataset. To make CC-based constraints more interpretable and comparable across studies, we recommend reporting: (i) a posterior summary of $H_0$ (central value and credible interval), (ii) a robustness diagnostic for sensitivity to individual measurements (such as MCDJ), and (iii) simulation-based validation metrics that test accuracy, calibration, and posterior quality (bias/RMSE, empirical 68\% and 95\% coverage, and log score at $H_{0,\mathrm{true}}$). Providing these elements makes method-to-method differences interpretable and enables like-for-like comparisons across analyses.

However, this study has several limitations: (i) The method performance comparison relies on two simulation approaches introduced in Subsection \ref{subsec:Simulation based on LCDM} and \ref{subsec:Simulation based on Gaussian Process}, which may not be universally applicable. (ii) The method performance results are derived from simulations using a prior, rather than actual observational data. (iii) The limited amount of $H(z)$ data in the CC dataset may have affected our results. (iv) This study is limited to two diagnostics, sensitivity to individual CC points and simulation-based performance, and we do not provide a comprehensive assessment of other method properties, such as dependence on modelling assumptions and method configuration choices. (v) We assume independent CC measurements (diagonal covariance), although correlated systematics may be present \citep{2020ApJ...898...82M} and could reduce the effective information content, typically broadening uncertainties. For consistency across methods, we adopt the standard diagonal-noise (independent-measurement) baseline in this comparison. Future work will extend this study by testing a wider range of simulation setups, assessing the impact of correlated CC systematics beyond the diagonal-noise approximation, and repeating the comparison with larger CC samples.


We performed a controlled, like-for-like comparison of EMCEE, GP, and MAF on the same CC dataset. We used MCDJ to quantify sensitivity to individual $H(z)$ data points, and controlled simulations to quantify method performance via validation metrics (bias/RMSE, empirical 68$\%$ and 95$\%$ coverage, and log score at $H_{0,\mathrm{true}}$). EMCEE is the recommended default for model-based CC-only $H_0$ inference. GP is most useful for model-independent reconstruction and as a robustness cross-check. MAF performs worst overall under the simulation-based validation metrics and is therefore not recommended as the default in the present CC-only setup.


\section*{Acknowledgements}

We thank the anonymous referee for the valuable comments and suggestions that helped improve this paper. We thank Yu-Chen Wang and Kang Jiao for their useful discussions. This work is supported by National Key R$\&$D Program of China (2023YFB4503305)，National SKA Program of China (2022SKA0110202), the China Manned Space Program with grant No. CMS-CSST-2025-A01, the National Natural Science Foundation of China (Grants No. 12373109, 61802428), and China Scholarship Council (File No.2306040042).

\section*{DATA AVAILABILITY}
All data included in this study are available upon request by contact with the corresponding author.

\nocite{*}

\bibliographystyle{apsrev4-2}
\bibliography{apssamp}%

\begin{thebibliography}{50}%
\makeatletter
\providecommand \@ifxundefined [1]{%
 \@ifx{#1\undefined}
}%
\providecommand \@ifnum [1]{%
 \ifnum #1\expandafter \@firstoftwo
 \else \expandafter \@secondoftwo
 \fi
}%
\providecommand \@ifx [1]{%
 \ifx #1\expandafter \@firstoftwo
 \else \expandafter \@secondoftwo
 \fi
}%
\providecommand \natexlab [1]{#1}%
\providecommand \enquote  [1]{``#1''}%
\providecommand \bibnamefont  [1]{#1}%
\providecommand \bibfnamefont [1]{#1}%
\providecommand \citenamefont [1]{#1}%
\providecommand \href@noop [0]{\@secondoftwo}%
\providecommand \href [0]{\begingroup \@sanitize@url \@href}%
\providecommand \@href[1]{\@@startlink{#1}\@@href}%
\providecommand \@@href[1]{\endgroup#1\@@endlink}%
\providecommand \@sanitize@url [0]{\catcode `\\12\catcode `\$12\catcode `\&12\catcode `\#12\catcode `\^12\catcode `\_12\catcode `\%12\relax}%
\providecommand \@@startlink[1]{}%
\providecommand \@@endlink[0]{}%
\providecommand \url  [0]{\begingroup\@sanitize@url \@url }%
\providecommand \@url [1]{\endgroup\@href {#1}{\urlprefix }}%
\providecommand \urlprefix  [0]{URL }%
\providecommand \Eprint [0]{\href }%
\providecommand \doibase [0]{https://doi.org/}%
\providecommand \selectlanguage [0]{\@gobble}%
\providecommand \bibinfo  [0]{\@secondoftwo}%
\providecommand \bibfield  [0]{\@secondoftwo}%
\providecommand \translation [1]{[#1]}%
\providecommand \BibitemOpen [0]{}%
\providecommand \bibitemStop [0]{}%
\providecommand \bibitemNoStop [0]{.\EOS\space}%
\providecommand \EOS [0]{\spacefactor3000\relax}%
\providecommand \BibitemShut  [1]{\csname bibitem#1\endcsname}%
\let\auto@bib@innerbib\@empty
\bibitem [{\citenamefont {{Planck Collaboration}}\ \emph {et~al.}(2020)\citenamefont {{Planck Collaboration}}, \citenamefont {{Aghanim}}, \citenamefont {{Akrami}}, \citenamefont {{Ashdown}}, \citenamefont {{Aumont}}, \citenamefont {{Baccigalupi}}, \citenamefont {{Ballardini}}, \citenamefont {{Banday}}, \citenamefont {{Barreiro}}, \citenamefont {{Bartolo}}, \citenamefont {{Basak}}, \citenamefont {{Battye}}, \citenamefont {{Benabed}}, \citenamefont {{Bernard}}, \citenamefont {{Bersanelli}}, \citenamefont {{Bielewicz}}, \citenamefont {{Bock}}, \citenamefont {{Bond}}, \citenamefont {{Borrill}}, \citenamefont {{Bouchet}}, \citenamefont {{Boulanger}}, \citenamefont {{Bucher}}, \citenamefont {{Burigana}}, \citenamefont {{Butler}}, \citenamefont {{Calabrese}}, \citenamefont {{Cardoso}}, \citenamefont {{Carron}}, \citenamefont {{Challinor}}, \citenamefont {{Chiang}}, \citenamefont {{Chluba}}, \citenamefont {{Colombo}}, \citenamefont {{Combet}}, \citenamefont {{Contreras}}, \citenamefont {{Crill}}, \citenamefont
  {{Cuttaia}}, \citenamefont {{de Bernardis}}, \citenamefont {{de Zotti}}, \citenamefont {{Delabrouille}}, \citenamefont {{Delouis}}, \citenamefont {{Di Valentino}}, \citenamefont {{Diego}}, \citenamefont {{Dor{\'e}}}, \citenamefont {{Douspis}}, \citenamefont {{Ducout}}, \citenamefont {{Dupac}}, \citenamefont {{Dusini}}, \citenamefont {{Efstathiou}}, \citenamefont {{Elsner}}, \citenamefont {{En{\ss}lin}}, \citenamefont {{Eriksen}}, \citenamefont {{Fantaye}}, \citenamefont {{Farhang}}, \citenamefont {{Fergusson}}, \citenamefont {{Fernandez-Cobos}}, \citenamefont {{Finelli}}, \citenamefont {{Forastieri}}, \citenamefont {{Frailis}}, \citenamefont {{Fraisse}}, \citenamefont {{Franceschi}}, \citenamefont {{Frolov}}, \citenamefont {{Galeotta}}, \citenamefont {{Galli}}, \citenamefont {{Ganga}}, \citenamefont {{G{\'e}nova-Santos}}, \citenamefont {{Gerbino}}, \citenamefont {{Ghosh}}, \citenamefont {{Gonz{\'a}lez-Nuevo}}, \citenamefont {{G{\'o}rski}}, \citenamefont {{Gratton}}, \citenamefont {{Gruppuso}}, \citenamefont
  {{Gudmundsson}}, \citenamefont {{Hamann}}, \citenamefont {{Handley}}, \citenamefont {{Hansen}}, \citenamefont {{Herranz}}, \citenamefont {{Hildebrandt}}, \citenamefont {{Hivon}}, \citenamefont {{Huang}}, \citenamefont {{Jaffe}}, \citenamefont {{Jones}}, \citenamefont {{Karakci}}, \citenamefont {{Keih{\"a}nen}}, \citenamefont {{Keskitalo}}, \citenamefont {{Kiiveri}}, \citenamefont {{Kim}}, \citenamefont {{Kisner}}, \citenamefont {{Knox}}, \citenamefont {{Krachmalnicoff}}, \citenamefont {{Kunz}}, \citenamefont {{Kurki-Suonio}}, \citenamefont {{Lagache}}, \citenamefont {{Lamarre}}, \citenamefont {{Lasenby}}, \citenamefont {{Lattanzi}}, \citenamefont {{Lawrence}}, \citenamefont {{Le Jeune}}, \citenamefont {{Lemos}}, \citenamefont {{Lesgourgues}}, \citenamefont {{Levrier}}, \citenamefont {{Lewis}}, \citenamefont {{Liguori}}, \citenamefont {{Lilje}}, \citenamefont {{Lilley}}, \citenamefont {{Lindholm}}, \citenamefont {{L{\'o}pez-Caniego}}, \citenamefont {{Lubin}}, \citenamefont {{Ma}}, \citenamefont
  {{Mac{\'\i}as-P{\'e}rez}}, \citenamefont {{Maggio}}, \citenamefont {{Maino}}, \citenamefont {{Mandolesi}}, \citenamefont {{Mangilli}}, \citenamefont {{Marcos-Caballero}}, \citenamefont {{Maris}}, \citenamefont {{Martin}}, \citenamefont {{Martinelli}}, \citenamefont {{Mart{\'\i}nez-Gonz{\'a}lez}}, \citenamefont {{Matarrese}}, \citenamefont {{Mauri}}, \citenamefont {{McEwen}}, \citenamefont {{Meinhold}}, \citenamefont {{Melchiorri}}, \citenamefont {{Mennella}}, \citenamefont {{Migliaccio}}, \citenamefont {{Millea}}, \citenamefont {{Mitra}}, \citenamefont {{Miville-Desch{\^e}nes}}, \citenamefont {{Molinari}}, \citenamefont {{Montier}}, \citenamefont {{Morgante}}, \citenamefont {{Moss}}, \citenamefont {{Natoli}}, \citenamefont {{N{\o}rgaard-Nielsen}}, \citenamefont {{Pagano}}, \citenamefont {{Paoletti}}, \citenamefont {{Partridge}}, \citenamefont {{Patanchon}}, \citenamefont {{Peiris}}, \citenamefont {{Perrotta}}, \citenamefont {{Pettorino}}, \citenamefont {{Piacentini}}, \citenamefont {{Polastri}},
  \citenamefont {{Polenta}}, \citenamefont {{Puget}}, \citenamefont {{Rachen}}, \citenamefont {{Reinecke}}, \citenamefont {{Remazeilles}}, \citenamefont {{Renzi}}, \citenamefont {{Rocha}}, \citenamefont {{Rosset}}, \citenamefont {{Roudier}}, \citenamefont {{Rubi{\~n}o-Mart{\'\i}n}}, \citenamefont {{Ruiz-Granados}}, \citenamefont {{Salvati}}, \citenamefont {{Sandri}}, \citenamefont {{Savelainen}}, \citenamefont {{Scott}}, \citenamefont {{Shellard}}, \citenamefont {{Sirignano}}, \citenamefont {{Sirri}}, \citenamefont {{Spencer}}, \citenamefont {{Sunyaev}}, \citenamefont {{Suur-Uski}}, \citenamefont {{Tauber}}, \citenamefont {{Tavagnacco}}, \citenamefont {{Tenti}}, \citenamefont {{Toffolatti}}, \citenamefont {{Tomasi}}, \citenamefont {{Trombetti}}, \citenamefont {{Valenziano}}, \citenamefont {{Valiviita}}, \citenamefont {{Van Tent}}, \citenamefont {{Vibert}}, \citenamefont {{Vielva}}, \citenamefont {{Villa}}, \citenamefont {{Vittorio}}, \citenamefont {{Wandelt}}, \citenamefont {{Wehus}}, \citenamefont {{White}},
  \citenamefont {{White}}, \citenamefont {{Zacchei}},\ and\ \citenamefont {{Zonca}}}]{2020A&A...641A...6P}%
  \BibitemOpen
  \bibfield  {author} {\bibinfo {author} {\bibnamefont {{Planck Collaboration}}}, \bibinfo {author} {\bibfnamefont {N.}~\bibnamefont {{Aghanim}}}, \bibinfo {author} {\bibfnamefont {Y.}~\bibnamefont {{Akrami}}}, \bibinfo {author} {\bibfnamefont {M.}~\bibnamefont {{Ashdown}}}, \bibinfo {author} {\bibfnamefont {J.}~\bibnamefont {{Aumont}}}, \bibinfo {author} {\bibfnamefont {C.}~\bibnamefont {{Baccigalupi}}}, \bibinfo {author} {\bibfnamefont {M.}~\bibnamefont {{Ballardini}}}, \bibinfo {author} {\bibfnamefont {A.~J.}\ \bibnamefont {{Banday}}}, \bibinfo {author} {\bibfnamefont {R.~B.}\ \bibnamefont {{Barreiro}}}, \bibinfo {author} {\bibfnamefont {N.}~\bibnamefont {{Bartolo}}}, \bibinfo {author} {\bibfnamefont {S.}~\bibnamefont {{Basak}}}, \bibinfo {author} {\bibfnamefont {R.}~\bibnamefont {{Battye}}}, \bibinfo {author} {\bibfnamefont {K.}~\bibnamefont {{Benabed}}}, \bibinfo {author} {\bibfnamefont {J.~P.}\ \bibnamefont {{Bernard}}}, \bibinfo {author} {\bibfnamefont {M.}~\bibnamefont {{Bersanelli}}}, \bibinfo
  {author} {\bibfnamefont {P.}~\bibnamefont {{Bielewicz}}}, \bibinfo {author} {\bibfnamefont {J.~J.}\ \bibnamefont {{Bock}}}, \bibinfo {author} {\bibfnamefont {J.~R.}\ \bibnamefont {{Bond}}}, \bibinfo {author} {\bibfnamefont {J.}~\bibnamefont {{Borrill}}}, \bibinfo {author} {\bibfnamefont {F.~R.}\ \bibnamefont {{Bouchet}}}, \bibinfo {author} {\bibfnamefont {F.}~\bibnamefont {{Boulanger}}}, \bibinfo {author} {\bibfnamefont {M.}~\bibnamefont {{Bucher}}}, \bibinfo {author} {\bibfnamefont {C.}~\bibnamefont {{Burigana}}}, \bibinfo {author} {\bibfnamefont {R.~C.}\ \bibnamefont {{Butler}}}, \bibinfo {author} {\bibfnamefont {E.}~\bibnamefont {{Calabrese}}}, \bibinfo {author} {\bibfnamefont {J.~F.}\ \bibnamefont {{Cardoso}}}, \bibinfo {author} {\bibfnamefont {J.}~\bibnamefont {{Carron}}}, \bibinfo {author} {\bibfnamefont {A.}~\bibnamefont {{Challinor}}}, \bibinfo {author} {\bibfnamefont {H.~C.}\ \bibnamefont {{Chiang}}}, \bibinfo {author} {\bibfnamefont {J.}~\bibnamefont {{Chluba}}}, \bibinfo {author} {\bibfnamefont
  {L.~P.~L.}\ \bibnamefont {{Colombo}}}, \bibinfo {author} {\bibfnamefont {C.}~\bibnamefont {{Combet}}}, \bibinfo {author} {\bibfnamefont {D.}~\bibnamefont {{Contreras}}}, \bibinfo {author} {\bibfnamefont {B.~P.}\ \bibnamefont {{Crill}}}, \bibinfo {author} {\bibfnamefont {F.}~\bibnamefont {{Cuttaia}}}, \bibinfo {author} {\bibfnamefont {P.}~\bibnamefont {{de Bernardis}}}, \bibinfo {author} {\bibfnamefont {G.}~\bibnamefont {{de Zotti}}}, \bibinfo {author} {\bibfnamefont {J.}~\bibnamefont {{Delabrouille}}}, \bibinfo {author} {\bibfnamefont {J.~M.}\ \bibnamefont {{Delouis}}}, \bibinfo {author} {\bibfnamefont {E.}~\bibnamefont {{Di Valentino}}}, \bibinfo {author} {\bibfnamefont {J.~M.}\ \bibnamefont {{Diego}}}, \bibinfo {author} {\bibfnamefont {O.}~\bibnamefont {{Dor{\'e}}}}, \bibinfo {author} {\bibfnamefont {M.}~\bibnamefont {{Douspis}}}, \bibinfo {author} {\bibfnamefont {A.}~\bibnamefont {{Ducout}}}, \bibinfo {author} {\bibfnamefont {X.}~\bibnamefont {{Dupac}}}, \bibinfo {author} {\bibfnamefont {S.}~\bibnamefont
  {{Dusini}}}, \bibinfo {author} {\bibfnamefont {G.}~\bibnamefont {{Efstathiou}}}, \bibinfo {author} {\bibfnamefont {F.}~\bibnamefont {{Elsner}}}, \bibinfo {author} {\bibfnamefont {T.~A.}\ \bibnamefont {{En{\ss}lin}}}, \bibinfo {author} {\bibfnamefont {H.~K.}\ \bibnamefont {{Eriksen}}}, \bibinfo {author} {\bibfnamefont {Y.}~\bibnamefont {{Fantaye}}}, \bibinfo {author} {\bibfnamefont {M.}~\bibnamefont {{Farhang}}}, \bibinfo {author} {\bibfnamefont {J.}~\bibnamefont {{Fergusson}}}, \bibinfo {author} {\bibfnamefont {R.}~\bibnamefont {{Fernandez-Cobos}}}, \bibinfo {author} {\bibfnamefont {F.}~\bibnamefont {{Finelli}}}, \bibinfo {author} {\bibfnamefont {F.}~\bibnamefont {{Forastieri}}}, \bibinfo {author} {\bibfnamefont {M.}~\bibnamefont {{Frailis}}}, \bibinfo {author} {\bibfnamefont {A.~A.}\ \bibnamefont {{Fraisse}}}, \bibinfo {author} {\bibfnamefont {E.}~\bibnamefont {{Franceschi}}}, \bibinfo {author} {\bibfnamefont {A.}~\bibnamefont {{Frolov}}}, \bibinfo {author} {\bibfnamefont {S.}~\bibnamefont {{Galeotta}}},
  \bibinfo {author} {\bibfnamefont {S.}~\bibnamefont {{Galli}}}, \bibinfo {author} {\bibfnamefont {K.}~\bibnamefont {{Ganga}}}, \bibinfo {author} {\bibfnamefont {R.~T.}\ \bibnamefont {{G{\'e}nova-Santos}}}, \bibinfo {author} {\bibfnamefont {M.}~\bibnamefont {{Gerbino}}}, \bibinfo {author} {\bibfnamefont {T.}~\bibnamefont {{Ghosh}}}, \bibinfo {author} {\bibfnamefont {J.}~\bibnamefont {{Gonz{\'a}lez-Nuevo}}}, \bibinfo {author} {\bibfnamefont {K.~M.}\ \bibnamefont {{G{\'o}rski}}}, \bibinfo {author} {\bibfnamefont {S.}~\bibnamefont {{Gratton}}}, \bibinfo {author} {\bibfnamefont {A.}~\bibnamefont {{Gruppuso}}}, \bibinfo {author} {\bibfnamefont {J.~E.}\ \bibnamefont {{Gudmundsson}}}, \bibinfo {author} {\bibfnamefont {J.}~\bibnamefont {{Hamann}}}, \bibinfo {author} {\bibfnamefont {W.}~\bibnamefont {{Handley}}}, \bibinfo {author} {\bibfnamefont {F.~K.}\ \bibnamefont {{Hansen}}}, \bibinfo {author} {\bibfnamefont {D.}~\bibnamefont {{Herranz}}}, \bibinfo {author} {\bibfnamefont {S.~R.}\ \bibnamefont {{Hildebrandt}}},
  \bibinfo {author} {\bibfnamefont {E.}~\bibnamefont {{Hivon}}}, \bibinfo {author} {\bibfnamefont {Z.}~\bibnamefont {{Huang}}}, \bibinfo {author} {\bibfnamefont {A.~H.}\ \bibnamefont {{Jaffe}}}, \bibinfo {author} {\bibfnamefont {W.~C.}\ \bibnamefont {{Jones}}}, \bibinfo {author} {\bibfnamefont {A.}~\bibnamefont {{Karakci}}}, \bibinfo {author} {\bibfnamefont {E.}~\bibnamefont {{Keih{\"a}nen}}}, \bibinfo {author} {\bibfnamefont {R.}~\bibnamefont {{Keskitalo}}}, \bibinfo {author} {\bibfnamefont {K.}~\bibnamefont {{Kiiveri}}}, \bibinfo {author} {\bibfnamefont {J.}~\bibnamefont {{Kim}}}, \bibinfo {author} {\bibfnamefont {T.~S.}\ \bibnamefont {{Kisner}}}, \bibinfo {author} {\bibfnamefont {L.}~\bibnamefont {{Knox}}}, \bibinfo {author} {\bibfnamefont {N.}~\bibnamefont {{Krachmalnicoff}}}, \bibinfo {author} {\bibfnamefont {M.}~\bibnamefont {{Kunz}}}, \bibinfo {author} {\bibfnamefont {H.}~\bibnamefont {{Kurki-Suonio}}}, \bibinfo {author} {\bibfnamefont {G.}~\bibnamefont {{Lagache}}}, \bibinfo {author} {\bibfnamefont
  {J.~M.}\ \bibnamefont {{Lamarre}}}, \bibinfo {author} {\bibfnamefont {A.}~\bibnamefont {{Lasenby}}}, \bibinfo {author} {\bibfnamefont {M.}~\bibnamefont {{Lattanzi}}}, \bibinfo {author} {\bibfnamefont {C.~R.}\ \bibnamefont {{Lawrence}}}, \bibinfo {author} {\bibfnamefont {M.}~\bibnamefont {{Le Jeune}}}, \bibinfo {author} {\bibfnamefont {P.}~\bibnamefont {{Lemos}}}, \bibinfo {author} {\bibfnamefont {J.}~\bibnamefont {{Lesgourgues}}}, \bibinfo {author} {\bibfnamefont {F.}~\bibnamefont {{Levrier}}}, \bibinfo {author} {\bibfnamefont {A.}~\bibnamefont {{Lewis}}}, \bibinfo {author} {\bibfnamefont {M.}~\bibnamefont {{Liguori}}}, \bibinfo {author} {\bibfnamefont {P.~B.}\ \bibnamefont {{Lilje}}}, \bibinfo {author} {\bibfnamefont {M.}~\bibnamefont {{Lilley}}}, \bibinfo {author} {\bibfnamefont {V.}~\bibnamefont {{Lindholm}}}, \bibinfo {author} {\bibfnamefont {M.}~\bibnamefont {{L{\'o}pez-Caniego}}}, \bibinfo {author} {\bibfnamefont {P.~M.}\ \bibnamefont {{Lubin}}}, \bibinfo {author} {\bibfnamefont {Y.~Z.}\ \bibnamefont
  {{Ma}}}, \bibinfo {author} {\bibfnamefont {J.~F.}\ \bibnamefont {{Mac{\'\i}as-P{\'e}rez}}}, \bibinfo {author} {\bibfnamefont {G.}~\bibnamefont {{Maggio}}}, \bibinfo {author} {\bibfnamefont {D.}~\bibnamefont {{Maino}}}, \bibinfo {author} {\bibfnamefont {N.}~\bibnamefont {{Mandolesi}}}, \bibinfo {author} {\bibfnamefont {A.}~\bibnamefont {{Mangilli}}}, \bibinfo {author} {\bibfnamefont {A.}~\bibnamefont {{Marcos-Caballero}}}, \bibinfo {author} {\bibfnamefont {M.}~\bibnamefont {{Maris}}}, \bibinfo {author} {\bibfnamefont {P.~G.}\ \bibnamefont {{Martin}}}, \bibinfo {author} {\bibfnamefont {M.}~\bibnamefont {{Martinelli}}}, \bibinfo {author} {\bibfnamefont {E.}~\bibnamefont {{Mart{\'\i}nez-Gonz{\'a}lez}}}, \bibinfo {author} {\bibfnamefont {S.}~\bibnamefont {{Matarrese}}}, \bibinfo {author} {\bibfnamefont {N.}~\bibnamefont {{Mauri}}}, \bibinfo {author} {\bibfnamefont {J.~D.}\ \bibnamefont {{McEwen}}}, \bibinfo {author} {\bibfnamefont {P.~R.}\ \bibnamefont {{Meinhold}}}, \bibinfo {author} {\bibfnamefont
  {A.}~\bibnamefont {{Melchiorri}}}, \bibinfo {author} {\bibfnamefont {A.}~\bibnamefont {{Mennella}}}, \bibinfo {author} {\bibfnamefont {M.}~\bibnamefont {{Migliaccio}}}, \bibinfo {author} {\bibfnamefont {M.}~\bibnamefont {{Millea}}}, \bibinfo {author} {\bibfnamefont {S.}~\bibnamefont {{Mitra}}}, \bibinfo {author} {\bibfnamefont {M.~A.}\ \bibnamefont {{Miville-Desch{\^e}nes}}}, \bibinfo {author} {\bibfnamefont {D.}~\bibnamefont {{Molinari}}}, \bibinfo {author} {\bibfnamefont {L.}~\bibnamefont {{Montier}}}, \bibinfo {author} {\bibfnamefont {G.}~\bibnamefont {{Morgante}}}, \bibinfo {author} {\bibfnamefont {A.}~\bibnamefont {{Moss}}}, \bibinfo {author} {\bibfnamefont {P.}~\bibnamefont {{Natoli}}}, \bibinfo {author} {\bibfnamefont {H.~U.}\ \bibnamefont {{N{\o}rgaard-Nielsen}}}, \bibinfo {author} {\bibfnamefont {L.}~\bibnamefont {{Pagano}}}, \bibinfo {author} {\bibfnamefont {D.}~\bibnamefont {{Paoletti}}}, \bibinfo {author} {\bibfnamefont {B.}~\bibnamefont {{Partridge}}}, \bibinfo {author} {\bibfnamefont
  {G.}~\bibnamefont {{Patanchon}}}, \bibinfo {author} {\bibfnamefont {H.~V.}\ \bibnamefont {{Peiris}}}, \bibinfo {author} {\bibfnamefont {F.}~\bibnamefont {{Perrotta}}}, \bibinfo {author} {\bibfnamefont {V.}~\bibnamefont {{Pettorino}}}, \bibinfo {author} {\bibfnamefont {F.}~\bibnamefont {{Piacentini}}}, \bibinfo {author} {\bibfnamefont {L.}~\bibnamefont {{Polastri}}}, \bibinfo {author} {\bibfnamefont {G.}~\bibnamefont {{Polenta}}}, \bibinfo {author} {\bibfnamefont {J.~L.}\ \bibnamefont {{Puget}}}, \bibinfo {author} {\bibfnamefont {J.~P.}\ \bibnamefont {{Rachen}}}, \bibinfo {author} {\bibfnamefont {M.}~\bibnamefont {{Reinecke}}}, \bibinfo {author} {\bibfnamefont {M.}~\bibnamefont {{Remazeilles}}}, \bibinfo {author} {\bibfnamefont {A.}~\bibnamefont {{Renzi}}}, \bibinfo {author} {\bibfnamefont {G.}~\bibnamefont {{Rocha}}}, \bibinfo {author} {\bibfnamefont {C.}~\bibnamefont {{Rosset}}}, \bibinfo {author} {\bibfnamefont {G.}~\bibnamefont {{Roudier}}}, \bibinfo {author} {\bibfnamefont {J.~A.}\ \bibnamefont
  {{Rubi{\~n}o-Mart{\'\i}n}}}, \bibinfo {author} {\bibfnamefont {B.}~\bibnamefont {{Ruiz-Granados}}}, \bibinfo {author} {\bibfnamefont {L.}~\bibnamefont {{Salvati}}}, \bibinfo {author} {\bibfnamefont {M.}~\bibnamefont {{Sandri}}}, \bibinfo {author} {\bibfnamefont {M.}~\bibnamefont {{Savelainen}}}, \bibinfo {author} {\bibfnamefont {D.}~\bibnamefont {{Scott}}}, \bibinfo {author} {\bibfnamefont {E.~P.~S.}\ \bibnamefont {{Shellard}}}, \bibinfo {author} {\bibfnamefont {C.}~\bibnamefont {{Sirignano}}}, \bibinfo {author} {\bibfnamefont {G.}~\bibnamefont {{Sirri}}}, \bibinfo {author} {\bibfnamefont {L.~D.}\ \bibnamefont {{Spencer}}}, \bibinfo {author} {\bibfnamefont {R.}~\bibnamefont {{Sunyaev}}}, \bibinfo {author} {\bibfnamefont {A.~S.}\ \bibnamefont {{Suur-Uski}}}, \bibinfo {author} {\bibfnamefont {J.~A.}\ \bibnamefont {{Tauber}}}, \bibinfo {author} {\bibfnamefont {D.}~\bibnamefont {{Tavagnacco}}}, \bibinfo {author} {\bibfnamefont {M.}~\bibnamefont {{Tenti}}}, \bibinfo {author} {\bibfnamefont {L.}~\bibnamefont
  {{Toffolatti}}}, \bibinfo {author} {\bibfnamefont {M.}~\bibnamefont {{Tomasi}}}, \bibinfo {author} {\bibfnamefont {T.}~\bibnamefont {{Trombetti}}}, \bibinfo {author} {\bibfnamefont {L.}~\bibnamefont {{Valenziano}}}, \bibinfo {author} {\bibfnamefont {J.}~\bibnamefont {{Valiviita}}}, \bibinfo {author} {\bibfnamefont {B.}~\bibnamefont {{Van Tent}}}, \bibinfo {author} {\bibfnamefont {L.}~\bibnamefont {{Vibert}}}, \bibinfo {author} {\bibfnamefont {P.}~\bibnamefont {{Vielva}}}, \bibinfo {author} {\bibfnamefont {F.}~\bibnamefont {{Villa}}}, \bibinfo {author} {\bibfnamefont {N.}~\bibnamefont {{Vittorio}}}, \bibinfo {author} {\bibfnamefont {B.~D.}\ \bibnamefont {{Wandelt}}}, \bibinfo {author} {\bibfnamefont {I.~K.}\ \bibnamefont {{Wehus}}}, \bibinfo {author} {\bibfnamefont {M.}~\bibnamefont {{White}}}, \bibinfo {author} {\bibfnamefont {S.~D.~M.}\ \bibnamefont {{White}}}, \bibinfo {author} {\bibfnamefont {A.}~\bibnamefont {{Zacchei}}},\ and\ \bibinfo {author} {\bibfnamefont {A.}~\bibnamefont {{Zonca}}},\ }\href
  {https://doi.org/10.1051/0004-6361/201833910} {\bibfield  {journal} {\bibinfo  {journal} {\aap}\ }\textbf {\bibinfo {volume} {641}},\ \bibinfo {eid} {A6} (\bibinfo {year} {2020})},\ \Eprint {https://arxiv.org/abs/1807.06209} {arXiv:1807.06209 [astro-ph.CO]} \BibitemShut {NoStop}%
\bibitem [{\citenamefont {{Riess}}\ \emph {et~al.}(2022)\citenamefont {{Riess}}, \citenamefont {{Yuan}}, \citenamefont {{Macri}}, \citenamefont {{Scolnic}}, \citenamefont {{Brout}}, \citenamefont {{Casertano}}, \citenamefont {{Jones}}, \citenamefont {{Murakami}}, \citenamefont {{Anand}}, \citenamefont {{Breuval}}, \citenamefont {{Brink}}, \citenamefont {{Filippenko}}, \citenamefont {{Hoffmann}}, \citenamefont {{Jha}}, \citenamefont {{D'arcy Kenworthy}}, \citenamefont {{Mackenty}}, \citenamefont {{Stahl}},\ and\ \citenamefont {{Zheng}}}]{2022ApJ...934L...7R}%
  \BibitemOpen
  \bibfield  {author} {\bibinfo {author} {\bibfnamefont {A.~G.}\ \bibnamefont {{Riess}}}, \bibinfo {author} {\bibfnamefont {W.}~\bibnamefont {{Yuan}}}, \bibinfo {author} {\bibfnamefont {L.~M.}\ \bibnamefont {{Macri}}}, \bibinfo {author} {\bibfnamefont {D.}~\bibnamefont {{Scolnic}}}, \bibinfo {author} {\bibfnamefont {D.}~\bibnamefont {{Brout}}}, \bibinfo {author} {\bibfnamefont {S.}~\bibnamefont {{Casertano}}}, \bibinfo {author} {\bibfnamefont {D.~O.}\ \bibnamefont {{Jones}}}, \bibinfo {author} {\bibfnamefont {Y.}~\bibnamefont {{Murakami}}}, \bibinfo {author} {\bibfnamefont {G.~S.}\ \bibnamefont {{Anand}}}, \bibinfo {author} {\bibfnamefont {L.}~\bibnamefont {{Breuval}}}, \bibinfo {author} {\bibfnamefont {T.~G.}\ \bibnamefont {{Brink}}}, \bibinfo {author} {\bibfnamefont {A.~V.}\ \bibnamefont {{Filippenko}}}, \bibinfo {author} {\bibfnamefont {S.}~\bibnamefont {{Hoffmann}}}, \bibinfo {author} {\bibfnamefont {S.~W.}\ \bibnamefont {{Jha}}}, \bibinfo {author} {\bibfnamefont {W.}~\bibnamefont {{D'arcy Kenworthy}}},
  \bibinfo {author} {\bibfnamefont {J.}~\bibnamefont {{Mackenty}}}, \bibinfo {author} {\bibfnamefont {B.~E.}\ \bibnamefont {{Stahl}}},\ and\ \bibinfo {author} {\bibfnamefont {W.}~\bibnamefont {{Zheng}}},\ }\href {https://doi.org/10.3847/2041-8213/ac5c5b} {\bibfield  {journal} {\bibinfo  {journal} {\apjl}\ }\textbf {\bibinfo {volume} {934}},\ \bibinfo {eid} {L7} (\bibinfo {year} {2022})},\ \Eprint {https://arxiv.org/abs/2112.04510} {arXiv:2112.04510 [astro-ph.CO]} \BibitemShut {NoStop}%
\bibitem [{\citenamefont {{Verde}}\ \emph {et~al.}(2019)\citenamefont {{Verde}}, \citenamefont {{Treu}},\ and\ \citenamefont {{Riess}}}]{2019NatAs...3..891V}%
  \BibitemOpen
  \bibfield  {author} {\bibinfo {author} {\bibfnamefont {L.}~\bibnamefont {{Verde}}}, \bibinfo {author} {\bibfnamefont {T.}~\bibnamefont {{Treu}}},\ and\ \bibinfo {author} {\bibfnamefont {A.~G.}\ \bibnamefont {{Riess}}},\ }\href {https://doi.org/10.1038/s41550-019-0902-0} {\bibfield  {journal} {\bibinfo  {journal} {Nature Astronomy}\ }\textbf {\bibinfo {volume} {3}},\ \bibinfo {pages} {891} (\bibinfo {year} {2019})},\ \Eprint {https://arxiv.org/abs/1907.10625} {arXiv:1907.10625 [astro-ph.CO]} \BibitemShut {NoStop}%
\bibitem [{\citenamefont {{Riess}}(2020)}]{2020NatRP...2...10R}%
  \BibitemOpen
  \bibfield  {author} {\bibinfo {author} {\bibfnamefont {A.~G.}\ \bibnamefont {{Riess}}},\ }\href {https://doi.org/10.1038/s42254-019-0137-0} {\bibfield  {journal} {\bibinfo  {journal} {Nature Reviews Physics}\ }\textbf {\bibinfo {volume} {2}},\ \bibinfo {pages} {10} (\bibinfo {year} {2020})},\ \Eprint {https://arxiv.org/abs/2001.03624} {arXiv:2001.03624 [astro-ph.CO]} \BibitemShut {NoStop}%
\bibitem [{\citenamefont {{Foreman-Mackey}}\ \emph {et~al.}(2013)\citenamefont {{Foreman-Mackey}}, \citenamefont {{Hogg}}, \citenamefont {{Lang}},\ and\ \citenamefont {{Goodman}}}]{2013PASP..125..306F}%
  \BibitemOpen
  \bibfield  {author} {\bibinfo {author} {\bibfnamefont {D.}~\bibnamefont {{Foreman-Mackey}}}, \bibinfo {author} {\bibfnamefont {D.~W.}\ \bibnamefont {{Hogg}}}, \bibinfo {author} {\bibfnamefont {D.}~\bibnamefont {{Lang}}},\ and\ \bibinfo {author} {\bibfnamefont {J.}~\bibnamefont {{Goodman}}},\ }\href {https://doi.org/10.1086/670067} {\bibfield  {journal} {\bibinfo  {journal} {\pasp}\ }\textbf {\bibinfo {volume} {125}},\ \bibinfo {pages} {306} (\bibinfo {year} {2013})},\ \Eprint {https://arxiv.org/abs/1202.3665} {arXiv:1202.3665 [astro-ph.IM]} \BibitemShut {NoStop}%
\bibitem [{\citenamefont {{Rasmussen}}\ and\ \citenamefont {{Williams}}(2006)}]{2006gpml.book.....R}%
  \BibitemOpen
  \bibfield  {author} {\bibinfo {author} {\bibfnamefont {C.~E.}\ \bibnamefont {{Rasmussen}}}\ and\ \bibinfo {author} {\bibfnamefont {C.~K.~I.}\ \bibnamefont {{Williams}}},\ }\href@noop {} {\emph {\bibinfo {title} {{Gaussian Processes for Machine Learning}}}}\ (\bibinfo {year} {2006})\BibitemShut {NoStop}%
\bibitem [{\citenamefont {{Seikel}}\ \emph {et~al.}(2012)\citenamefont {{Seikel}}, \citenamefont {{Clarkson}},\ and\ \citenamefont {{Smith}}}]{2012JCAP...06..036S}%
  \BibitemOpen
  \bibfield  {author} {\bibinfo {author} {\bibfnamefont {M.}~\bibnamefont {{Seikel}}}, \bibinfo {author} {\bibfnamefont {C.}~\bibnamefont {{Clarkson}}},\ and\ \bibinfo {author} {\bibfnamefont {M.}~\bibnamefont {{Smith}}},\ }\href {https://doi.org/10.1088/1475-7516/2012/06/036} {\bibfield  {journal} {\bibinfo  {journal} {\jcap}\ }\textbf {\bibinfo {volume} {2012}},\ \bibinfo {eid} {036} (\bibinfo {year} {2012})},\ \Eprint {https://arxiv.org/abs/1204.2832} {arXiv:1204.2832 [astro-ph.CO]} \BibitemShut {NoStop}%
\bibitem [{\citenamefont {{Papamakarios}}\ \emph {et~al.}(2017)\citenamefont {{Papamakarios}}, \citenamefont {{Pavlakou}},\ and\ \citenamefont {{Murray}}}]{2017arXiv170507057P}%
  \BibitemOpen
  \bibfield  {author} {\bibinfo {author} {\bibfnamefont {G.}~\bibnamefont {{Papamakarios}}}, \bibinfo {author} {\bibfnamefont {T.}~\bibnamefont {{Pavlakou}}},\ and\ \bibinfo {author} {\bibfnamefont {I.}~\bibnamefont {{Murray}}},\ }\href {https://doi.org/10.48550/arXiv.1705.07057} {\bibfield  {journal} {\bibinfo  {journal} {Proceedings of the 31st International Conference on Neural Information Processing Systems}\ ,\ \bibinfo {eid} {arXiv:1705.07057}} (\bibinfo {year} {2017})},\ \Eprint {https://arxiv.org/abs/1705.07057} {arXiv:1705.07057 [stat.ML]} \BibitemShut {NoStop}%
\bibitem [{\citenamefont {{Ryan}}(2021)}]{2021JCAP...08..051R}%
  \BibitemOpen
  \bibfield  {author} {\bibinfo {author} {\bibfnamefont {J.}~\bibnamefont {{Ryan}}},\ }\href {https://doi.org/10.1088/1475-7516/2021/08/051} {\bibfield  {journal} {\bibinfo  {journal} {\jcap}\ }\textbf {\bibinfo {volume} {2021}},\ \bibinfo {eid} {051} (\bibinfo {year} {2021})},\ \Eprint {https://arxiv.org/abs/2102.08457} {arXiv:2102.08457 [astro-ph.CO]} \BibitemShut {NoStop}%
\bibitem [{\citenamefont {{G{\'o}mez-Valent}}\ and\ \citenamefont {{Amendola}}(2018)}]{2018JCAP...04..051G}%
  \BibitemOpen
  \bibfield  {author} {\bibinfo {author} {\bibfnamefont {A.}~\bibnamefont {{G{\'o}mez-Valent}}}\ and\ \bibinfo {author} {\bibfnamefont {L.}~\bibnamefont {{Amendola}}},\ }\href {https://doi.org/10.1088/1475-7516/2018/04/051} {\bibfield  {journal} {\bibinfo  {journal} {\jcap}\ }\textbf {\bibinfo {volume} {2018}},\ \bibinfo {eid} {051} (\bibinfo {year} {2018})},\ \Eprint {https://arxiv.org/abs/1802.01505} {arXiv:1802.01505 [astro-ph.CO]} \BibitemShut {NoStop}%
\bibitem [{\citenamefont {{Wang}}\ \emph {et~al.}(2021)\citenamefont {{Wang}}, \citenamefont {{Xie}}, \citenamefont {{Zhang}}, \citenamefont {{Huang}}, \citenamefont {{Zhang}},\ and\ \citenamefont {{Liu}}}]{2021ApJS..254...43W}%
  \BibitemOpen
  \bibfield  {author} {\bibinfo {author} {\bibfnamefont {Y.-C.}\ \bibnamefont {{Wang}}}, \bibinfo {author} {\bibfnamefont {Y.-B.}\ \bibnamefont {{Xie}}}, \bibinfo {author} {\bibfnamefont {T.-J.}\ \bibnamefont {{Zhang}}}, \bibinfo {author} {\bibfnamefont {H.-C.}\ \bibnamefont {{Huang}}}, \bibinfo {author} {\bibfnamefont {T.}~\bibnamefont {{Zhang}}},\ and\ \bibinfo {author} {\bibfnamefont {K.}~\bibnamefont {{Liu}}},\ }\href {https://doi.org/10.3847/1538-4365/abf8aa} {\bibfield  {journal} {\bibinfo  {journal} {\apjs}\ }\textbf {\bibinfo {volume} {254}},\ \bibinfo {eid} {43} (\bibinfo {year} {2021})},\ \Eprint {https://arxiv.org/abs/2005.10628} {arXiv:2005.10628 [astro-ph.CO]} \BibitemShut {NoStop}%
\bibitem [{\citenamefont {{Jimenez}}\ and\ \citenamefont {{Loeb}}(2002)}]{2002ApJ...573...37J}%
  \BibitemOpen
  \bibfield  {author} {\bibinfo {author} {\bibfnamefont {R.}~\bibnamefont {{Jimenez}}}\ and\ \bibinfo {author} {\bibfnamefont {A.}~\bibnamefont {{Loeb}}},\ }\href {https://doi.org/10.1086/340549} {\bibfield  {journal} {\bibinfo  {journal} {\apj}\ }\textbf {\bibinfo {volume} {573}},\ \bibinfo {pages} {37} (\bibinfo {year} {2002})},\ \Eprint {https://arxiv.org/abs/astro-ph/0106145} {arXiv:astro-ph/0106145 [astro-ph]} \BibitemShut {NoStop}%
\bibitem [{\citenamefont {{Moresco}}\ \emph {et~al.}(2020)\citenamefont {{Moresco}}, \citenamefont {{Jimenez}}, \citenamefont {{Verde}}, \citenamefont {{Cimatti}},\ and\ \citenamefont {{Pozzetti}}}]{2020ApJ...898...82M}%
  \BibitemOpen
  \bibfield  {author} {\bibinfo {author} {\bibfnamefont {M.}~\bibnamefont {{Moresco}}}, \bibinfo {author} {\bibfnamefont {R.}~\bibnamefont {{Jimenez}}}, \bibinfo {author} {\bibfnamefont {L.}~\bibnamefont {{Verde}}}, \bibinfo {author} {\bibfnamefont {A.}~\bibnamefont {{Cimatti}}},\ and\ \bibinfo {author} {\bibfnamefont {L.}~\bibnamefont {{Pozzetti}}},\ }\href {https://doi.org/10.3847/1538-4357/ab9eb0} {\bibfield  {journal} {\bibinfo  {journal} {\apj}\ }\textbf {\bibinfo {volume} {898}},\ \bibinfo {eid} {82} (\bibinfo {year} {2020})},\ \Eprint {https://arxiv.org/abs/2003.07362} {arXiv:2003.07362 [astro-ph.GA]} \BibitemShut {NoStop}%
\bibitem [{\citenamefont {{Jiao}}\ \emph {et~al.}(2023)\citenamefont {{Jiao}}, \citenamefont {{Borghi}}, \citenamefont {{Moresco}},\ and\ \citenamefont {{Zhang}}}]{2023ApJS..265...48J}%
  \BibitemOpen
  \bibfield  {author} {\bibinfo {author} {\bibfnamefont {K.}~\bibnamefont {{Jiao}}}, \bibinfo {author} {\bibfnamefont {N.}~\bibnamefont {{Borghi}}}, \bibinfo {author} {\bibfnamefont {M.}~\bibnamefont {{Moresco}}},\ and\ \bibinfo {author} {\bibfnamefont {T.-J.}\ \bibnamefont {{Zhang}}},\ }\href {https://doi.org/10.3847/1538-4365/acbc77} {\bibfield  {journal} {\bibinfo  {journal} {\apjs}\ }\textbf {\bibinfo {volume} {265}},\ \bibinfo {eid} {48} (\bibinfo {year} {2023})},\ \Eprint {https://arxiv.org/abs/2205.05701} {arXiv:2205.05701 [astro-ph.CO]} \BibitemShut {NoStop}%
\bibitem [{\citenamefont {{Zhang}}\ \emph {et~al.}(2014)\citenamefont {{Zhang}}, \citenamefont {{Zhang}}, \citenamefont {{Yuan}}, \citenamefont {{Liu}}, \citenamefont {{Zhang}},\ and\ \citenamefont {{Sun}}}]{2014RAA....14.1221Z}%
  \BibitemOpen
  \bibfield  {author} {\bibinfo {author} {\bibfnamefont {C.}~\bibnamefont {{Zhang}}}, \bibinfo {author} {\bibfnamefont {H.}~\bibnamefont {{Zhang}}}, \bibinfo {author} {\bibfnamefont {S.}~\bibnamefont {{Yuan}}}, \bibinfo {author} {\bibfnamefont {S.}~\bibnamefont {{Liu}}}, \bibinfo {author} {\bibfnamefont {T.-J.}\ \bibnamefont {{Zhang}}},\ and\ \bibinfo {author} {\bibfnamefont {Y.-C.}\ \bibnamefont {{Sun}}},\ }\href {https://doi.org/10.1088/1674-4527/14/10/002} {\bibfield  {journal} {\bibinfo  {journal} {Research in Astronomy and Astrophysics}\ }\textbf {\bibinfo {volume} {14}},\ \bibinfo {eid} {1221-1233} (\bibinfo {year} {2014})},\ \Eprint {https://arxiv.org/abs/1207.4541} {arXiv:1207.4541 [astro-ph.CO]} \BibitemShut {NoStop}%
\bibitem [{\citenamefont {{Simon}}\ \emph {et~al.}(2005)\citenamefont {{Simon}}, \citenamefont {{Verde}},\ and\ \citenamefont {{Jimenez}}}]{2005PhRvD..71l3001S}%
  \BibitemOpen
  \bibfield  {author} {\bibinfo {author} {\bibfnamefont {J.}~\bibnamefont {{Simon}}}, \bibinfo {author} {\bibfnamefont {L.}~\bibnamefont {{Verde}}},\ and\ \bibinfo {author} {\bibfnamefont {R.}~\bibnamefont {{Jimenez}}},\ }\href {https://doi.org/10.1103/PhysRevD.71.123001} {\bibfield  {journal} {\bibinfo  {journal} {\prd}\ }\textbf {\bibinfo {volume} {71}},\ \bibinfo {eid} {123001} (\bibinfo {year} {2005})},\ \Eprint {https://arxiv.org/abs/astro-ph/0412269} {arXiv:astro-ph/0412269 [astro-ph]} \BibitemShut {NoStop}%
\bibitem [{\citenamefont {{Moresco}}\ \emph {et~al.}(2012)\citenamefont {{Moresco}}, \citenamefont {{Cimatti}}, \citenamefont {{Jimenez}}, \citenamefont {{Pozzetti}}, \citenamefont {{Zamorani}}, \citenamefont {{Bolzonella}}, \citenamefont {{Dunlop}}, \citenamefont {{Lamareille}}, \citenamefont {{Mignoli}}, \citenamefont {{Pearce}}, \citenamefont {{Rosati}}, \citenamefont {{Stern}}, \citenamefont {{Verde}}, \citenamefont {{Zucca}}, \citenamefont {{Carollo}}, \citenamefont {{Contini}}, \citenamefont {{Kneib}}, \citenamefont {{Le F{\`e}vre}}, \citenamefont {{Lilly}}, \citenamefont {{Mainieri}}, \citenamefont {{Renzini}}, \citenamefont {{Scodeggio}}, \citenamefont {{Balestra}}, \citenamefont {{Gobat}}, \citenamefont {{McLure}}, \citenamefont {{Bardelli}}, \citenamefont {{Bongiorno}}, \citenamefont {{Caputi}}, \citenamefont {{Cucciati}}, \citenamefont {{de la Torre}}, \citenamefont {{de Ravel}}, \citenamefont {{Franzetti}}, \citenamefont {{Garilli}}, \citenamefont {{Iovino}}, \citenamefont {{Kampczyk}},
  \citenamefont {{Knobel}}, \citenamefont {{Kova{\v{c}}}}, \citenamefont {{Le Borgne}}, \citenamefont {{Le Brun}}, \citenamefont {{Maier}}, \citenamefont {{Pell{\'o}}}, \citenamefont {{Peng}}, \citenamefont {{Perez-Montero}}, \citenamefont {{Presotto}}, \citenamefont {{Silverman}}, \citenamefont {{Tanaka}}, \citenamefont {{Tasca}}, \citenamefont {{Tresse}}, \citenamefont {{Vergani}}, \citenamefont {{Almaini}}, \citenamefont {{Barnes}}, \citenamefont {{Bordoloi}}, \citenamefont {{Bradshaw}}, \citenamefont {{Cappi}}, \citenamefont {{Chuter}}, \citenamefont {{Cirasuolo}}, \citenamefont {{Coppa}}, \citenamefont {{Diener}}, \citenamefont {{Foucaud}}, \citenamefont {{Hartley}}, \citenamefont {{Kamionkowski}}, \citenamefont {{Koekemoer}}, \citenamefont {{L{\'o}pez-Sanjuan}}, \citenamefont {{McCracken}}, \citenamefont {{Nair}}, \citenamefont {{Oesch}}, \citenamefont {{Stanford}},\ and\ \citenamefont {{Welikala}}}]{2012JCAP...08..006M}%
  \BibitemOpen
  \bibfield  {author} {\bibinfo {author} {\bibfnamefont {M.}~\bibnamefont {{Moresco}}}, \bibinfo {author} {\bibfnamefont {A.}~\bibnamefont {{Cimatti}}}, \bibinfo {author} {\bibfnamefont {R.}~\bibnamefont {{Jimenez}}}, \bibinfo {author} {\bibfnamefont {L.}~\bibnamefont {{Pozzetti}}}, \bibinfo {author} {\bibfnamefont {G.}~\bibnamefont {{Zamorani}}}, \bibinfo {author} {\bibfnamefont {M.}~\bibnamefont {{Bolzonella}}}, \bibinfo {author} {\bibfnamefont {J.}~\bibnamefont {{Dunlop}}}, \bibinfo {author} {\bibfnamefont {F.}~\bibnamefont {{Lamareille}}}, \bibinfo {author} {\bibfnamefont {M.}~\bibnamefont {{Mignoli}}}, \bibinfo {author} {\bibfnamefont {H.}~\bibnamefont {{Pearce}}}, \bibinfo {author} {\bibfnamefont {P.}~\bibnamefont {{Rosati}}}, \bibinfo {author} {\bibfnamefont {D.}~\bibnamefont {{Stern}}}, \bibinfo {author} {\bibfnamefont {L.}~\bibnamefont {{Verde}}}, \bibinfo {author} {\bibfnamefont {E.}~\bibnamefont {{Zucca}}}, \bibinfo {author} {\bibfnamefont {C.~M.}\ \bibnamefont {{Carollo}}}, \bibinfo {author}
  {\bibfnamefont {T.}~\bibnamefont {{Contini}}}, \bibinfo {author} {\bibfnamefont {J.~P.}\ \bibnamefont {{Kneib}}}, \bibinfo {author} {\bibfnamefont {O.}~\bibnamefont {{Le F{\`e}vre}}}, \bibinfo {author} {\bibfnamefont {S.~J.}\ \bibnamefont {{Lilly}}}, \bibinfo {author} {\bibfnamefont {V.}~\bibnamefont {{Mainieri}}}, \bibinfo {author} {\bibfnamefont {A.}~\bibnamefont {{Renzini}}}, \bibinfo {author} {\bibfnamefont {M.}~\bibnamefont {{Scodeggio}}}, \bibinfo {author} {\bibfnamefont {I.}~\bibnamefont {{Balestra}}}, \bibinfo {author} {\bibfnamefont {R.}~\bibnamefont {{Gobat}}}, \bibinfo {author} {\bibfnamefont {R.}~\bibnamefont {{McLure}}}, \bibinfo {author} {\bibfnamefont {S.}~\bibnamefont {{Bardelli}}}, \bibinfo {author} {\bibfnamefont {A.}~\bibnamefont {{Bongiorno}}}, \bibinfo {author} {\bibfnamefont {K.}~\bibnamefont {{Caputi}}}, \bibinfo {author} {\bibfnamefont {O.}~\bibnamefont {{Cucciati}}}, \bibinfo {author} {\bibfnamefont {S.}~\bibnamefont {{de la Torre}}}, \bibinfo {author} {\bibfnamefont
  {L.}~\bibnamefont {{de Ravel}}}, \bibinfo {author} {\bibfnamefont {P.}~\bibnamefont {{Franzetti}}}, \bibinfo {author} {\bibfnamefont {B.}~\bibnamefont {{Garilli}}}, \bibinfo {author} {\bibfnamefont {A.}~\bibnamefont {{Iovino}}}, \bibinfo {author} {\bibfnamefont {P.}~\bibnamefont {{Kampczyk}}}, \bibinfo {author} {\bibfnamefont {C.}~\bibnamefont {{Knobel}}}, \bibinfo {author} {\bibfnamefont {K.}~\bibnamefont {{Kova{\v{c}}}}}, \bibinfo {author} {\bibfnamefont {J.~F.}\ \bibnamefont {{Le Borgne}}}, \bibinfo {author} {\bibfnamefont {V.}~\bibnamefont {{Le Brun}}}, \bibinfo {author} {\bibfnamefont {C.}~\bibnamefont {{Maier}}}, \bibinfo {author} {\bibfnamefont {R.}~\bibnamefont {{Pell{\'o}}}}, \bibinfo {author} {\bibfnamefont {Y.}~\bibnamefont {{Peng}}}, \bibinfo {author} {\bibfnamefont {E.}~\bibnamefont {{Perez-Montero}}}, \bibinfo {author} {\bibfnamefont {V.}~\bibnamefont {{Presotto}}}, \bibinfo {author} {\bibfnamefont {J.~D.}\ \bibnamefont {{Silverman}}}, \bibinfo {author} {\bibfnamefont {M.}~\bibnamefont
  {{Tanaka}}}, \bibinfo {author} {\bibfnamefont {L.~A.~M.}\ \bibnamefont {{Tasca}}}, \bibinfo {author} {\bibfnamefont {L.}~\bibnamefont {{Tresse}}}, \bibinfo {author} {\bibfnamefont {D.}~\bibnamefont {{Vergani}}}, \bibinfo {author} {\bibfnamefont {O.}~\bibnamefont {{Almaini}}}, \bibinfo {author} {\bibfnamefont {L.}~\bibnamefont {{Barnes}}}, \bibinfo {author} {\bibfnamefont {R.}~\bibnamefont {{Bordoloi}}}, \bibinfo {author} {\bibfnamefont {E.}~\bibnamefont {{Bradshaw}}}, \bibinfo {author} {\bibfnamefont {A.}~\bibnamefont {{Cappi}}}, \bibinfo {author} {\bibfnamefont {R.}~\bibnamefont {{Chuter}}}, \bibinfo {author} {\bibfnamefont {M.}~\bibnamefont {{Cirasuolo}}}, \bibinfo {author} {\bibfnamefont {G.}~\bibnamefont {{Coppa}}}, \bibinfo {author} {\bibfnamefont {C.}~\bibnamefont {{Diener}}}, \bibinfo {author} {\bibfnamefont {S.}~\bibnamefont {{Foucaud}}}, \bibinfo {author} {\bibfnamefont {W.}~\bibnamefont {{Hartley}}}, \bibinfo {author} {\bibfnamefont {M.}~\bibnamefont {{Kamionkowski}}}, \bibinfo {author}
  {\bibfnamefont {A.~M.}\ \bibnamefont {{Koekemoer}}}, \bibinfo {author} {\bibfnamefont {C.}~\bibnamefont {{L{\'o}pez-Sanjuan}}}, \bibinfo {author} {\bibfnamefont {H.~J.}\ \bibnamefont {{McCracken}}}, \bibinfo {author} {\bibfnamefont {P.}~\bibnamefont {{Nair}}}, \bibinfo {author} {\bibfnamefont {P.}~\bibnamefont {{Oesch}}}, \bibinfo {author} {\bibfnamefont {A.}~\bibnamefont {{Stanford}}},\ and\ \bibinfo {author} {\bibfnamefont {N.}~\bibnamefont {{Welikala}}},\ }\href {https://doi.org/10.1088/1475-7516/2012/08/006} {\bibfield  {journal} {\bibinfo  {journal} {\jcap}\ }\textbf {\bibinfo {volume} {2012}},\ \bibinfo {eid} {006} (\bibinfo {year} {2012})},\ \Eprint {https://arxiv.org/abs/1201.3609} {arXiv:1201.3609 [astro-ph.CO]} \BibitemShut {NoStop}%
\bibitem [{\citenamefont {{Moresco}}\ \emph {et~al.}(2016)\citenamefont {{Moresco}}, \citenamefont {{Pozzetti}}, \citenamefont {{Cimatti}}, \citenamefont {{Jimenez}}, \citenamefont {{Maraston}}, \citenamefont {{Verde}}, \citenamefont {{Thomas}}, \citenamefont {{Citro}}, \citenamefont {{Tojeiro}},\ and\ \citenamefont {{Wilkinson}}}]{2016JCAP...05..014M}%
  \BibitemOpen
  \bibfield  {author} {\bibinfo {author} {\bibfnamefont {M.}~\bibnamefont {{Moresco}}}, \bibinfo {author} {\bibfnamefont {L.}~\bibnamefont {{Pozzetti}}}, \bibinfo {author} {\bibfnamefont {A.}~\bibnamefont {{Cimatti}}}, \bibinfo {author} {\bibfnamefont {R.}~\bibnamefont {{Jimenez}}}, \bibinfo {author} {\bibfnamefont {C.}~\bibnamefont {{Maraston}}}, \bibinfo {author} {\bibfnamefont {L.}~\bibnamefont {{Verde}}}, \bibinfo {author} {\bibfnamefont {D.}~\bibnamefont {{Thomas}}}, \bibinfo {author} {\bibfnamefont {A.}~\bibnamefont {{Citro}}}, \bibinfo {author} {\bibfnamefont {R.}~\bibnamefont {{Tojeiro}}},\ and\ \bibinfo {author} {\bibfnamefont {D.}~\bibnamefont {{Wilkinson}}},\ }\href {https://doi.org/10.1088/1475-7516/2016/05/014} {\bibfield  {journal} {\bibinfo  {journal} {\jcap}\ }\textbf {\bibinfo {volume} {2016}},\ \bibinfo {eid} {014} (\bibinfo {year} {2016})},\ \Eprint {https://arxiv.org/abs/1601.01701} {arXiv:1601.01701 [astro-ph.CO]} \BibitemShut {NoStop}%
\bibitem [{\citenamefont {{Ratsimbazafy}}\ \emph {et~al.}(2017)\citenamefont {{Ratsimbazafy}}, \citenamefont {{Loubser}}, \citenamefont {{Crawford}}, \citenamefont {{Cress}}, \citenamefont {{Bassett}}, \citenamefont {{Nichol}},\ and\ \citenamefont {{V{\"a}is{\"a}nen}}}]{2017MNRAS.467.3239R}%
  \BibitemOpen
  \bibfield  {author} {\bibinfo {author} {\bibfnamefont {A.~L.}\ \bibnamefont {{Ratsimbazafy}}}, \bibinfo {author} {\bibfnamefont {S.~I.}\ \bibnamefont {{Loubser}}}, \bibinfo {author} {\bibfnamefont {S.~M.}\ \bibnamefont {{Crawford}}}, \bibinfo {author} {\bibfnamefont {C.~M.}\ \bibnamefont {{Cress}}}, \bibinfo {author} {\bibfnamefont {B.~A.}\ \bibnamefont {{Bassett}}}, \bibinfo {author} {\bibfnamefont {R.~C.}\ \bibnamefont {{Nichol}}},\ and\ \bibinfo {author} {\bibfnamefont {P.}~\bibnamefont {{V{\"a}is{\"a}nen}}},\ }\href {https://doi.org/10.1093/mnras/stx301} {\bibfield  {journal} {\bibinfo  {journal} {\mnras}\ }\textbf {\bibinfo {volume} {467}},\ \bibinfo {pages} {3239} (\bibinfo {year} {2017})},\ \Eprint {https://arxiv.org/abs/1702.00418} {arXiv:1702.00418 [astro-ph.CO]} \BibitemShut {NoStop}%
\bibitem [{\citenamefont {{Stern}}\ \emph {et~al.}(2010)\citenamefont {{Stern}}, \citenamefont {{Jimenez}}, \citenamefont {{Verde}}, \citenamefont {{Kamionkowski}},\ and\ \citenamefont {{Stanford}}}]{2010JCAP...02..008S}%
  \BibitemOpen
  \bibfield  {author} {\bibinfo {author} {\bibfnamefont {D.}~\bibnamefont {{Stern}}}, \bibinfo {author} {\bibfnamefont {R.}~\bibnamefont {{Jimenez}}}, \bibinfo {author} {\bibfnamefont {L.}~\bibnamefont {{Verde}}}, \bibinfo {author} {\bibfnamefont {M.}~\bibnamefont {{Kamionkowski}}},\ and\ \bibinfo {author} {\bibfnamefont {S.~A.}\ \bibnamefont {{Stanford}}},\ }\href {https://doi.org/10.1088/1475-7516/2010/02/008} {\bibfield  {journal} {\bibinfo  {journal} {\jcap}\ }\textbf {\bibinfo {volume} {2010}},\ \bibinfo {eid} {008} (\bibinfo {year} {2010})},\ \Eprint {https://arxiv.org/abs/0907.3149} {arXiv:0907.3149 [astro-ph.CO]} \BibitemShut {NoStop}%
\bibitem [{\citenamefont {{Tomasetti}}\ \emph {et~al.}(2023)\citenamefont {{Tomasetti}}, \citenamefont {{Moresco}}, \citenamefont {{Borghi}}, \citenamefont {{Jiao}}, \citenamefont {{Cimatti}}, \citenamefont {{Pozzetti}}, \citenamefont {{Carnall}}, \citenamefont {{McLure}},\ and\ \citenamefont {{Pentericci}}}]{2023AA...679A..96T}%
  \BibitemOpen
  \bibfield  {author} {\bibinfo {author} {\bibfnamefont {E.}~\bibnamefont {{Tomasetti}}}, \bibinfo {author} {\bibfnamefont {M.}~\bibnamefont {{Moresco}}}, \bibinfo {author} {\bibfnamefont {N.}~\bibnamefont {{Borghi}}}, \bibinfo {author} {\bibfnamefont {K.}~\bibnamefont {{Jiao}}}, \bibinfo {author} {\bibfnamefont {A.}~\bibnamefont {{Cimatti}}}, \bibinfo {author} {\bibfnamefont {L.}~\bibnamefont {{Pozzetti}}}, \bibinfo {author} {\bibfnamefont {A.~C.}\ \bibnamefont {{Carnall}}}, \bibinfo {author} {\bibfnamefont {R.~J.}\ \bibnamefont {{McLure}}},\ and\ \bibinfo {author} {\bibfnamefont {L.}~\bibnamefont {{Pentericci}}},\ }\href {https://doi.org/10.1051/0004-6361/202346992} {\bibfield  {journal} {\bibinfo  {journal} {\aap}\ }\textbf {\bibinfo {volume} {679}},\ \bibinfo {eid} {A96} (\bibinfo {year} {2023})},\ \Eprint {https://arxiv.org/abs/2305.16387} {arXiv:2305.16387 [astro-ph.CO]} \BibitemShut {NoStop}%
\bibitem [{\citenamefont {{Moresco}}(2015)}]{2015MNRAS.450L..16M}%
  \BibitemOpen
  \bibfield  {author} {\bibinfo {author} {\bibfnamefont {M.}~\bibnamefont {{Moresco}}},\ }\href {https://doi.org/10.1093/mnrasl/slv037} {\bibfield  {journal} {\bibinfo  {journal} {\mnras}\ }\textbf {\bibinfo {volume} {450}},\ \bibinfo {pages} {L16} (\bibinfo {year} {2015})},\ \Eprint {https://arxiv.org/abs/1503.01116} {arXiv:1503.01116 [astro-ph.CO]} \BibitemShut {NoStop}%
\bibitem [{\citenamefont {{Ma}}\ and\ \citenamefont {{Zhang}}(2011)}]{2011ApJ...730...74M}%
  \BibitemOpen
  \bibfield  {author} {\bibinfo {author} {\bibfnamefont {C.}~\bibnamefont {{Ma}}}\ and\ \bibinfo {author} {\bibfnamefont {T.-J.}\ \bibnamefont {{Zhang}}},\ }\href {https://doi.org/10.1088/0004-637X/730/2/74} {\bibfield  {journal} {\bibinfo  {journal} {\apj}\ }\textbf {\bibinfo {volume} {730}},\ \bibinfo {eid} {74} (\bibinfo {year} {2011})},\ \Eprint {https://arxiv.org/abs/1007.3787} {arXiv:1007.3787 [astro-ph.CO]} \BibitemShut {NoStop}%
\bibitem [{\citenamefont {{Zhang}}\ \emph {et~al.}(2023)\citenamefont {{Zhang}}, \citenamefont {{Wang}}, \citenamefont {{Zhang}},\ and\ \citenamefont {{Zhang}}}]{2023ApJS..266...27Z}%
  \BibitemOpen
  \bibfield  {author} {\bibinfo {author} {\bibfnamefont {H.}~\bibnamefont {{Zhang}}}, \bibinfo {author} {\bibfnamefont {Y.-C.}\ \bibnamefont {{Wang}}}, \bibinfo {author} {\bibfnamefont {T.-J.}\ \bibnamefont {{Zhang}}},\ and\ \bibinfo {author} {\bibfnamefont {T.}~\bibnamefont {{Zhang}}},\ }\href {https://doi.org/10.3847/1538-4365/accb92} {\bibfield  {journal} {\bibinfo  {journal} {\apjs}\ }\textbf {\bibinfo {volume} {266}},\ \bibinfo {eid} {27} (\bibinfo {year} {2023})},\ \Eprint {https://arxiv.org/abs/2304.03911} {arXiv:2304.03911 [astro-ph.CO]} \BibitemShut {NoStop}%
\bibitem [{\citenamefont {{Sun}}\ \emph {et~al.}(2021)\citenamefont {{Sun}}, \citenamefont {{Jiao}},\ and\ \citenamefont {{Zhang}}}]{2021ApJ...915..123S}%
  \BibitemOpen
  \bibfield  {author} {\bibinfo {author} {\bibfnamefont {W.}~\bibnamefont {{Sun}}}, \bibinfo {author} {\bibfnamefont {K.}~\bibnamefont {{Jiao}}},\ and\ \bibinfo {author} {\bibfnamefont {T.-J.}\ \bibnamefont {{Zhang}}},\ }\href {https://doi.org/10.3847/1538-4357/ac05b8} {\bibfield  {journal} {\bibinfo  {journal} {\apj}\ }\textbf {\bibinfo {volume} {915}},\ \bibinfo {eid} {123} (\bibinfo {year} {2021})},\ \Eprint {https://arxiv.org/abs/2105.12618} {arXiv:2105.12618 [astro-ph.CO]} \BibitemShut {NoStop}%
\bibitem [{\citenamefont {{Germain}}\ \emph {et~al.}(2015)\citenamefont {{Germain}}, \citenamefont {{Gregor}}, \citenamefont {{Murray}},\ and\ \citenamefont {{Larochelle}}}]{2015arXiv150203509G}%
  \BibitemOpen
  \bibfield  {author} {\bibinfo {author} {\bibfnamefont {M.}~\bibnamefont {{Germain}}}, \bibinfo {author} {\bibfnamefont {K.}~\bibnamefont {{Gregor}}}, \bibinfo {author} {\bibfnamefont {I.}~\bibnamefont {{Murray}}},\ and\ \bibinfo {author} {\bibfnamefont {H.}~\bibnamefont {{Larochelle}}},\ }\href {https://doi.org/10.48550/arXiv.1502.03509} {\bibfield  {journal} {\bibinfo  {journal} {arXiv e-prints}\ ,\ \bibinfo {eid} {arXiv:1502.03509}} (\bibinfo {year} {2015})},\ \Eprint {https://arxiv.org/abs/1502.03509} {arXiv:1502.03509 [cs.LG]} \BibitemShut {NoStop}%
\bibitem [{\citenamefont {{Uria}}\ \emph {et~al.}(2016)\citenamefont {{Uria}}, \citenamefont {{C{\^o}t{\'e}}}, \citenamefont {{Gregor}}, \citenamefont {{Murray}},\ and\ \citenamefont {{Larochelle}}}]{2016arXiv160502226U}%
  \BibitemOpen
  \bibfield  {author} {\bibinfo {author} {\bibfnamefont {B.}~\bibnamefont {{Uria}}}, \bibinfo {author} {\bibfnamefont {M.-A.}\ \bibnamefont {{C{\^o}t{\'e}}}}, \bibinfo {author} {\bibfnamefont {K.}~\bibnamefont {{Gregor}}}, \bibinfo {author} {\bibfnamefont {I.}~\bibnamefont {{Murray}}},\ and\ \bibinfo {author} {\bibfnamefont {H.}~\bibnamefont {{Larochelle}}},\ }\href {https://doi.org/10.5555/2946645.3053487} {\bibfield  {journal} {\bibinfo  {journal} {The Journal of Machine Learning Research}\ ,\ \bibinfo {eid} {arXiv:1605.02226}} (\bibinfo {year} {2016})},\ \Eprint {https://arxiv.org/abs/1605.02226} {arXiv:1605.02226 [cs.LG]} \BibitemShut {NoStop}%
\bibitem [{\citenamefont {{Jimenez Rezende}}\ and\ \citenamefont {{Mohamed}}(2015)}]{2015arXiv150505770J}%
  \BibitemOpen
  \bibfield  {author} {\bibinfo {author} {\bibfnamefont {D.}~\bibnamefont {{Jimenez Rezende}}}\ and\ \bibinfo {author} {\bibfnamefont {S.}~\bibnamefont {{Mohamed}}},\ }\href {https://doi.org/10.48550/arXiv.1505.05770} {\bibfield  {journal} {\bibinfo  {journal} {arXiv e-prints}\ ,\ \bibinfo {eid} {arXiv:1505.05770}} (\bibinfo {year} {2015})},\ \Eprint {https://arxiv.org/abs/1505.05770} {arXiv:1505.05770 [stat.ML]} \BibitemShut {NoStop}%
\bibitem [{\citenamefont {{Chen}}\ \emph {et~al.}(2023)\citenamefont {{Chen}}, \citenamefont {{Wang}}, \citenamefont {{Zhang}},\ and\ \citenamefont {{Zhang}}}]{2023PhRvD.107f3517C}%
  \BibitemOpen
  \bibfield  {author} {\bibinfo {author} {\bibfnamefont {J.-F.}\ \bibnamefont {{Chen}}}, \bibinfo {author} {\bibfnamefont {Y.-C.}\ \bibnamefont {{Wang}}}, \bibinfo {author} {\bibfnamefont {T.}~\bibnamefont {{Zhang}}},\ and\ \bibinfo {author} {\bibfnamefont {T.-J.}\ \bibnamefont {{Zhang}}},\ }\href {https://doi.org/10.1103/PhysRevD.107.063517} {\bibfield  {journal} {\bibinfo  {journal} {\prd}\ }\textbf {\bibinfo {volume} {107}},\ \bibinfo {eid} {063517} (\bibinfo {year} {2023})},\ \Eprint {https://arxiv.org/abs/2211.05064} {arXiv:2211.05064 [astro-ph.CO]} \BibitemShut {NoStop}%
\bibitem [{\citenamefont {Shao}\ and\ \citenamefont {Tu}(1995)}]{ShaoTu1995}%
  \BibitemOpen
  \bibfield  {author} {\bibinfo {author} {\bibfnamefont {J.}~\bibnamefont {Shao}}\ and\ \bibinfo {author} {\bibfnamefont {D.}~\bibnamefont {Tu}},\ }\href {https://doi.org/10.1007/978-1-4612-0795-5} {\emph {\bibinfo {title} {The Jackknife and Bootstrap}}},\ Springer Series in Statistics\ (\bibinfo  {publisher} {Springer},\ \bibinfo {address} {New York, NY},\ \bibinfo {year} {1995})\BibitemShut {NoStop}%
\bibitem [{\citenamefont {{Astropy Collaboration}}\ \emph {et~al.}(2018)\citenamefont {{Astropy Collaboration}}, \citenamefont {{Price-Whelan}}, \citenamefont {{Sip{\H{o}}cz}}, \citenamefont {{G{\"u}nther}}, \citenamefont {{Lim}}, \citenamefont {{Crawford}}, \citenamefont {{Conseil}}, \citenamefont {{Shupe}}, \citenamefont {{Craig}}, \citenamefont {{Dencheva}}, \citenamefont {{Ginsburg}}, \citenamefont {{VanderPlas}}, \citenamefont {{Bradley}}, \citenamefont {{P{\'e}rez-Su{\'a}rez}}, \citenamefont {{de Val-Borro}}, \citenamefont {{Aldcroft}}, \citenamefont {{Cruz}}, \citenamefont {{Robitaille}}, \citenamefont {{Tollerud}}, \citenamefont {{Ardelean}}, \citenamefont {{Babej}}, \citenamefont {{Bach}}, \citenamefont {{Bachetti}}, \citenamefont {{Bakanov}}, \citenamefont {{Bamford}}, \citenamefont {{Barentsen}}, \citenamefont {{Barmby}}, \citenamefont {{Baumbach}}, \citenamefont {{Berry}}, \citenamefont {{Biscani}}, \citenamefont {{Boquien}}, \citenamefont {{Bostroem}}, \citenamefont {{Bouma}}, \citenamefont
  {{Brammer}}, \citenamefont {{Bray}}, \citenamefont {{Breytenbach}}, \citenamefont {{Buddelmeijer}}, \citenamefont {{Burke}}, \citenamefont {{Calderone}}, \citenamefont {{Cano Rodr{\'\i}guez}}, \citenamefont {{Cara}}, \citenamefont {{Cardoso}}, \citenamefont {{Cheedella}}, \citenamefont {{Copin}}, \citenamefont {{Corrales}}, \citenamefont {{Crichton}}, \citenamefont {{D'Avella}}, \citenamefont {{Deil}}, \citenamefont {{Depagne}}, \citenamefont {{Dietrich}}, \citenamefont {{Donath}}, \citenamefont {{Droettboom}}, \citenamefont {{Earl}}, \citenamefont {{Erben}}, \citenamefont {{Fabbro}}, \citenamefont {{Ferreira}}, \citenamefont {{Finethy}}, \citenamefont {{Fox}}, \citenamefont {{Garrison}}, \citenamefont {{Gibbons}}, \citenamefont {{Goldstein}}, \citenamefont {{Gommers}}, \citenamefont {{Greco}}, \citenamefont {{Greenfield}}, \citenamefont {{Groener}}, \citenamefont {{Grollier}}, \citenamefont {{Hagen}}, \citenamefont {{Hirst}}, \citenamefont {{Homeier}}, \citenamefont {{Horton}}, \citenamefont
  {{Hosseinzadeh}}, \citenamefont {{Hu}}, \citenamefont {{Hunkeler}}, \citenamefont {{Ivezi{\'c}}}, \citenamefont {{Jain}}, \citenamefont {{Jenness}}, \citenamefont {{Kanarek}}, \citenamefont {{Kendrew}}, \citenamefont {{Kern}}, \citenamefont {{Kerzendorf}}, \citenamefont {{Khvalko}}, \citenamefont {{King}}, \citenamefont {{Kirkby}}, \citenamefont {{Kulkarni}}, \citenamefont {{Kumar}}, \citenamefont {{Lee}}, \citenamefont {{Lenz}}, \citenamefont {{Littlefair}}, \citenamefont {{Ma}}, \citenamefont {{Macleod}}, \citenamefont {{Mastropietro}}, \citenamefont {{McCully}}, \citenamefont {{Montagnac}}, \citenamefont {{Morris}}, \citenamefont {{Mueller}}, \citenamefont {{Mumford}}, \citenamefont {{Muna}}, \citenamefont {{Murphy}}, \citenamefont {{Nelson}}, \citenamefont {{Nguyen}}, \citenamefont {{Ninan}}, \citenamefont {{N{\"o}the}}, \citenamefont {{Ogaz}}, \citenamefont {{Oh}}, \citenamefont {{Parejko}}, \citenamefont {{Parley}}, \citenamefont {{Pascual}}, \citenamefont {{Patil}}, \citenamefont {{Patil}},
  \citenamefont {{Plunkett}}, \citenamefont {{Prochaska}}, \citenamefont {{Rastogi}}, \citenamefont {{Reddy Janga}}, \citenamefont {{Sabater}}, \citenamefont {{Sakurikar}}, \citenamefont {{Seifert}}, \citenamefont {{Sherbert}}, \citenamefont {{Sherwood-Taylor}}, \citenamefont {{Shih}}, \citenamefont {{Sick}}, \citenamefont {{Silbiger}}, \citenamefont {{Singanamalla}}, \citenamefont {{Singer}}, \citenamefont {{Sladen}}, \citenamefont {{Sooley}}, \citenamefont {{Sornarajah}}, \citenamefont {{Streicher}}, \citenamefont {{Teuben}}, \citenamefont {{Thomas}}, \citenamefont {{Tremblay}}, \citenamefont {{Turner}}, \citenamefont {{Terr{\'o}n}}, \citenamefont {{van Kerkwijk}}, \citenamefont {{de la Vega}}, \citenamefont {{Watkins}}, \citenamefont {{Weaver}}, \citenamefont {{Whitmore}}, \citenamefont {{Woillez}}, \citenamefont {{Zabalza}},\ and\ \citenamefont {{Astropy Contributors}}}]{2018AJ....156..123A}%
  \BibitemOpen
  \bibfield  {author} {\bibinfo {author} {\bibnamefont {{Astropy Collaboration}}}, \bibinfo {author} {\bibfnamefont {A.~M.}\ \bibnamefont {{Price-Whelan}}}, \bibinfo {author} {\bibfnamefont {B.~M.}\ \bibnamefont {{Sip{\H{o}}cz}}}, \bibinfo {author} {\bibfnamefont {H.~M.}\ \bibnamefont {{G{\"u}nther}}}, \bibinfo {author} {\bibfnamefont {P.~L.}\ \bibnamefont {{Lim}}}, \bibinfo {author} {\bibfnamefont {S.~M.}\ \bibnamefont {{Crawford}}}, \bibinfo {author} {\bibfnamefont {S.}~\bibnamefont {{Conseil}}}, \bibinfo {author} {\bibfnamefont {D.~L.}\ \bibnamefont {{Shupe}}}, \bibinfo {author} {\bibfnamefont {M.~W.}\ \bibnamefont {{Craig}}}, \bibinfo {author} {\bibfnamefont {N.}~\bibnamefont {{Dencheva}}}, \bibinfo {author} {\bibfnamefont {A.}~\bibnamefont {{Ginsburg}}}, \bibinfo {author} {\bibfnamefont {J.~T.}\ \bibnamefont {{VanderPlas}}}, \bibinfo {author} {\bibfnamefont {L.~D.}\ \bibnamefont {{Bradley}}}, \bibinfo {author} {\bibfnamefont {D.}~\bibnamefont {{P{\'e}rez-Su{\'a}rez}}}, \bibinfo {author} {\bibfnamefont
  {M.}~\bibnamefont {{de Val-Borro}}}, \bibinfo {author} {\bibfnamefont {T.~L.}\ \bibnamefont {{Aldcroft}}}, \bibinfo {author} {\bibfnamefont {K.~L.}\ \bibnamefont {{Cruz}}}, \bibinfo {author} {\bibfnamefont {T.~P.}\ \bibnamefont {{Robitaille}}}, \bibinfo {author} {\bibfnamefont {E.~J.}\ \bibnamefont {{Tollerud}}}, \bibinfo {author} {\bibfnamefont {C.}~\bibnamefont {{Ardelean}}}, \bibinfo {author} {\bibfnamefont {T.}~\bibnamefont {{Babej}}}, \bibinfo {author} {\bibfnamefont {Y.~P.}\ \bibnamefont {{Bach}}}, \bibinfo {author} {\bibfnamefont {M.}~\bibnamefont {{Bachetti}}}, \bibinfo {author} {\bibfnamefont {A.~V.}\ \bibnamefont {{Bakanov}}}, \bibinfo {author} {\bibfnamefont {S.~P.}\ \bibnamefont {{Bamford}}}, \bibinfo {author} {\bibfnamefont {G.}~\bibnamefont {{Barentsen}}}, \bibinfo {author} {\bibfnamefont {P.}~\bibnamefont {{Barmby}}}, \bibinfo {author} {\bibfnamefont {A.}~\bibnamefont {{Baumbach}}}, \bibinfo {author} {\bibfnamefont {K.~L.}\ \bibnamefont {{Berry}}}, \bibinfo {author} {\bibfnamefont
  {F.}~\bibnamefont {{Biscani}}}, \bibinfo {author} {\bibfnamefont {M.}~\bibnamefont {{Boquien}}}, \bibinfo {author} {\bibfnamefont {K.~A.}\ \bibnamefont {{Bostroem}}}, \bibinfo {author} {\bibfnamefont {L.~G.}\ \bibnamefont {{Bouma}}}, \bibinfo {author} {\bibfnamefont {G.~B.}\ \bibnamefont {{Brammer}}}, \bibinfo {author} {\bibfnamefont {E.~M.}\ \bibnamefont {{Bray}}}, \bibinfo {author} {\bibfnamefont {H.}~\bibnamefont {{Breytenbach}}}, \bibinfo {author} {\bibfnamefont {H.}~\bibnamefont {{Buddelmeijer}}}, \bibinfo {author} {\bibfnamefont {D.~J.}\ \bibnamefont {{Burke}}}, \bibinfo {author} {\bibfnamefont {G.}~\bibnamefont {{Calderone}}}, \bibinfo {author} {\bibfnamefont {J.~L.}\ \bibnamefont {{Cano Rodr{\'\i}guez}}}, \bibinfo {author} {\bibfnamefont {M.}~\bibnamefont {{Cara}}}, \bibinfo {author} {\bibfnamefont {J.~V.~M.}\ \bibnamefont {{Cardoso}}}, \bibinfo {author} {\bibfnamefont {S.}~\bibnamefont {{Cheedella}}}, \bibinfo {author} {\bibfnamefont {Y.}~\bibnamefont {{Copin}}}, \bibinfo {author} {\bibfnamefont
  {L.}~\bibnamefont {{Corrales}}}, \bibinfo {author} {\bibfnamefont {D.}~\bibnamefont {{Crichton}}}, \bibinfo {author} {\bibfnamefont {D.}~\bibnamefont {{D'Avella}}}, \bibinfo {author} {\bibfnamefont {C.}~\bibnamefont {{Deil}}}, \bibinfo {author} {\bibfnamefont {{\'E}.}~\bibnamefont {{Depagne}}}, \bibinfo {author} {\bibfnamefont {J.~P.}\ \bibnamefont {{Dietrich}}}, \bibinfo {author} {\bibfnamefont {A.}~\bibnamefont {{Donath}}}, \bibinfo {author} {\bibfnamefont {M.}~\bibnamefont {{Droettboom}}}, \bibinfo {author} {\bibfnamefont {N.}~\bibnamefont {{Earl}}}, \bibinfo {author} {\bibfnamefont {T.}~\bibnamefont {{Erben}}}, \bibinfo {author} {\bibfnamefont {S.}~\bibnamefont {{Fabbro}}}, \bibinfo {author} {\bibfnamefont {L.~A.}\ \bibnamefont {{Ferreira}}}, \bibinfo {author} {\bibfnamefont {T.}~\bibnamefont {{Finethy}}}, \bibinfo {author} {\bibfnamefont {R.~T.}\ \bibnamefont {{Fox}}}, \bibinfo {author} {\bibfnamefont {L.~H.}\ \bibnamefont {{Garrison}}}, \bibinfo {author} {\bibfnamefont {S.~L.~J.}\ \bibnamefont
  {{Gibbons}}}, \bibinfo {author} {\bibfnamefont {D.~A.}\ \bibnamefont {{Goldstein}}}, \bibinfo {author} {\bibfnamefont {R.}~\bibnamefont {{Gommers}}}, \bibinfo {author} {\bibfnamefont {J.~P.}\ \bibnamefont {{Greco}}}, \bibinfo {author} {\bibfnamefont {P.}~\bibnamefont {{Greenfield}}}, \bibinfo {author} {\bibfnamefont {A.~M.}\ \bibnamefont {{Groener}}}, \bibinfo {author} {\bibfnamefont {F.}~\bibnamefont {{Grollier}}}, \bibinfo {author} {\bibfnamefont {A.}~\bibnamefont {{Hagen}}}, \bibinfo {author} {\bibfnamefont {P.}~\bibnamefont {{Hirst}}}, \bibinfo {author} {\bibfnamefont {D.}~\bibnamefont {{Homeier}}}, \bibinfo {author} {\bibfnamefont {A.~J.}\ \bibnamefont {{Horton}}}, \bibinfo {author} {\bibfnamefont {G.}~\bibnamefont {{Hosseinzadeh}}}, \bibinfo {author} {\bibfnamefont {L.}~\bibnamefont {{Hu}}}, \bibinfo {author} {\bibfnamefont {J.~S.}\ \bibnamefont {{Hunkeler}}}, \bibinfo {author} {\bibfnamefont {{\v{Z}}.}~\bibnamefont {{Ivezi{\'c}}}}, \bibinfo {author} {\bibfnamefont {A.}~\bibnamefont {{Jain}}},
  \bibinfo {author} {\bibfnamefont {T.}~\bibnamefont {{Jenness}}}, \bibinfo {author} {\bibfnamefont {G.}~\bibnamefont {{Kanarek}}}, \bibinfo {author} {\bibfnamefont {S.}~\bibnamefont {{Kendrew}}}, \bibinfo {author} {\bibfnamefont {N.~S.}\ \bibnamefont {{Kern}}}, \bibinfo {author} {\bibfnamefont {W.~E.}\ \bibnamefont {{Kerzendorf}}}, \bibinfo {author} {\bibfnamefont {A.}~\bibnamefont {{Khvalko}}}, \bibinfo {author} {\bibfnamefont {J.}~\bibnamefont {{King}}}, \bibinfo {author} {\bibfnamefont {D.}~\bibnamefont {{Kirkby}}}, \bibinfo {author} {\bibfnamefont {A.~M.}\ \bibnamefont {{Kulkarni}}}, \bibinfo {author} {\bibfnamefont {A.}~\bibnamefont {{Kumar}}}, \bibinfo {author} {\bibfnamefont {A.}~\bibnamefont {{Lee}}}, \bibinfo {author} {\bibfnamefont {D.}~\bibnamefont {{Lenz}}}, \bibinfo {author} {\bibfnamefont {S.~P.}\ \bibnamefont {{Littlefair}}}, \bibinfo {author} {\bibfnamefont {Z.}~\bibnamefont {{Ma}}}, \bibinfo {author} {\bibfnamefont {D.~M.}\ \bibnamefont {{Macleod}}}, \bibinfo {author} {\bibfnamefont
  {M.}~\bibnamefont {{Mastropietro}}}, \bibinfo {author} {\bibfnamefont {C.}~\bibnamefont {{McCully}}}, \bibinfo {author} {\bibfnamefont {S.}~\bibnamefont {{Montagnac}}}, \bibinfo {author} {\bibfnamefont {B.~M.}\ \bibnamefont {{Morris}}}, \bibinfo {author} {\bibfnamefont {M.}~\bibnamefont {{Mueller}}}, \bibinfo {author} {\bibfnamefont {S.~J.}\ \bibnamefont {{Mumford}}}, \bibinfo {author} {\bibfnamefont {D.}~\bibnamefont {{Muna}}}, \bibinfo {author} {\bibfnamefont {N.~A.}\ \bibnamefont {{Murphy}}}, \bibinfo {author} {\bibfnamefont {S.}~\bibnamefont {{Nelson}}}, \bibinfo {author} {\bibfnamefont {G.~H.}\ \bibnamefont {{Nguyen}}}, \bibinfo {author} {\bibfnamefont {J.~P.}\ \bibnamefont {{Ninan}}}, \bibinfo {author} {\bibfnamefont {M.}~\bibnamefont {{N{\"o}the}}}, \bibinfo {author} {\bibfnamefont {S.}~\bibnamefont {{Ogaz}}}, \bibinfo {author} {\bibfnamefont {S.}~\bibnamefont {{Oh}}}, \bibinfo {author} {\bibfnamefont {J.~K.}\ \bibnamefont {{Parejko}}}, \bibinfo {author} {\bibfnamefont {N.}~\bibnamefont {{Parley}}},
  \bibinfo {author} {\bibfnamefont {S.}~\bibnamefont {{Pascual}}}, \bibinfo {author} {\bibfnamefont {R.}~\bibnamefont {{Patil}}}, \bibinfo {author} {\bibfnamefont {A.~A.}\ \bibnamefont {{Patil}}}, \bibinfo {author} {\bibfnamefont {A.~L.}\ \bibnamefont {{Plunkett}}}, \bibinfo {author} {\bibfnamefont {J.~X.}\ \bibnamefont {{Prochaska}}}, \bibinfo {author} {\bibfnamefont {T.}~\bibnamefont {{Rastogi}}}, \bibinfo {author} {\bibfnamefont {V.}~\bibnamefont {{Reddy Janga}}}, \bibinfo {author} {\bibfnamefont {J.}~\bibnamefont {{Sabater}}}, \bibinfo {author} {\bibfnamefont {P.}~\bibnamefont {{Sakurikar}}}, \bibinfo {author} {\bibfnamefont {M.}~\bibnamefont {{Seifert}}}, \bibinfo {author} {\bibfnamefont {L.~E.}\ \bibnamefont {{Sherbert}}}, \bibinfo {author} {\bibfnamefont {H.}~\bibnamefont {{Sherwood-Taylor}}}, \bibinfo {author} {\bibfnamefont {A.~Y.}\ \bibnamefont {{Shih}}}, \bibinfo {author} {\bibfnamefont {J.}~\bibnamefont {{Sick}}}, \bibinfo {author} {\bibfnamefont {M.~T.}\ \bibnamefont {{Silbiger}}}, \bibinfo
  {author} {\bibfnamefont {S.}~\bibnamefont {{Singanamalla}}}, \bibinfo {author} {\bibfnamefont {L.~P.}\ \bibnamefont {{Singer}}}, \bibinfo {author} {\bibfnamefont {P.~H.}\ \bibnamefont {{Sladen}}}, \bibinfo {author} {\bibfnamefont {K.~A.}\ \bibnamefont {{Sooley}}}, \bibinfo {author} {\bibfnamefont {S.}~\bibnamefont {{Sornarajah}}}, \bibinfo {author} {\bibfnamefont {O.}~\bibnamefont {{Streicher}}}, \bibinfo {author} {\bibfnamefont {P.}~\bibnamefont {{Teuben}}}, \bibinfo {author} {\bibfnamefont {S.~W.}\ \bibnamefont {{Thomas}}}, \bibinfo {author} {\bibfnamefont {G.~R.}\ \bibnamefont {{Tremblay}}}, \bibinfo {author} {\bibfnamefont {J.~E.~H.}\ \bibnamefont {{Turner}}}, \bibinfo {author} {\bibfnamefont {V.}~\bibnamefont {{Terr{\'o}n}}}, \bibinfo {author} {\bibfnamefont {M.~H.}\ \bibnamefont {{van Kerkwijk}}}, \bibinfo {author} {\bibfnamefont {A.}~\bibnamefont {{de la Vega}}}, \bibinfo {author} {\bibfnamefont {L.~L.}\ \bibnamefont {{Watkins}}}, \bibinfo {author} {\bibfnamefont {B.~A.}\ \bibnamefont {{Weaver}}},
  \bibinfo {author} {\bibfnamefont {J.~B.}\ \bibnamefont {{Whitmore}}}, \bibinfo {author} {\bibfnamefont {J.}~\bibnamefont {{Woillez}}}, \bibinfo {author} {\bibfnamefont {V.}~\bibnamefont {{Zabalza}}},\ and\ \bibinfo {author} {\bibnamefont {{Astropy Contributors}}},\ }\href {https://doi.org/10.3847/1538-3881/aabc4f} {\bibfield  {journal} {\bibinfo  {journal} {\aj}\ }\textbf {\bibinfo {volume} {156}},\ \bibinfo {eid} {123} (\bibinfo {year} {2018})},\ \Eprint {https://arxiv.org/abs/1801.02634} {arXiv:1801.02634 [astro-ph.IM]} \BibitemShut {NoStop}%
\bibitem [{\citenamefont {{Astropy Collaboration}}\ \emph {et~al.}(2013)\citenamefont {{Astropy Collaboration}}, \citenamefont {{Robitaille}}, \citenamefont {{Tollerud}}, \citenamefont {{Greenfield}}, \citenamefont {{Droettboom}}, \citenamefont {{Bray}}, \citenamefont {{Aldcroft}}, \citenamefont {{Davis}}, \citenamefont {{Ginsburg}}, \citenamefont {{Price-Whelan}}, \citenamefont {{Kerzendorf}}, \citenamefont {{Conley}}, \citenamefont {{Crighton}}, \citenamefont {{Barbary}}, \citenamefont {{Muna}}, \citenamefont {{Ferguson}}, \citenamefont {{Grollier}}, \citenamefont {{Parikh}}, \citenamefont {{Nair}}, \citenamefont {{Unther}}, \citenamefont {{Deil}}, \citenamefont {{Woillez}}, \citenamefont {{Conseil}}, \citenamefont {{Kramer}}, \citenamefont {{Turner}}, \citenamefont {{Singer}}, \citenamefont {{Fox}}, \citenamefont {{Weaver}}, \citenamefont {{Zabalza}}, \citenamefont {{Edwards}}, \citenamefont {{Azalee Bostroem}}, \citenamefont {{Burke}}, \citenamefont {{Casey}}, \citenamefont {{Crawford}}, \citenamefont
  {{Dencheva}}, \citenamefont {{Ely}}, \citenamefont {{Jenness}}, \citenamefont {{Labrie}}, \citenamefont {{Lim}}, \citenamefont {{Pierfederici}}, \citenamefont {{Pontzen}}, \citenamefont {{Ptak}}, \citenamefont {{Refsdal}}, \citenamefont {{Servillat}},\ and\ \citenamefont {{Streicher}}}]{2013A&A...558A..33A}%
  \BibitemOpen
  \bibfield  {author} {\bibinfo {author} {\bibnamefont {{Astropy Collaboration}}}, \bibinfo {author} {\bibfnamefont {T.~P.}\ \bibnamefont {{Robitaille}}}, \bibinfo {author} {\bibfnamefont {E.~J.}\ \bibnamefont {{Tollerud}}}, \bibinfo {author} {\bibfnamefont {P.}~\bibnamefont {{Greenfield}}}, \bibinfo {author} {\bibfnamefont {M.}~\bibnamefont {{Droettboom}}}, \bibinfo {author} {\bibfnamefont {E.}~\bibnamefont {{Bray}}}, \bibinfo {author} {\bibfnamefont {T.}~\bibnamefont {{Aldcroft}}}, \bibinfo {author} {\bibfnamefont {M.}~\bibnamefont {{Davis}}}, \bibinfo {author} {\bibfnamefont {A.}~\bibnamefont {{Ginsburg}}}, \bibinfo {author} {\bibfnamefont {A.~M.}\ \bibnamefont {{Price-Whelan}}}, \bibinfo {author} {\bibfnamefont {W.~E.}\ \bibnamefont {{Kerzendorf}}}, \bibinfo {author} {\bibfnamefont {A.}~\bibnamefont {{Conley}}}, \bibinfo {author} {\bibfnamefont {N.}~\bibnamefont {{Crighton}}}, \bibinfo {author} {\bibfnamefont {K.}~\bibnamefont {{Barbary}}}, \bibinfo {author} {\bibfnamefont {D.}~\bibnamefont {{Muna}}},
  \bibinfo {author} {\bibfnamefont {H.}~\bibnamefont {{Ferguson}}}, \bibinfo {author} {\bibfnamefont {F.}~\bibnamefont {{Grollier}}}, \bibinfo {author} {\bibfnamefont {M.~M.}\ \bibnamefont {{Parikh}}}, \bibinfo {author} {\bibfnamefont {P.~H.}\ \bibnamefont {{Nair}}}, \bibinfo {author} {\bibfnamefont {H.~M.}\ \bibnamefont {{Unther}}}, \bibinfo {author} {\bibfnamefont {C.}~\bibnamefont {{Deil}}}, \bibinfo {author} {\bibfnamefont {J.}~\bibnamefont {{Woillez}}}, \bibinfo {author} {\bibfnamefont {S.}~\bibnamefont {{Conseil}}}, \bibinfo {author} {\bibfnamefont {R.}~\bibnamefont {{Kramer}}}, \bibinfo {author} {\bibfnamefont {J.~E.~H.}\ \bibnamefont {{Turner}}}, \bibinfo {author} {\bibfnamefont {L.}~\bibnamefont {{Singer}}}, \bibinfo {author} {\bibfnamefont {R.}~\bibnamefont {{Fox}}}, \bibinfo {author} {\bibfnamefont {B.~A.}\ \bibnamefont {{Weaver}}}, \bibinfo {author} {\bibfnamefont {V.}~\bibnamefont {{Zabalza}}}, \bibinfo {author} {\bibfnamefont {Z.~I.}\ \bibnamefont {{Edwards}}}, \bibinfo {author} {\bibfnamefont
  {K.}~\bibnamefont {{Azalee Bostroem}}}, \bibinfo {author} {\bibfnamefont {D.~J.}\ \bibnamefont {{Burke}}}, \bibinfo {author} {\bibfnamefont {A.~R.}\ \bibnamefont {{Casey}}}, \bibinfo {author} {\bibfnamefont {S.~M.}\ \bibnamefont {{Crawford}}}, \bibinfo {author} {\bibfnamefont {N.}~\bibnamefont {{Dencheva}}}, \bibinfo {author} {\bibfnamefont {J.}~\bibnamefont {{Ely}}}, \bibinfo {author} {\bibfnamefont {T.}~\bibnamefont {{Jenness}}}, \bibinfo {author} {\bibfnamefont {K.}~\bibnamefont {{Labrie}}}, \bibinfo {author} {\bibfnamefont {P.~L.}\ \bibnamefont {{Lim}}}, \bibinfo {author} {\bibfnamefont {F.}~\bibnamefont {{Pierfederici}}}, \bibinfo {author} {\bibfnamefont {A.}~\bibnamefont {{Pontzen}}}, \bibinfo {author} {\bibfnamefont {A.}~\bibnamefont {{Ptak}}}, \bibinfo {author} {\bibfnamefont {B.}~\bibnamefont {{Refsdal}}}, \bibinfo {author} {\bibfnamefont {M.}~\bibnamefont {{Servillat}}},\ and\ \bibinfo {author} {\bibfnamefont {O.}~\bibnamefont {{Streicher}}},\ }\href {https://doi.org/10.1051/0004-6361/201322068}
  {\bibfield  {journal} {\bibinfo  {journal} {\aap}\ }\textbf {\bibinfo {volume} {558}},\ \bibinfo {eid} {A33} (\bibinfo {year} {2013})},\ \Eprint {https://arxiv.org/abs/1307.6212} {arXiv:1307.6212 [astro-ph.IM]} \BibitemShut {NoStop}%
\bibitem [{\citenamefont {{Bertin}}\ and\ \citenamefont {{Arnouts}}(1996)}]{1996AAS..117..393B}%
  \BibitemOpen
  \bibfield  {author} {\bibinfo {author} {\bibfnamefont {E.}~\bibnamefont {{Bertin}}}\ and\ \bibinfo {author} {\bibfnamefont {S.}~\bibnamefont {{Arnouts}}},\ }\href {https://doi.org/10.1051/aas:1996164} {\bibfield  {journal} {\bibinfo  {journal} {\aaps}\ }\textbf {\bibinfo {volume} {117}},\ \bibinfo {pages} {393} (\bibinfo {year} {1996})}\BibitemShut {NoStop}%
\bibitem [{\citenamefont {{Cloutier}}\ \emph {et~al.}(2018)\citenamefont {{Cloutier}}, \citenamefont {{Doyon}}, \citenamefont {{Bouchy}},\ and\ \citenamefont {{H{\'e}brard}}}]{2018AJ....156...82C}%
  \BibitemOpen
  \bibfield  {author} {\bibinfo {author} {\bibfnamefont {R.}~\bibnamefont {{Cloutier}}}, \bibinfo {author} {\bibfnamefont {R.}~\bibnamefont {{Doyon}}}, \bibinfo {author} {\bibfnamefont {F.}~\bibnamefont {{Bouchy}}},\ and\ \bibinfo {author} {\bibfnamefont {G.}~\bibnamefont {{H{\'e}brard}}},\ }\href {https://doi.org/10.3847/1538-3881/aacea9} {\bibfield  {journal} {\bibinfo  {journal} {\aj}\ }\textbf {\bibinfo {volume} {156}},\ \bibinfo {eid} {82} (\bibinfo {year} {2018})},\ \Eprint {https://arxiv.org/abs/1807.01263} {arXiv:1807.01263 [astro-ph.EP]} \BibitemShut {NoStop}%
\bibitem [{\citenamefont {{Corrales}}(2015)}]{2015ApJ...805...23C}%
  \BibitemOpen
  \bibfield  {author} {\bibinfo {author} {\bibfnamefont {L.}~\bibnamefont {{Corrales}}},\ }\href {https://doi.org/10.1088/0004-637X/805/1/23} {\bibfield  {journal} {\bibinfo  {journal} {\apj}\ }\textbf {\bibinfo {volume} {805}},\ \bibinfo {eid} {23} (\bibinfo {year} {2015})},\ \Eprint {https://arxiv.org/abs/1503.01475} {arXiv:1503.01475 [astro-ph.HE]} \BibitemShut {NoStop}%
\bibitem [{\citenamefont {{Ferland}}\ \emph {et~al.}(2013)\citenamefont {{Ferland}}, \citenamefont {{Porter}}, \citenamefont {{van Hoof}}, \citenamefont {{Williams}}, \citenamefont {{Abel}}, \citenamefont {{Lykins}}, \citenamefont {{Shaw}}, \citenamefont {{Henney}},\ and\ \citenamefont {{Stancil}}}]{2013RMxAA..49..137F}%
  \BibitemOpen
  \bibfield  {author} {\bibinfo {author} {\bibfnamefont {G.~J.}\ \bibnamefont {{Ferland}}}, \bibinfo {author} {\bibfnamefont {R.~L.}\ \bibnamefont {{Porter}}}, \bibinfo {author} {\bibfnamefont {P.~A.~M.}\ \bibnamefont {{van Hoof}}}, \bibinfo {author} {\bibfnamefont {R.~J.~R.}\ \bibnamefont {{Williams}}}, \bibinfo {author} {\bibfnamefont {N.~P.}\ \bibnamefont {{Abel}}}, \bibinfo {author} {\bibfnamefont {M.~L.}\ \bibnamefont {{Lykins}}}, \bibinfo {author} {\bibfnamefont {G.}~\bibnamefont {{Shaw}}}, \bibinfo {author} {\bibfnamefont {W.~J.}\ \bibnamefont {{Henney}}},\ and\ \bibinfo {author} {\bibfnamefont {P.~C.}\ \bibnamefont {{Stancil}}},\ }\href@noop {} {\bibfield  {journal} {\bibinfo  {journal} {\rmxaa}\ }\textbf {\bibinfo {volume} {49}},\ \bibinfo {pages} {137} (\bibinfo {year} {2013})},\ \Eprint {https://arxiv.org/abs/1302.4485} {arXiv:1302.4485 [astro-ph.GA]} \BibitemShut {NoStop}%
\bibitem [{\citenamefont {{Hanisch}}\ and\ \citenamefont {{Biemesderfer}}(1989)}]{1989BAAS...21..780H}%
  \BibitemOpen
  \bibfield  {author} {\bibinfo {author} {\bibfnamefont {R.~J.}\ \bibnamefont {{Hanisch}}}\ and\ \bibinfo {author} {\bibfnamefont {C.~D.}\ \bibnamefont {{Biemesderfer}}},\ }in\ \href@noop {} {\emph {\bibinfo {booktitle} {\baas}}}\ (\bibinfo {year} {1989})\ p.\ \bibinfo {pages} {780}\BibitemShut {NoStop}%
\bibitem [{\citenamefont {{Lamport}}(1994)}]{lamport94}%
  \BibitemOpen
  \bibfield  {author} {\bibinfo {author} {\bibfnamefont {L.}~\bibnamefont {{Lamport}}},\ }\href@noop {} {\emph {\bibinfo {title} {{LaTeX: A Document Preparation System}}}},\ \bibinfo {edition} {2nd}\ ed.\ (\bibinfo  {publisher} {Addison-Wesley Professional},\ \bibinfo {year} {1994})\BibitemShut {NoStop}%
\bibitem [{\citenamefont {{Li}}\ \emph {et~al.}(2018)\citenamefont {{Li}}, \citenamefont {{Zhang}}, \citenamefont {{Peter}}, \citenamefont {{Chitta}}, \citenamefont {{Su}}, \citenamefont {{Song}}, \citenamefont {{Xia}},\ and\ \citenamefont {{Hou}}}]{2018ApJ...868L..33L}%
  \BibitemOpen
  \bibfield  {author} {\bibinfo {author} {\bibfnamefont {L.}~\bibnamefont {{Li}}}, \bibinfo {author} {\bibfnamefont {J.}~\bibnamefont {{Zhang}}}, \bibinfo {author} {\bibfnamefont {H.}~\bibnamefont {{Peter}}}, \bibinfo {author} {\bibfnamefont {L.~P.}\ \bibnamefont {{Chitta}}}, \bibinfo {author} {\bibfnamefont {J.}~\bibnamefont {{Su}}}, \bibinfo {author} {\bibfnamefont {H.}~\bibnamefont {{Song}}}, \bibinfo {author} {\bibfnamefont {C.}~\bibnamefont {{Xia}}},\ and\ \bibinfo {author} {\bibfnamefont {Y.}~\bibnamefont {{Hou}}},\ }\href {https://doi.org/10.3847/2041-8213/aaf167} {\bibfield  {journal} {\bibinfo  {journal} {\apj}\ }\textbf {\bibinfo {volume} {868}},\ \bibinfo {eid} {L33} (\bibinfo {year} {2018})},\ \Eprint {https://arxiv.org/abs/1811.08553} {arXiv:1811.08553 [astro-ph.SR]} \BibitemShut {NoStop}%
\bibitem [{\citenamefont {{Pr{\v{s}}a}}\ \emph {et~al.}(2016)\citenamefont {{Pr{\v{s}}a}}, \citenamefont {{Harmanec}}, \citenamefont {{Torres}}, \citenamefont {{Mamajek}}, \citenamefont {{Asplund}}, \citenamefont {{Capitaine}}, \citenamefont {{Christensen-Dalsgaard}}, \citenamefont {{Depagne}}, \citenamefont {{Haberreiter}},\ and\ \citenamefont {{Hekker}}}]{2016AJ....152...41P}%
  \BibitemOpen
  \bibfield  {author} {\bibinfo {author} {\bibfnamefont {A.}~\bibnamefont {{Pr{\v{s}}a}}}, \bibinfo {author} {\bibfnamefont {P.}~\bibnamefont {{Harmanec}}}, \bibinfo {author} {\bibfnamefont {G.}~\bibnamefont {{Torres}}}, \bibinfo {author} {\bibfnamefont {E.}~\bibnamefont {{Mamajek}}}, \bibinfo {author} {\bibfnamefont {M.}~\bibnamefont {{Asplund}}}, \bibinfo {author} {\bibfnamefont {N.}~\bibnamefont {{Capitaine}}}, \bibinfo {author} {\bibfnamefont {J.}~\bibnamefont {{Christensen-Dalsgaard}}}, \bibinfo {author} {\bibfnamefont {{\'E}.}~\bibnamefont {{Depagne}}}, \bibinfo {author} {\bibfnamefont {M.}~\bibnamefont {{Haberreiter}}},\ and\ \bibinfo {author} {\bibfnamefont {S.}~\bibnamefont {{Hekker}}},\ }\href {https://doi.org/10.3847/0004-6256/152/2/41} {\bibfield  {journal} {\bibinfo  {journal} {\aj}\ }\textbf {\bibinfo {volume} {152}},\ \bibinfo {eid} {41} (\bibinfo {year} {2016})},\ \Eprint {https://arxiv.org/abs/1605.09788} {arXiv:1605.09788 [astro-ph.SR]} \BibitemShut {NoStop}%
\bibitem [{\citenamefont {{Schwarz}}\ \emph {et~al.}(2011)\citenamefont {{Schwarz}}, \citenamefont {{Ness}}, \citenamefont {{Osborne}}, \citenamefont {{Page}}, \citenamefont {{Evans}}, \citenamefont {{Beardmore}}, \citenamefont {{Walter}}, \citenamefont {{Helton}}, \citenamefont {{Woodward}}, \citenamefont {{Bode}}, \citenamefont {{Starrfield}},\ and\ \citenamefont {{Drake}}}]{2011ApJS..197...31S}%
  \BibitemOpen
  \bibfield  {author} {\bibinfo {author} {\bibfnamefont {G.~J.}\ \bibnamefont {{Schwarz}}}, \bibinfo {author} {\bibfnamefont {J.-U.}\ \bibnamefont {{Ness}}}, \bibinfo {author} {\bibfnamefont {J.~P.}\ \bibnamefont {{Osborne}}}, \bibinfo {author} {\bibfnamefont {K.~L.}\ \bibnamefont {{Page}}}, \bibinfo {author} {\bibfnamefont {P.~A.}\ \bibnamefont {{Evans}}}, \bibinfo {author} {\bibfnamefont {A.~P.}\ \bibnamefont {{Beardmore}}}, \bibinfo {author} {\bibfnamefont {F.~M.}\ \bibnamefont {{Walter}}}, \bibinfo {author} {\bibfnamefont {L.~A.}\ \bibnamefont {{Helton}}}, \bibinfo {author} {\bibfnamefont {C.~E.}\ \bibnamefont {{Woodward}}}, \bibinfo {author} {\bibfnamefont {M.}~\bibnamefont {{Bode}}}, \bibinfo {author} {\bibfnamefont {S.}~\bibnamefont {{Starrfield}}},\ and\ \bibinfo {author} {\bibfnamefont {J.~J.}\ \bibnamefont {{Drake}}},\ }\href {https://doi.org/10.1088/0067-0049/197/2/31} {\bibfield  {journal} {\bibinfo  {journal} {\apjs}\ }\textbf {\bibinfo {volume} {197}},\ \bibinfo {eid} {31} (\bibinfo {year}
  {2011})},\ \Eprint {https://arxiv.org/abs/1110.6224} {arXiv:1110.6224 [astro-ph.SR]} \BibitemShut {NoStop}%
\bibitem [{\citenamefont {{Vogt}}\ \emph {et~al.}(2014)\citenamefont {{Vogt}}, \citenamefont {{Dopita}}, \citenamefont {{Kewley}}, \citenamefont {{Sutherland}}, \citenamefont {{Scharw{\"a}chter}}, \citenamefont {{Basurah}}, \citenamefont {{Ali}},\ and\ \citenamefont {{Amer}}}]{2014ApJ...793..127V}%
  \BibitemOpen
  \bibfield  {author} {\bibinfo {author} {\bibfnamefont {F.~P.~A.}\ \bibnamefont {{Vogt}}}, \bibinfo {author} {\bibfnamefont {M.~A.}\ \bibnamefont {{Dopita}}}, \bibinfo {author} {\bibfnamefont {L.~J.}\ \bibnamefont {{Kewley}}}, \bibinfo {author} {\bibfnamefont {R.~S.}\ \bibnamefont {{Sutherland}}}, \bibinfo {author} {\bibfnamefont {J.}~\bibnamefont {{Scharw{\"a}chter}}}, \bibinfo {author} {\bibfnamefont {H.~M.}\ \bibnamefont {{Basurah}}}, \bibinfo {author} {\bibfnamefont {A.}~\bibnamefont {{Ali}}},\ and\ \bibinfo {author} {\bibfnamefont {M.~A.}\ \bibnamefont {{Amer}}},\ }\href {https://doi.org/10.1088/0004-637X/793/2/127} {\bibfield  {journal} {\bibinfo  {journal} {\apj}\ }\textbf {\bibinfo {volume} {793}},\ \bibinfo {eid} {127} (\bibinfo {year} {2014})},\ \Eprint {https://arxiv.org/abs/1406.5186} {arXiv:1406.5186 [astro-ph.GA]} \BibitemShut {NoStop}%
\bibitem [{\citenamefont {{Niu}}\ \emph {et~al.}(2023)\citenamefont {{Niu}}, \citenamefont {{Chen}},\ and\ \citenamefont {{Zhang}}}]{2023arXiv230504752N}%
  \BibitemOpen
  \bibfield  {author} {\bibinfo {author} {\bibfnamefont {J.}~\bibnamefont {{Niu}}}, \bibinfo {author} {\bibfnamefont {Y.}~\bibnamefont {{Chen}}},\ and\ \bibinfo {author} {\bibfnamefont {T.-J.}\ \bibnamefont {{Zhang}}},\ }\href {https://doi.org/10.48550/arXiv.2305.04752} {\bibfield  {journal} {\bibinfo  {journal} {arXiv e-prints}\ ,\ \bibinfo {eid} {arXiv:2305.04752}} (\bibinfo {year} {2023})},\ \Eprint {https://arxiv.org/abs/2305.04752} {arXiv:2305.04752 [astro-ph.CO]} \BibitemShut {NoStop}%
\bibitem [{\citenamefont {{Niu}}\ and\ \citenamefont {{Zhang}}(2023)}]{2023PDU....3901147N}%
  \BibitemOpen
  \bibfield  {author} {\bibinfo {author} {\bibfnamefont {J.}~\bibnamefont {{Niu}}}\ and\ \bibinfo {author} {\bibfnamefont {T.-J.}\ \bibnamefont {{Zhang}}},\ }\href {https://doi.org/10.1016/j.dark.2022.101147} {\bibfield  {journal} {\bibinfo  {journal} {Physics of the Dark Universe}\ }\textbf {\bibinfo {volume} {39}},\ \bibinfo {eid} {101147} (\bibinfo {year} {2023})},\ \Eprint {https://arxiv.org/abs/2204.10597} {arXiv:2204.10597 [astro-ph.CO]} \BibitemShut {NoStop}%
\bibitem [{\citenamefont {Jimenez~Rezende}\ and\ \citenamefont {Mohamed}(2015)}]{articleRezende}%
  \BibitemOpen
  \bibfield  {author} {\bibinfo {author} {\bibfnamefont {D.}~\bibnamefont {Jimenez~Rezende}}\ and\ \bibinfo {author} {\bibfnamefont {S.}~\bibnamefont {Mohamed}},\ }\href@noop {} {\bibfield  {journal} {\bibinfo  {journal} {Proceedings of the 32nd International Conference on Machine Learning}\ } (\bibinfo {year} {2015})}\BibitemShut {NoStop}%
\bibitem [{\citenamefont {Endo}\ \emph {et~al.}(2019)\citenamefont {Endo}, \citenamefont {{van Leeuwen}},\ and\ \citenamefont {Baguelin}}]{ENDO2019100363}%
  \BibitemOpen
  \bibfield  {author} {\bibinfo {author} {\bibfnamefont {A.}~\bibnamefont {Endo}}, \bibinfo {author} {\bibfnamefont {E.}~\bibnamefont {{van Leeuwen}}},\ and\ \bibinfo {author} {\bibfnamefont {M.}~\bibnamefont {Baguelin}},\ }\href {https://doi.org/https://doi.org/10.1016/j.epidem.2019.100363} {\bibfield  {journal} {\bibinfo  {journal} {Epidemics}\ }\textbf {\bibinfo {volume} {29}},\ \bibinfo {pages} {100363} (\bibinfo {year} {2019})}\BibitemShut {NoStop}%
\bibitem [{\citenamefont {Thomas}(2022)}]{Economics2022Lux}%
  \BibitemOpen
  \bibfield  {author} {\bibinfo {author} {\bibfnamefont {L.}~\bibnamefont {Thomas}},\ }\href {https://doi.org/10.1007/s10614-021-10155-0} {\bibfield  {journal} {\bibinfo  {journal} {Computational Economics}\ }\textbf {\bibinfo {volume} {60}},\ \bibinfo {pages} {451} (\bibinfo {year} {2022})}\BibitemShut {NoStop}%
\bibitem [{\citenamefont {{Li}}\ \emph {et~al.}(2021)\citenamefont {{Li}}, \citenamefont {{Rao}}, \citenamefont {{Hassaine}}, \citenamefont {{Ramakrishnan}}, \citenamefont {{Canoy}}, \citenamefont {{Salimi-Khorshidi}}, \citenamefont {{Mamouei}}, \citenamefont {{Lukasiewicz}},\ and\ \citenamefont {{Rahimi}}}]{Li2021}%
  \BibitemOpen
  \bibfield  {author} {\bibinfo {author} {\bibfnamefont {Y.}~\bibnamefont {{Li}}}, \bibinfo {author} {\bibfnamefont {S.}~\bibnamefont {{Rao}}}, \bibinfo {author} {\bibfnamefont {A.}~\bibnamefont {{Hassaine}}}, \bibinfo {author} {\bibfnamefont {R.}~\bibnamefont {{Ramakrishnan}}}, \bibinfo {author} {\bibfnamefont {D.}~\bibnamefont {{Canoy}}}, \bibinfo {author} {\bibfnamefont {G.}~\bibnamefont {{Salimi-Khorshidi}}}, \bibinfo {author} {\bibfnamefont {M.}~\bibnamefont {{Mamouei}}}, \bibinfo {author} {\bibfnamefont {T.}~\bibnamefont {{Lukasiewicz}}},\ and\ \bibinfo {author} {\bibfnamefont {K.}~\bibnamefont {{Rahimi}}},\ }\href {https://doi.org/10.1038/s41598-021-00144-6} {\bibfield  {journal} {\bibinfo  {journal} {Scientific Reports}\ }\textbf {\bibinfo {volume} {11}},\ \bibinfo {pages} {20685} (\bibinfo {year} {2021})}\BibitemShut {NoStop}%
\bibitem [{\citenamefont {Rostam}\ \emph {et~al.}(2020)\citenamefont {Rostam}, \citenamefont {Nagamune},\ and\ \citenamefont {Grebenyuk}}]{ROSTAM2020149}%
  \BibitemOpen
  \bibfield  {author} {\bibinfo {author} {\bibfnamefont {M.}~\bibnamefont {Rostam}}, \bibinfo {author} {\bibfnamefont {R.}~\bibnamefont {Nagamune}},\ and\ \bibinfo {author} {\bibfnamefont {V.}~\bibnamefont {Grebenyuk}},\ }\href {https://doi.org/https://doi.org/10.1016/j.jprocont.2020.06.006} {\bibfield  {journal} {\bibinfo  {journal} {Journal of Process Control}\ }\textbf {\bibinfo {volume} {92}},\ \bibinfo {pages} {149} (\bibinfo {year} {2020})}\BibitemShut {NoStop}%
\bibitem [{\citenamefont {{Racine}}\ \emph {et~al.}(2016)\citenamefont {{Racine}}, \citenamefont {{Jewell}}, \citenamefont {{Eriksen}},\ and\ \citenamefont {{Wehus}}}]{2016ApJ...820...31R}%
  \BibitemOpen
  \bibfield  {author} {\bibinfo {author} {\bibfnamefont {B.}~\bibnamefont {{Racine}}}, \bibinfo {author} {\bibfnamefont {J.~B.}\ \bibnamefont {{Jewell}}}, \bibinfo {author} {\bibfnamefont {H.~K.}\ \bibnamefont {{Eriksen}}},\ and\ \bibinfo {author} {\bibfnamefont {I.~K.}\ \bibnamefont {{Wehus}}},\ }\href {https://doi.org/10.3847/0004-637X/820/1/31} {\bibfield  {journal} {\bibinfo  {journal} {\apj}\ }\textbf {\bibinfo {volume} {820}},\ \bibinfo {eid} {31} (\bibinfo {year} {2016})},\ \Eprint {https://arxiv.org/abs/1512.06619} {arXiv:1512.06619 [astro-ph.CO]} \BibitemShut {NoStop}%
\end{thebibliography}%


\appendix

\section{Robustness to the subset size in MCDJ}
\label{app:MCDJ_22}

In Section~\ref{subsec:Monte Carlo delete d jackknife} we applied the MCDJ resampling scheme by selecting 26 out of the 33 CC data points in each realization. This choice retains about 80\% of the sample and provides a compromise between keeping enough data to constrain $H_0$ well and deleting a non-negligible number of points so that the effect of data removal is clearly visible. To test whether our conclusions depend sensitively on this particular subset size, we performed an additional MCDJ run in which only 22 of the 33 CC points are retained in each realization, using 500 realizations for each method.

For this 22/33 setup we computed the same three summary statistics as in Table~2: the absolute difference between the mode of the resampled $H_0$ distribution and the full-data value, $\Delta H_{0,\mathrm{mode-CC}}$, the corresponding difference for the median, $\Delta H_{0,\mathrm{median-CC}}$, the width of the central $2\sigma$ range, $\Delta H_{0,2\sigma}$. The results are listed in Table~\ref{tab:MCDJ_test_22_33}. Compared to the 26/33 case, the numerical values of these quantities change only moderately, reflecting the stronger deletion of data and the smaller number of realizations. However, the qualitative pattern is the same: for all three metrics we still find $\Delta H_{0}^{\mathrm{EMCEE}} < \Delta H_{0}^{\mathrm{MAF}} < \Delta H_{0}^{\mathrm{GP}}$. Thus, GP remains the most sensitive to the removal of CC points, MAF is intermediate, and EMCEE is the least sensitive, , as already found in Section~\ref{subsec:Monte Carlo delete d jackknife} for the 26/33 case. This additional test confirms that our ranking of the three methods by sensitivity to CC data is robust to the precise fraction of points retained in the MCDJ resampling.


\begin{table}
\centering
\caption{Comparing results using MCDJ with 22/33 CC points (500 realizations). Column definitions are the same as in Table~\ref{tab:MCDJ_33}.}
\label{tab:MCDJ_test_22_33}
\begin{tabular}{lccc}
\hline
Methods & $\Delta H_{0,\mathrm{mode-CC}}$
        & $\Delta H_{0,\mathrm{median-CC}}$
        & $\Delta H_{0,2\sigma}$ \\
\hline
EMCEE & 0.55 & 0.20 & 8.15 \\
GP    & 2.91 & 1.50 & 14.03 \\
MAF   & 1.05 & 0.58 & 8.66 \\
\hline
\end{tabular}
\end{table}


\section{Robustness to the selection fraction in the redshift split}
\label{app:2redshift_27}

In Section~\ref{subsec:Seperate to different areas} we quantified the relative sensitivity of the posterior central value of $H_0$ to low- and high-redshift CC data by applying the MCDJ scheme to Partition 2, selecting 13 of 17 low-$z$ points (or 13 of 16 high-$z$ points) in each realization. This corresponds to retaining 29 of the 33 CC points and led to the $\Delta H_{0,1\sigma,\mathrm{L-H}}$ and $\Delta H_{0,2\sigma,\mathrm{L-H}}$ values listed in Table~\ref{tab:2_separate}, from which we concluded that EMCEE and GP are more sensitive to the high-redshift region, while MAF is more sensitive to the low-redshift region.

To test how strongly this conclusion depends on the chosen resampling fraction, we performed an additional set of MCDJ runs with a more aggressive deletion in each region. In the low-$z$ case we selected 11 of the 17 low-$z$ points in each realization while keeping all high-$z$ points fixed; in the high-$z$ case we selected 11 of the 16 high-$z$ points while keeping all low-$z$ points fixed. For each configuration we generated 500 realizations and recomputed the 1$\sigma$ and 2$\sigma$ ranges of $H_0$ in the low- and high-redshift subsets and their differences $\Delta H_{0,1\sigma,\mathrm{L-H}}$ and $\Delta H_{0,2\sigma,\mathrm{L-H}}$.

The results are summarised in Table~\ref{tab:2_separate_27}. Compared with Table~\ref{tab:2_separate}, the absolute values of $\Delta H_{0,1\sigma,\mathrm{L-H}}$ and $\Delta H_{0,2\sigma,\mathrm{L-H}}$ move closer to zero, as expected when fewer CC points are deleted. Nevertheless, the 1$\sigma$ indicators retain the same qualitative pattern: EMCEE and GP show larger sensitivity to the high-$z$ region (negative $\Delta H_{0,1\sigma,\mathrm{L-H}}$), while MAF shows slightly larger sensitivity to the low-$z$ region (positive $\Delta H_{0,1\sigma,\mathrm{L-H}}$). The 2$\sigma$ indicators become weaker and, in some cases, nearly symmetric around zero, reflecting the increased influence of statistical fluctuations in the distribution tails when the perturbation is small. Overall, this additional test indicates that the main redshift–dependence found in Section~\ref{subsec:Seperate to different areas} does not rely on the exact resampling fraction, although the strength of the effect decreases as fewer data points are removed.

\begin{table}
    \centering
    \caption{Comparison of $H_0$ sensitivity between two redshift regions for the additional MCDJ runs with 27/33 CC points (500 realizations). Column definitions are the same as in Table~\ref{tab:2_separate}.}
    \label{tab:2_separate_27}
    \begin{tabular}{ccc}
        \hline
        Methods & $\Delta H_{0,1\sigma,\mathrm{L-H}}$ & $\Delta H_{0,2\sigma,\mathrm{L-H}}$ \\
        \hline
        EMCEE & -1.09 & -0.12 \\
        GP & -2.57 & 2.36 \\
        MAF & 0.25 & -0.05 \\
        \hline
    \end{tabular}
\end{table}


\section{Additional sanity checks for the GP bias}
\label{app:SanityChecks}

To further examine the bias in the GP posterior central values of $H_0$, we compare it with realization-by-realization low-$z$ fluctuations of the simulated mock-data points, using the normalized residual defined in equation~(\ref{eq:normalized_residual}). For each realization, we define the mean low-$z$ normalized residual as
\begin{equation}
\bar{r}_{\mathrm{low}} \equiv \frac{1}{n}\sum_{i=1}^{n} r_i,
\label{eq:meanNormalizesResidual}
\end{equation}
where $n$ is the number of low-$z$ points included in the subset. In the present analysis, we take $n=5$, corresponding to the five smallest-redshift points in each realization.

\begin{figure}
    \centering
    \includegraphics[width=8cm]{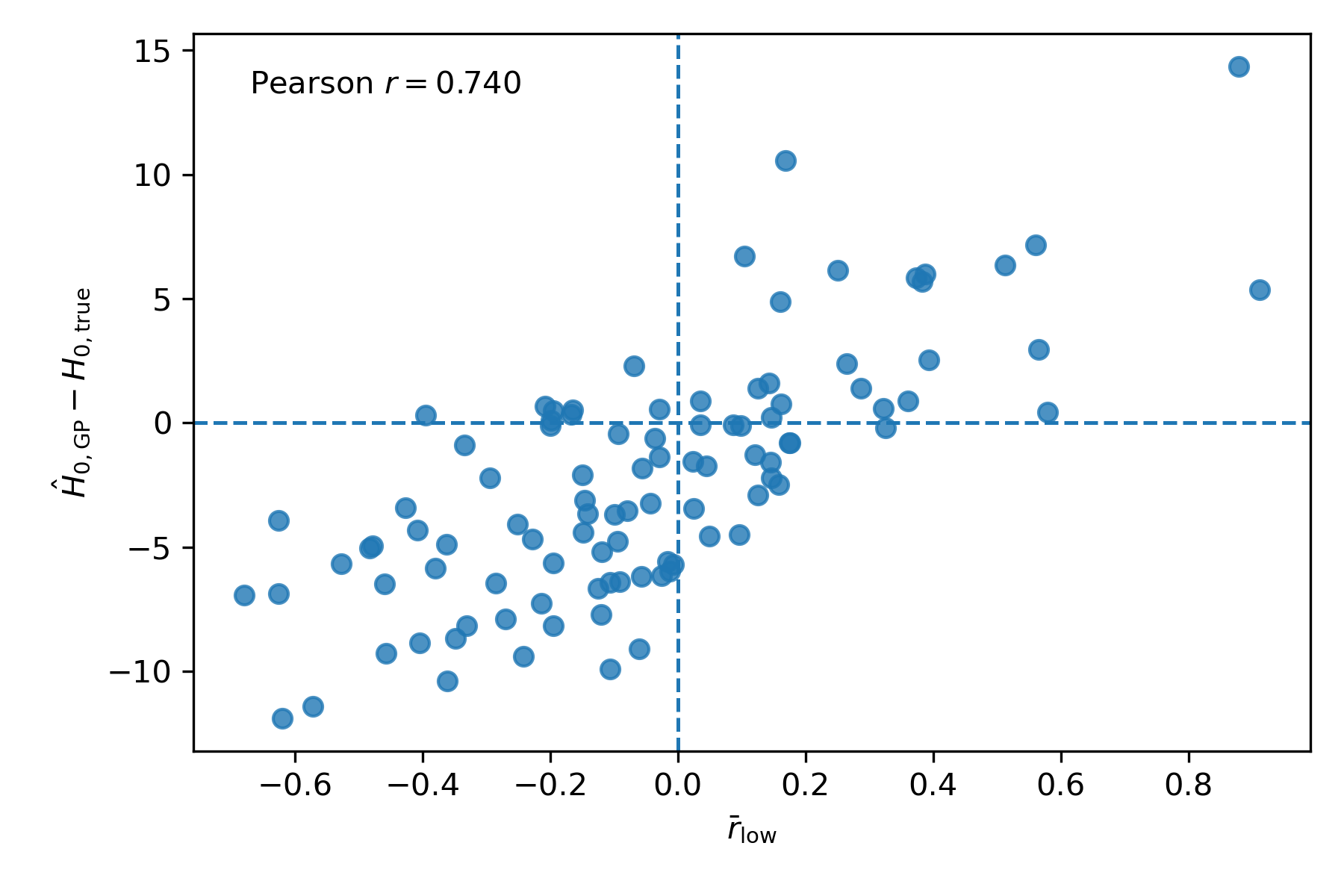}
    \caption{Relation between GP bias and mean low-$z$ normalized residual $\bar{r}_{\mathrm{low}}$ in the $\Lambda$CDM-based simulations. Here $\hat{H}_{0,\mathrm{GP}}$ denotes the GP posterior central value of $H_0$ for a given realization, and $H_{0,\mathrm{true}}$ is the input value used to generate the mocks. In each realization, the low-$z$ subset is defined as the five smallest-redshift points.}   \label{fig:LCDMbased_sanitycheck_A2_GP_bias_vs_lowz_r}
\end{figure}

\begin{figure}
    \centering
    \includegraphics[width=8cm]{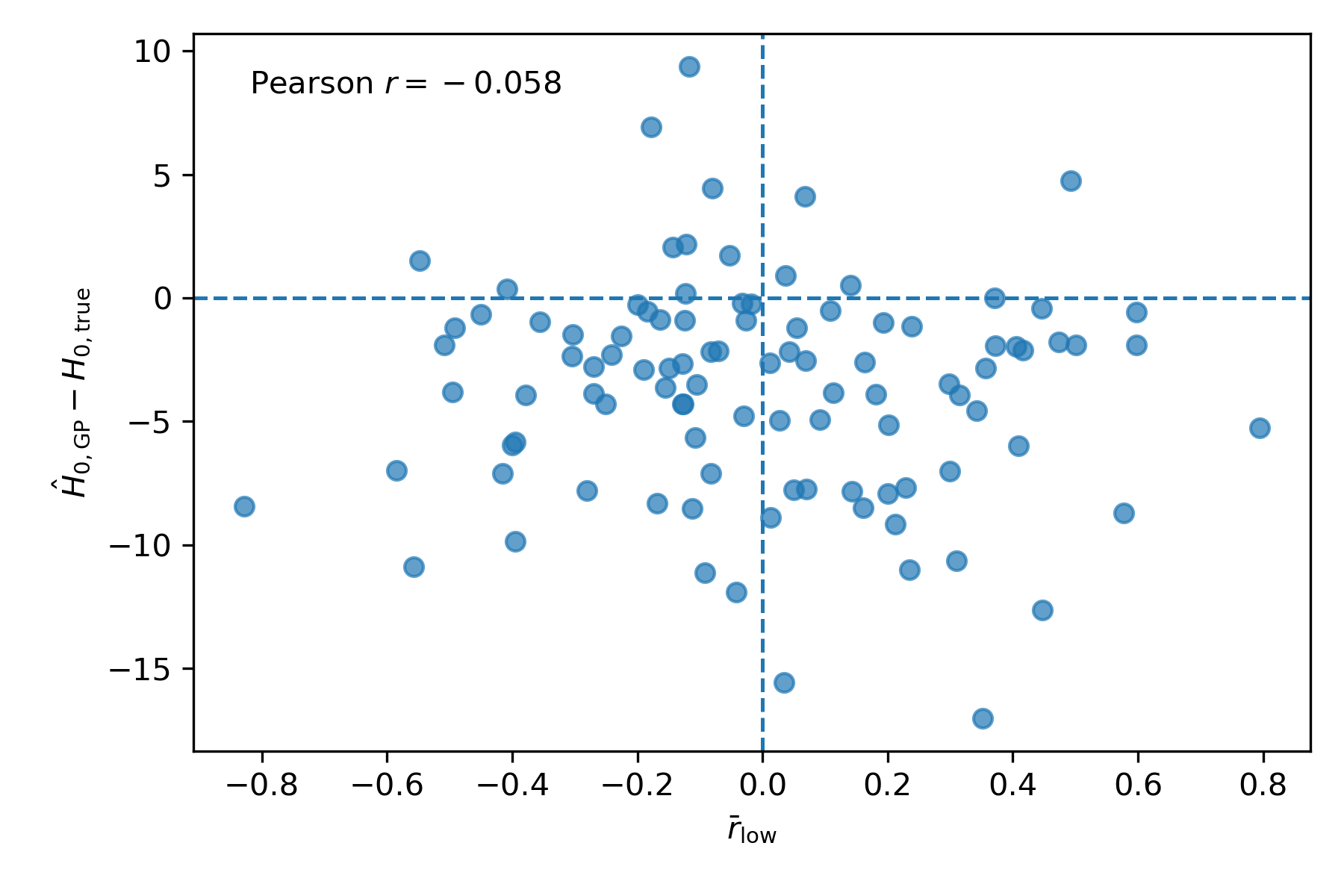}
    \caption{Relation between GP bias and mean low-$z$ normalized residual $\bar{r}_{\mathrm{low}}$ in the GP-based simulations. Here $\hat{H}_{0,\mathrm{GP}}$ denotes the GP posterior central value of $H_0$ for a given realization, and $H_{0,\mathrm{true}}$ is the input value used to generate the mocks. In each realization, the low-$z$ subset is defined as the five smallest-redshift points.}    \label{fig:GPbased_sanitycheck_A2_GP_bias_vs_lowz_r}
\end{figure}

Fig.~\ref{fig:LCDMbased_sanitycheck_A2_GP_bias_vs_lowz_r} shows that, for the $\Lambda$CDM-based mocks, the GP bias is positively correlated with $\bar{r}_{\mathrm{low}}$. This indicates that, in that simulation framework, the GP extrapolation to $z=0$ is sensitive to realization-by-realization low-$z$ fluctuations. Fig.~\ref{fig:GPbased_sanitycheck_A2_GP_bias_vs_lowz_r} shows that, for the GP-based mocks, the negative GP bias persists, but without an equally strong correlation with $\bar{r}_{\mathrm{low}}$. Taken together, these appendix results support the conclusion in Section~\ref{sec:discussions and conclusions} that the GP bias is robust across the two simulation frameworks, but it does not manifest in the same way in the $\Lambda$CDM-based and GP-based simulations.


\label{lastpage}
\end{CJK*}
\end{document}